\documentclass[11pt]{article}
\usepackage{graphicx}
\usepackage{amsmath}
\usepackage{amsthm}

\DeclareFontFamily{U}{mathx}{\hyphenchar\font45}
\DeclareFontShape{U}{mathx}{m}{n}{<-> mathx10}{}
\DeclareSymbolFont{mathx}{U}{mathx}{m}{n}
\DeclareMathAccent{\widebar}{0}{mathx}{"73}

\usepackage{amssymb}
\usepackage{amsfonts}
\usepackage{mathtools}
\usepackage[english]{babel}
\usepackage{footnote}
\makesavenoteenv{tabular}
\usepackage{floatpag}
\usepackage{graphicx,psfrag,epsf}
\usepackage{enumerate}
\usepackage[noend]{algpseudocode}
\usepackage{natbib}
\usepackage{url} 
\usepackage{comment}
\usepackage{booktabs}   
\usepackage{bm}
\usepackage{enumitem}

\newcommand{\rmnum}[1]{\romannumeral #1}
\usepackage{lineno}
\usepackage{algorithm}
\usepackage{algpseudocode}
\usepackage{caption}
\usepackage{subcaption}
\usepackage{multirow}
\usepackage{threeparttable}
\usepackage{wasysym}
\usepackage{framed}
\usepackage{pgfgantt,rotating}
\usepackage{multirow}
\usepackage{blindtext}
\usepackage{times}
\usepackage[above]{placeins}

\usepackage{float}
\usepackage[outdir=./]{epstopdf}
\usepackage{verbatim} 

\usetikzlibrary{arrows,shapes.arrows,shapes.geometric,shapes.multipart,backgrounds,decorations.pathmorphing,positioning,fit,automata}
\tikzset{
	>=stealth',
	true/.style={
		rectangle,
		draw=black, very thick,
		text width=6.5em,
		minimum height=2em,
		text centered,
		fill=gray, opacity = 0.5},
	punkt/.style={
		rectangle,
		rounded corners,
		draw=black, very thick,
		text width=6.5em,
		minimum height=2em,
		text centered},
	est/.style={
		circle,
		draw=black, very thick,
		text centered},
			estblue/.style={
		circle,
		draw=blue, very thick,
		text centered},
	shade/.style={
		circle,
		draw=white, very thick, fill=gray!50,
		text centered},
	weight/.style={
		circle,
		draw=black, very thick,
		text width=6.5em,
		minimum height=2em,
		text centered},
	pil/.style={
		->,
		thick,
		shorten <=2pt,
		shorten >=2pt,},
	pilred/.style={
		->, red,
		thick,
		shorten <=2pt,
		shorten >=2pt,},
	pilblue/.style={
		->,blue,
		thick,
		shorten <=2pt,
		shorten >=2pt,},
	pilgray/.style={
		dashed,gray,
		thick,
  linewidth=0.3mm,
		shorten <=2pt,
		shorten >=2pt,},
	double/.style={
		<->,
		thick,
		shorten <=2pt,
		shorten >=2pt,},
	dash/.style={
		dashed,
		thick,
		shorten <=2pt,
		shorten >=2pt,},
	dashdouble/.style={
		<->,
		dashed,
		thick,
		shorten <=2pt,
		shorten >=2pt,}
}
\DeclareMathOperator*{\argmin}{arg\,min}

\newtheorem{example}{Example} 
\newtheorem{theorem}{Theorem}
\newtheorem{lemma}{Lemma} 
\newtheorem{proposition}{Proposition} 
\newtheorem{remark}{Remark}

\newtheorem{definition}{Definition}
\newtheorem{observation}{Observation}

\newtheorem{assumption}{Assumption}

\newcommand{\ind}{\perp \!\!\! \perp }

\newcommand{\betaz}{\beta}

\def\T{{ \mathrm{\scriptscriptstyle T} }}

\usepackage{graphicx}
\usepackage{enumerate}
\usepackage{natbib}
\usepackage{url} 

\newcommand{\blind}{1}

\usepackage{xr}
\makeatletter

\newcommand*{\addFileDependency}[1]{
\typeout{(#1)}
\@addtofilelist{#1}
\IfFileExists{#1}{}{\typeout{No file #1.}}
}\makeatother

\begin{document}

\def\spacingset#1{\renewcommand{\baselinestretch}%
{#1}\small\normalsize} \spacingset{1}


\if1\blind
{
  \title{\bf \Large{The synthetic instrument: From sparse association to sparse causation}}
   \date{}
   \author{}
  \maketitle

\begin{center}
   \vspace{-10pt}
  \author{\large Dingke Tang$^{1}$, Dehan Kong$^{2}$, and Linbo Wang$^{2}$\footnote{Address for correspondence: Linbo Wang,      Department of Statistical Sciences, University of Toronto, 700 University Avenue, 9th Floor,
Toronto, ON, Canada, M5G 1Z5 \\
Email: linbo.wang@utoronto.ca} \\
  \vspace{10pt}
  $^{1}$Department of Mathematics and Statistics, University of Ottawa, Ottawa, Ontario, Canada\\
  $^{2}$ Department of Statistical Sciences, University of Toronto, Toronto, Ontario, Canada}\\ \vspace{20pt}

\end{center}

} \fi

\if0\blind
{
  \bigskip
  \bigskip
  \bigskip
  \begin{center}
    {\LARGE\bf }
\end{center}
  \medskip
} \fi

\vspace{-2em}

\spacingset{1.5} 
\begin{abstract}
In many observational studies, researchers are often interested in the effects of multiple exposures on a single outcome. Standard approaches for high-dimensional data, such as the Lasso, assume that the associations between the exposures and the outcome are sparse. However, these methods do not estimate causal effects in the presence of unmeasured confounding. In this paper, we consider an alternative approach that assumes the causal effects under consideration are sparse. We show that under sparse causation, causal effects are identifiable even with unmeasured confounding. Our proposal is built around a novel device called the synthetic instrument, which, in contrast to standard instrumental variables, can be constructed directly from the observed exposures. We demonstrate that, under the assumption of sparse causation, the problem of causal effect estimation can be formulated as an $\ell_0$-penalization problem and solved efficiently using off-the-shelf software. Simulations show that our approach outperforms state-of-the-art methods in both low- and high-dimensional settings. We further illustrate our method using a mouse obesity dataset.
\end{abstract}

\noindent%

\spacingset{1.5} 

\section{Introduction}
\label{sec:intro}

Sparsity is a common assumption in the modern statistical learning literature, as it facilitates variable selection in models and enhances the interpretability of parameter estimates. For example, the Lasso \citep{tibshirani1996regression} assumes sparse associations between a single outcome and potentially high-dimensional predictors; in other words, only a small subset of predictors have nonzero associations with the outcome. \cite{hastie2009elements} summarizes the philosophy behind such methods as the ``bet on sparsity'' principle: use a procedure that performs well in sparse settings, since no procedure performs well in dense ones. Methods like the Lasso perform well under sparse associations and, as a result, have gained significant popularity in recent decades.  

Importantly, the ``bet on sparsity'' principle does not restrict the types of problems to which sparsity may apply. Beyond sparse associations, a growing body of literature emphasizes \emph{sparse causation}, where only a fraction of exposures exert nonzero causal effects on the outcome \citep[e.g.,][]{spirtes1991algorithm,claassen2013learning,wang2017confounder,miao2023identifying,zhou2024promises}. This assumption is often more interpretable and plausible in real data applications. For example, suppose we are interested in the relationship between gene expression and a phenotype such as lung cancer. Biological evidence suggests that only a small proportion of genes may influence the risk of lung cancer \citep[e.g.,][]{kanwal2017familial}. However, this does not imply sparse association, since unmeasured confounding may induce spurious correlations between many genes and the phenotype.  

To illustrate, consider a linear structural model \citep{pearl2013linear} with a $p$-dimensional exposure vector $X = (X_1,\ldots,X_p)^\T$, an outcome $Y$, and a $q$-dimensional latent variable $U$:  
\begin{flalign}
    X &= \Lambda U + \epsilon_{x}, \label{eqn:model1} \\
    Y &= X^\T \betaz + U^\T \gamma + \epsilon_{y}, \label{eqn:model-lineary} 
\end{flalign}  
where $\Lambda \in \mathbb{R}^{p\times q}$, $\betaz \in \mathbb{R}^p$, and $\gamma \in \mathbb{R}^q$ are coefficient vectors, and $\epsilon_x = (\epsilon_1,\ldots,\epsilon_p)^\T$, $\epsilon_y$, and $U$ are mutually uncorrelated. Under this model, spurious correlations induced by unmeasured confounding, given by $\text{Cov}(X)^{-1}\Lambda\gamma$, are typically dense. Consequently, the overall association between $X$ and $Y$ is dense, even when the causal effect $\beta$ itself is sparse.

Identification and estimation of the causal parameter $\beta$ are nontrivial due to the presence of unmeasured confounding by $U$. The contributions of this paper are twofold. First, under an additional plurality condition, we establish that the parameter $\beta$ in model \eqref{eqn:model2} is identifiable \emph{if and only if} $\|\betaz\|_0 < p - q$. This sparsity assumption is both necessary and sufficient, representing a significant improvement over conditions previously introduced in the literature; see Section~\ref{sec:related-works} for details. Remarkably, in contrast to many other identification assumptions in causal inference, this assumption can be consistently tested from data.  

Second, we develop a two-stage synthetic regularized regression approach for estimating $\beta$, with a first stage based on ordinary least squares and a second stage using $\ell_0$-penalized regression. The key technique behind our results is a novel device, which we term the \emph{synthetic instrument}. Unlike standard instrumental variables, the synthetic instrument is constructed from a subset of exposures, enabling identification of causal effects without requiring exogenous variables. Our procedure enjoys Lasso-type theoretical guarantees in both low- and high-dimensional settings.

\subsection{Related works}
\label{sec:related-works}
Our proposal is related to recent work on multivariate hidden confounding. \cite{cevid2020spectral} and \cite{guo2022doubly} propose a spectral deconfounding method for estimating $\beta$ in a high-dimensional model. Their method assumes a dense confounding structure, which is feasible only in a high-dimensional regime where $p$ tends to infinity with the sample size and the magnitudes of spurious associations tend to zero. \cite{bing2022adaptive} consider a more general setup than the one we study, in which they also allow the outcome to be multivariate. However, they aim to identify the projection of $\beta$ onto a related space rather than the causal parameter $\beta$ itself. \cite{chandrasekaran2010latent} study a related problem under the assumption that $(X, Y)$ are normally distributed. Under this assumption, they not only identify the effect of $X$ on $Y$ but also recover the covariance among components of $X$ conditional on $U$.

The estimation problem for $\beta$ can be framed within the context of causal inference with unmeasured confounding. Currently, the most popular approach in practice is the instrumental variable (IV) framework, which uses information from an exogenous variable known as an IV to identify causal effects \citep[e.g.,][]{angrist1996identification, wang2018bounded, pfister2022identifiability}. Another approach that has gained attention recently is the proximal causal inference framework \citep{tchetgen2024introduction}, which uses information from ancillary variables, known as negative control exposures and outcomes, to remove bias due to unmeasured confounding. Compared with these frameworks, our approach does not rely on the collection of additional ancillary variables, which can be challenging in many practical settings. Instead, we rely on the availability of multiple exposures and the sparsity assumption for identification and estimation.

Recently, a strand of literature has sought to identify the causal effects of multiple exposures. \cite{wang2019blessings} popularized this setting by proposing the so-called deconfounder method, which first obtains an estimate $\widehat{U}$ of the unmeasured confounder and then adjusts for $\widehat{U}$ using standard regression methods. However, it has been pointed out that in this setting, without further assumptions, the causal effect $\beta$ is not identifiable \citep{d2019multi, ogburn2020counterexamples}. \cite{kong2022identifiability} show that under model \eqref{eqn:model1} and a binary choice model for the outcome with a non-probit link, the causal effects are identifiable. Their identification results, however, apply only to binary outcomes and do not lead to straightforward estimation procedures. \cite{miao2023identifying} consider a similar setting to \eqref{eqn:model1} and \eqref{eqn:model2}, showing that the causal effect is identifiable if $\Vert \beta \Vert_0 \leq (p-q)/2$. Their sparsity constraint is significantly stronger than ours, especially when the number of exposures is large relative to the number of latent confounders. \cite{miao2023identifying} also develop a robust linear regression-based estimator for $\beta$. In contrast to our estimator, their estimator is consistent only in the low-dimensional regime where $p$ is fixed and $\Vert \beta \Vert_0 \leq p/2 - q + 1$. Furthermore, their estimator for $\beta$ is not sparse and therefore cannot be used for selecting treatments with nonzero effects.

Our results also connect to recent literature on multiply robust causal identification \citep[e.g.,][]{sun2021semiparametric}, as we show identification in the union of many causal models. This contrasts with the extensive literature on multiply robust estimators under the same causal model \citep[e.g.,][]{wang2018bounded} and on improved doubly robust estimators that are consistent under multiple working models for two components of the likelihood \citep[e.g.,][]{han2013estimation}.
\subsection{Outline of this paper}
The rest of this article is organized as follows. In Section \ref{sec:framework}, we introduce the setup and background. In Section \ref{sec:identification}, we describe our identification strategy using the synthetic instrument method. In Section \ref{sec:esti}, we present our estimation procedure and provide theoretical justifications. We also discuss extensions to nonlinear outcome models. Simulation studies in Section \ref{sec:simulation} compare our proposal with several state-of-the-art methods in finite-sample performance. In Section \ref{sec:data}, we apply our method to mouse obesity data. We conclude with a brief discussion in Section \ref{sec:discussion}.

The proposed method is implemented in an {\tt R} package, available at \url{https://github.com/dingketang/syntheticIV}.
\section{Framework, notation, and identifiability}\label{sec:framework}

\subsection{The model}
\label{sec:the_model}
We assume that we observe  \( n \) independent samples from the joint distribution of  \( (X, Y) \). Consider structural model \eqref{eqn:model1} and  
\begin{align}
    Y &= X^\T \beta + g(U) + \epsilon_y\label{eqn:model2}.
\end{align}
Here,  \( g(U) : \mathbb{R}^q \to \mathbb{R} \) is a measurable function encoding the effects of unmeasured confounders  \( U \) on the outcome  \( Y \). We do not assume knowledge of the functional form of  \( g(U) \) because  \( U \) is unmeasured, making it implausible to specify the exact form of  \( g(\cdot) \).

We start with a linear outcome model where the treatment effect is linear in  \( X \). In Section \ref{sec:nonlinear_main}, we will consider a nonlinear treatment effect model, where the relationship between treatment  \( X \) and outcome  \( Y \) is represented by a potentially nonlinear function  \( f(X; \beta) \).

We consider both low- and high-dimensional settings, where  \( p \) may be smaller or larger than the sample size  \( n \). Let  \( \dot{\beta} \) denote the true value of  \( \beta \) in model \eqref{eqn:model2}. Without loss of generality, we assume all the variables in \eqref{eqn:model1} and \eqref{eqn:model2} are centered,  \( \text{Cov}(U) = I_q \), and  \( \mathbb{E}(g(U)) = 0 \).

We maintain the following conditions throughout the article.
\begin{enumerate}[label={A\arabic*}]
    \item \label{A1} (Invertibility) Any  \( q \times q \) submatrix of  \( \text{Cov}^{-1}(X)\Lambda \) is invertible.
    \item \label{A2}  \( \Lambda \) is identifiable up to a rotation.
\end{enumerate}
Condition \ref{A1} is a regularity condition commonly assumed in the literature \citep[e.g.,][Theorem 3]{miao2023identifying}. However, it may be relaxed in our setting. For example, if certain treatment effects are unconfounded after normalization, so that specific rows of \( \text{Cov}^{-1}(X)\Lambda \) are zero, then even though Condition \ref{A1} is violated, our proposed method can still be used to identify and estimate the treatment effects. See Sections \ref{sec:discussA1} and \ref{sec:weakivandidentification} of the supplementary material for further details on how the algorithm identifies treatment effects under relaxed versions of \ref{A1}.

Condition \ref{A2} has been discussed extensively in the factor model literature. One classical result is Proposition \ref{prop:loading identifiability}, which is a direct corollary of \citet[Theorem 5.1]{anderson1956statistical}.

\begin{proposition}
\label{prop:loading identifiability}
Under models \eqref{eqn:model1}, \eqref{eqn:model2}, and Condition \ref{A1}, if  \( p \geq 2q+1 \) and  \( D = \text{Cov}(\epsilon_x) \) is a diagonal matrix, then  \( \Lambda \) is identifiable up to a rotation.
\end{proposition}

We note that the condition that  \( D \) be a diagonal matrix is a classical assumption in the factor analysis literature. However, Condition \ref{A2}, and hence our algorithm, may still hold even if  \( D \) is not diagonal. For example, under the assumption that  \( D \) is sparse, the covariance structure  \( \text{Cov}(X) = D + \Lambda \Lambda^\T \) implies a sparse plus low-rank decomposition. This allows for the identification of the low-rank component  \( \Lambda \Lambda^\T \), as established in \citet[][Corollary 3]{chandrasekaran2011rank}, which leads to identifying  \( \Lambda \) up to a rotation. In another example, in the high-dimensional setting where  \( p \rightarrow \infty \), it is possible to identify  \( \Lambda^* \in \mathbb{R}^{p \times q} \), whose columns correspond to the top  \( q \) eigenvalues of  \( \text{Cov}(X) \). Under additional boundedness assumptions on the correlation matrix  \( D \) and the coefficient matrix  \( \Lambda \), one can show that there exists a matrix  \( O \in \mathbb{R}^{q \times q} \) such that the  \( \ell_2 \)-norm between each column of  \( \Lambda O \) and  \( \Lambda^* \) converges to zero as  \( p \) tends to infinity. See \citet[][Proposition 2.2, Theorem 3.3]{fan2013large}, \citet[][Theorem 2]{bai2003inferential}, and \citet[][Theorem 1]{shen2016general} for more details.

\subsection{Identifiability of the causal effect $\beta$}
\label{sec:identifiability}
In this section, we discuss the identifiability of the causal parameter  \( \beta \) in \eqref{eqn:model2}. We illustrate the key ideas using the specific example where  \( p = 3 \) and  \( q = 1 \) in models \eqref{eqn:model1} and \eqref{eqn:model2}. Figure \ref{fig:three cases} provides graphical illustrations.

First, note that without additional assumptions,  \( \beta \) is generally not identifiable due to unmeasured confounding by  \( U \). To see this, observe that under models \eqref{eqn:model1} and \eqref{eqn:model2}, we have 
\begin{equation}
\label{eqn:idenbeta}
    \text{Cov}(X_j, Y) = \beta_1 \text{Cov}(X_j, X_1) + \beta_2 \text{Cov}(X_j, X_2) + \beta_3 \text{Cov}(X_j, X_3) + \gamma \Lambda_j, \quad j = 1, 2, 3,
\end{equation}
where  \( \Lambda_j \) is the  \( j \)th element of  \( \Lambda \in \mathbb{R}^{p \times 1} \) and  \( \gamma = \mathbb{E}(U g(U)) \). Since there are three equations in \eqref{eqn:idenbeta} but four unknown parameters,  \( \beta_1, \beta_2, \beta_3, \gamma \), the causal parameters  \( \beta \) are not identifiable from these equations.

One possible approach to identifying  \( \beta \) is to assume prior knowledge about certain elements of  \( \beta \). For instance, in Figure \ref{fig:p=3,s=2}, it is assumed that  \( \beta_2 = 0 \), meaning that  \( X_2 \) has no causal effect on the outcome  \( Y \). In this scenario, it is straightforward to see from \eqref{eqn:idenbeta} that under Conditions \ref{A1} and \ref{A2},  \( \beta_1 \),  \( \beta_3 \), and  \( |\gamma| \) are identifiable.

\begin{figure}[!htbp]
    \centering{
    \begin{subfigure}{0.3\textwidth}
  \centering
  \scalebox{0.75}{
  \begin{tikzpicture}[->,>=stealth',shorten >=1pt,auto,node distance=2.3cm,
     semithick, scale=0.50]
     pre/.style={-,>=stealth,semithick,blue,ultra thick,line width = 1.5pt}]
     \tikzstyle{every state}=[fill=none,draw=black,text=black]
    \node[shade] (U)                    {$U$};
     \node[] (X2) [below of = U] {$X_2$};
     \node[] (X1) [left  = 1cm of X2] {$X_1$};
     \node[] (X3) [right  = 0.8cm of X2] {$X_3$};
     \node[] (Y) [below of =  X2] {{$Y$}};
    \path  (U) edge [left] node {$\Lambda_1$} (X1);
    \path (U) edge [left] node {$\Lambda_2$} (X2);
    \path (U) edge node {$\Lambda_3$} (X3);
    \path (X1) edge [left] node {${\beta_1}$} (Y);
    \path (X2) edge [left] node {${\beta_2}$} (Y);
    \path (X3) edge [] node {${\beta_3}$}  (Y);
    \draw (0.8,0) arc (90:-90:4.5);
     \node[] (m) [right = 2.2cm of X2] {${\gamma}$};
     (U) edge [bend left = 90] node {} (Y);
    \end{tikzpicture}
    }
    \caption{No additional assumptions.}
     \label{fig:p=3,s=3}
  \end{subfigure}
  \hfill
  \begin{subfigure}{0.3\textwidth}
  \centering
    \scalebox{0.75}{\begin{tikzpicture}[->,>=stealth',shorten >=1pt,auto,node distance=2.3cm,
     semithick, scale=0.50]
     pre/.style={-,>=stealth,semithick,blue,ultra thick,line width = 1.5pt}]
     \tikzstyle{every state}=[fill=none,draw=black,text=black]
    \node[shade] (U)                    {$U$};
     \node[] (X2) [below of = U] {$X_2$};
     \node[] (X1) [left  = 1cm of X2] {$X_1$};
     \node[] (e1) [left  = 0.5cm of X1] {$\epsilon_1$};
     \node[] (X3) [right  = 0.8cm of X2] {$X_3$};
     \node[] (Y) [below of =  X2] {{$Y$}};
    \path  (U) edge [left] node {$\Lambda_1$} (X1);
    \path  (e1) edge [left] node {} (X1);
    \path (U) edge [left] node {$\Lambda_2$} (X2);
    \path (U) edge node {$\Lambda_3$} (X3);
    \path (X1) edge [left] node {${\beta_1}$} (Y);
    \path (X3) edge [] node {${\beta_3}$}  (Y);
    \draw (0.8,0) arc (90:-90:4.5);
     \node[] (m) [right = 2.2cm of X2] {${\gamma}$};
     (U) edge [bend left = 90] node {} (Y);
    \end{tikzpicture}}
    \caption{Assume $\beta_2=0$.}
    \label{fig:p=3,s=2}
  \end{subfigure}  
  \hfill
  \begin{subfigure}{0.3\textwidth} 
  \centering  
  \scalebox{0.75}{
  \begin{tikzpicture}[->,>=stealth',shorten >=1pt,auto,node distance=2.3cm,
     semithick, scale=0.50]
     pre/.style={-,>=stealth,semithick,blue,ultra thick,line width = 1.5pt}]
     \tikzstyle{every state}=[fill=none,draw=black,text=black]
    \node[shade] (U)                    {$U$};
     \node[] (X2) [below of = U] {$X_2$};
     \node[] (X1) [left  = 1cm of X2] {$X_1$};
     \node[] (X3) [right  = 0.8cm of X2] {$X_3$};
     \node[] (Y) [below of =  X2] {{$Y$}};
    \path  (U) edge [left] node {$\Lambda_1$} (X1);
    \path (U) edge [left] node {$\Lambda_2$} (X2);
    \path (U) edge node {$\Lambda_3$} (X3);
\path 
      (X1) edge [dashed,left] node {$ {\beta_1}$} (Y)
     (X2) edge [dashed,left] node {$ {\beta_2}$} (Y)
     (X3) edge [dashed] node {$ {\beta_3}$}  (Y);
    \draw (0.8,0) arc (90:-90:4.5);
     \node[] (m) [right = 2.2cm of X2] {${\gamma}$};
     (U) edge [bend left = 90] node {} (Y);
    \end{tikzpicture}
    }
    \caption{Assume $||\beta||_0\leq 1$.}
    \label{fig:p=3,s=1}
  \end{subfigure}
  }
    \caption{Causal diagrams corresponding to models \eqref{eqn:model1} and \eqref{eqn:model2}: $p = 3, q=1.$ 
    }
    \label{fig:three cases}
\end{figure}

In practice, however, it is often difficult to know which exposures have zero causal effects \emph{a priori}. In this paper, we instead consider the following sparsity assumption; see Figure \ref{fig:three cases}c for an illustration.

\begin{enumerate}[label={A3}]
    \item \label{A3} (Sparsity)  \( \Vert \dot{\beta} \Vert_0 \leq p - q - 1 \), where  \( \dot{\beta} \) denotes the true value of  \( \beta \).
\end{enumerate}

\begin{remark}
   Condition \ref{A3} is significantly less restrictive than similar assumptions in the existing literature used for causal effect identification in this context. For example, \cite{miao2023identifying} assumed  \( \Vert \dot{\beta} \Vert_0 \leq (p-q)/2 \).
\end{remark}

\subsection{Instrumental variable }
\label{sec:iv}
The method of instrumental variables is a widely used approach for estimating causal relationships when unmeasured confounders exist between the exposure  \( X \) and the outcome  \( Y \). Suppose we have an exogenous variable  \( Z \). For simplicity, assume that the relationships among the random variables are linear and follow the structural equation models:
\begin{equation*}
    \begin{split}
        Y &= \beta X + \gamma U + \pi Z + \epsilon_y,\\
        X &= \alpha_z Z + \Lambda U + \epsilon_x.
    \end{split}
\end{equation*}
For  \( Z \) to be a valid instrumental variable, the following assumptions are commonly made \citep[e.g.,][]{wang2018bounded}:  
 \( \pi = 0 \) (exclusion restriction),  \( \alpha_z \neq 0 \) (instrumental relevance), and  \( \text{Cov}(U, Z) = 0 \) (unconfoundedness). Under these assumptions, one can consistently estimate  \( \beta \) via a two-stage least squares estimator: first, obtain the predicted exposure  \( \widehat{\mathbb{E}}(X \mid Z) \) by linearly regressing  \( X \) on  \( Z \), and then regress  \( Y \) on  \( \widehat{\mathbb{E}}(X \mid Z) \) to obtain an estimate of  \( \beta \). Here,  \( \mathbb{E}(X \mid Z) \) refers to the conditional expectation of  \( X \) given  \( Z \), and  \( \widehat{\mathbb{E}}(X \mid Z) \) refers to its estimator obtained through linear regression.

\section{Identifying causal effects via the synthetic instrument}
\label{sec:identification}

\subsection{A new identification approach via voting}
\label{sec:voting}

We now present a new identification strategy for  \( \beta \) under the sparsity condition \ref{A3}. Consider the scenario depicted in Figure \ref{fig:p=3,s=1}, where  \( \dot{\beta}_1 = \dot{\beta}_2 = 0 \) but  \( \dot{\beta}_3 \neq 0 \). We assume that this information is unavailable to the analyst. Instead, the analyst relies on Condition \ref{A3}, assuming that  \( \Vert \dot{\beta} \Vert_0 \leq 1 \).

To explain the identification strategy, it is helpful to consider a voting analogy; see also \cite{zhou2014statistical} and \cite{guo2018confidence} for similar approaches in different contexts. Suppose the analyst consults three experts, and expert $j$ hypothesizes that $\beta_j = 0$. Based on this hypothesis, one can identify the other elements in $\beta$ using the approach described in Section \ref{sec:identifiability}. Specifically, for $j=1,2,3$, let $\widetilde{\beta}^{(j)}$ (and $|\widetilde{\gamma}^{(j)}|$) solve \eqref{eqn:idenbeta} assuming $\beta_j = 0$. Table \ref{tab:voter} summarizes these solutions. Note that the hypotheses by experts 1 and 2 are both correct, so we have $\widetilde{\beta}^{(1)} = \widetilde{\beta}^{(2)} = \betaz$ under Conditions \ref{A1}--\ref{A2}. On the other hand, the hypothesis postulated by expert 3 is incorrect. Therefore, in general, $\widetilde{\beta}^{(3)} \neq \betaz$. To decide among these three experts, we compare the solutions $\widetilde{\beta}^{(j)}$ and find their mode, defined as 
 \(
\beta_{\text{mode}} = \mathop{\arg\max}\limits_{\beta\in \mathbb{R}^3} |\{j: \widetilde{\beta}^{(j)} = \beta \}|,
\) 
where $|\mathcal{S}|$ denotes the cardinality of a set $\mathcal{S}$. One can easily see from Table \ref{tab:voter} that $\beta_{\text{mode}} = \dot{\beta}$.

\begin{table}[!htbp]
    \centering
    \caption{A voting analogy of our identification approach for $\beta$. Note $\widetilde{\beta}^{(j)} = \left( \widetilde{\beta}^{(j)}_1, \widetilde{\beta}^{(j)}_2, \widetilde{\beta}^{(j)}_3 \right)$  denotes the solution to equation \eqref{eqn:idenbeta} under the hypothesis that $\beta_j = 0$ }
    \begin{tabular}{ccccc}
   Expert index $j$ & Expert hypothesis     &  \multicolumn{3}{c}{Solution to the identification equation \eqref{eqn:idenbeta}} \\
    && $\widetilde{\beta}_1^{(j)}$ 
    & $\widetilde{\beta}_2^{(j)}$  & $\widetilde{\beta}_3^{(j)}$ \\
    \midrule
    $j=1$ & $\beta_1 = 0$  & 0  &  0 & $\dot{\beta}_3$ \\
   $j=2$ & $\beta_2 = 0$  & 0 &  0  &  $\dot{\beta}_3$ \\
   $j=3$ & $\beta_3 = 0$   & non-zero &  non-zero  &  0 \\
    \end{tabular}
    \label{tab:voter}
\end{table}

\subsection{The synthetic instrument}
\label{sec:siv}
On the surface, one may follow the identification strategy described in Section \ref{sec:voting} to estimate $\beta$. However, in the general case where $q > 1$, each expert would hypothesize that exactly $q$ elements of $\beta$ are zero. In total, there are $C_p^q$ different hypotheses. Several challenges arise when the data are moderate to high-dimensional, so that $p$ and $q$ are not small.
\begin{enumerate}
    \item [(\rmnum1)] One needs to solve the empirical version of equation \eqref{eqn:idenbeta} $C_p^q$ times. This could be computationally expensive.
    \item[(\rmnum2)] Finding the mode of $C_p^q$ $p$-dimensional estimates is a non-trivial statistical problem.
\end{enumerate}

To overcome these challenges, we introduce a new device, called the synthetic instrumental variable (SIV) method. As we shall see later, the SIV method has significant advantages in terms of both computational efficiency and identifiability for $\beta$.

\begin{remark}
  Other approaches that use the voting analogy for identification \citep[e.g.,][]{zhou2014statistical,guo2018confidence} face the same challenges we present here. It is only due to the special structure of our problem that we are able to develop a method that bypasses the model selection step and addresses these challenges. 
\end{remark}

In the following, we first introduce the SIV in the context of Figure \ref{fig:p=3,s=2}, where it is assumed that $\beta_2 = 0$. Note from Figure \ref{fig:p=3,s=2} that the error term $\epsilon_1$ serves as an instrumental variable for estimating the effect parameter $\beta_1$. However, $\epsilon_1$ is not observable. Instead, note that \eqref{eqn:model1} implies  
\begin{equation}
\label{eqn:siv} 
\begin{split}
    X_1 &= \Lambda_1 U + \epsilon_1, \\ 
    X_2 &= \Lambda_2 U + \epsilon_2,
\end{split}
\end{equation}
where $\Lambda_1$ and $\Lambda_2$ are identified up to the same sign flip, so that $\Lambda_1/\Lambda_2$ is identifiable. Eliminating $U$ from \eqref{eqn:siv}, we obtain 
 \(
X_1 - {\Lambda_2}X_2/{\Lambda_1} = \epsilon_1 - {\Lambda_1}\epsilon_2/{\Lambda_1},
\) 
which depends only on the error terms $\epsilon_1$ and $\epsilon_2$. Since $\epsilon_2$ is also uncorrelated with $U$, it is not difficult to see from Figure \ref{fig:p=3,s=2} that $SIV_1^{(2)} = X_1 - {\Lambda_1}X_2/{\Lambda_2}$ satisfies the conditions for an instrumental variable for identifying $\beta_1$ described in Section \ref{sec:iv}, hence the name synthetic instrument. In contrast to a standard instrumental variable, the synthetic instrument is directly constructed as a linear combination of the exposures, so there is no need to measure additional exogenous variables.

To identify $\beta_3$, one can similarly define $SIV_3^{(2)} = X_3 - {\Lambda_3}X_2/{\Lambda_2}$. Let $SIV^{(2)} = \left(SIV_1^{(2)}, SIV_3^{(2)}\right)$. One can then obtain $(\beta_1,\beta_3)$ using the so-called synthetic two-stage least squares:
\begin{enumerate}
    \item Fit a linear regression of $X = (X_1, X_2, X_3)$ on $SIV^{(2)} = (SIV_1^{(2)}, SIV_3^{(2)})$ and obtain $\widetilde{X} = \widehat{\mathbb{E}}[X\mid SIV^{(2)}]$ through the fitted values of the linear regression.
    \item Fit a linear regression of $Y$ on $\widetilde{X}$, fixing $\beta_2 = 0$, and obtain the coefficients $\widetilde{\beta}_1$ and $\widetilde{\beta}_3$.
\end{enumerate}

\subsection{Voting with the synthetic instrument}
\label{sec:voting2}
Now consider applying the synthetic instrument to the case in Figure \ref{fig:p=3,s=1}, where the analyst does not have prior information on which exposure has zero effect on the outcome. Instead, we assume the sparsity condition that $\|\beta\|_0 \leq 1$. 

By combining the voting procedure in Section \ref{sec:voting} with the synthetic two-stage least squares method in Section \ref{sec:siv}, we arrive at Algorithm \ref{algorithm:naive-voting} for the estimation of $\beta$.

\begin{algorithm}[!htbp]
    \caption{A naive voting procedure with synthetic two-stage least squares}
    \label{algorithm:naive-voting}
    \begin{enumerate}
        \item For $j=1,2,3$, fit a linear regression of $X$ on $SIV^{(j)}$ and obtain $\widetilde{X}^{(j)} = \widehat{\mathbb{E}}\left[X \mid SIV^{(j)}\right]$;
        \item Fit a linear regression of $Y$ on $\widetilde{X}^{(j)}$, fixing $\beta_j = 0$, and obtain the coefficients $\widetilde{\beta}^{(j)}$;
        \item Find the mode among $\widetilde{\beta}^{(j)}, \, j=1,2,3.$
    \end{enumerate}
\end{algorithm}

On the surface, similar to the problems described at the beginning of Section \ref{sec:siv}, voting with the synthetic instrument still involves fitting three different regressions and comparing three vectors $\widetilde{\beta}^{(j)}$. We now make two key observations regarding the properties of the synthetic instrument, which allow us to simplify Algorithm \ref{algorithm:naive-voting} into a two-stage regression procedure.

\begin{observation}
\label{observation:1} 
Let $\Lambda = (\Lambda_1, \Lambda_2, \Lambda_3)$. For $j=1,2,3$, $SIV^{(j)} \in \mathbb{R}^2$ span the same linear space $\{\lambda^\T X : \lambda \in \Lambda^\perp \}$. As a result, $\widetilde{X}^{(j)} = \mathbb{E}\left[X \mid SIV^{(j)}\right]$ does not depend on the choice of $j$, so that one only needs to run Step 1 of Algorithm \ref{algorithm:naive-voting} once.
\end{observation}

\begin{observation}
\label{observation:2}    
From Table \ref{tab:voter}, we observe that $\|\widetilde{\beta}^{(1)}\|_0 = \|\widetilde{\beta}^{(2)}\|_0 = 1$, while $\|\widetilde{\beta}^{(3)}\|_0 = 2$. Recall that the true value is $\dot{\beta} = \widetilde{\beta}^{(1)} = \widetilde{\beta}^{(2)}$. 
Instead of calculating $\widetilde{\beta}^{(j)}, \, j=1,2,3,$ separately for each $j$, Steps 2 and 3 in Algorithm \ref{algorithm:naive-voting} can be replaced with the following penalized regression:
\[
\beta^{SIV} = \argmin_{\beta \in \mathbb{R}^3} \| Y - \widetilde{X}^\T \beta \|_2^2 \quad \text{subject to } \|\beta\|_0 \leq 1,
\]
where, due to Observation \ref{observation:1}, $\widetilde{X}^{(1)} = \widetilde{X}^{(2)} = \widetilde{X}^{(3)} \equiv \widetilde{X}$.
\end{observation}

With these observations, Algorithm \ref{algorithm:naive-voting} simplifies to a two-step regularized regression procedure.

\subsection{Synthetic two-stage regularized regression}
\label{sec:2srr}
We now formally introduce the synthetic two-stage regularized regression for the general case. Motivated by Observation \ref{observation:1}, we provide the following definition of the synthetic instrument.

\begin{definition}[Synthetic Instrument] 
\label{def:si}
Define 
\[
SIV =  B_{\Lambda^\perp}^\T X \in \mathbb{R}^{p-q},
\]
where $B_{\Lambda^{\perp}} \in \mathbb{R}^{p \times (p-q)}$ is a semi-orthogonal matrix whose column space is orthogonal to the column space of $\Lambda \in \mathbb{R}^{p \times q}$.
\end{definition}

The following proposition confirms that the $SIV$ are valid instruments.

\begin{proposition}
\label{prop:validIV}
Under models \eqref{eqn:model1}, \eqref{eqn:model2}, and Condition \ref{A2}, the $SIV$ given by Definition \ref{def:si} serve as valid instrumental variables for estimating the treatment effects of $X$ on $Y$.
\end{proposition}

To identify the causal parameter $\beta$ in the general case, we introduce the following plurality condition \ref{A4}.

\begin{enumerate}[label=A4]
    \item \label{A4} (Plurality rule) Let $C^*$ be a subset of $\{1,2,\ldots,p\}$ with cardinality $q$, and suppose that $\dot{\beta}_{C^*} \neq 0$. The synthetic two-stage least squares coefficient obtained by assuming $\beta_{C^*} = 0$ is given by
     \(
    \widetilde{\beta}^{C^*} = \underset{\beta\in \mathbb{R}^p: \beta_{C^*}=0}{\argmin} \, \mathbb{E}\bigl(Y - \widetilde{X}^\T \beta\bigr)^2,
    \) 
    where $\widetilde{X} = \widehat{\mathbb{E}}(X \mid SIV)$.
    The plurality rule assumes that 
     \(
    \max\limits_{\beta\in \mathbb{R}^p} \, |\{C^*: \widetilde{\beta}^{C^*} = \beta\}| \leq q.
    \)
\end{enumerate}


In Condition \ref{A4}, each $C^*$ corresponds to an expert who makes the incorrect hypothesis that $\beta_{C^*} = 0$. Let $s = \Vert \dot{\beta} \Vert_0$. In general, there are $C_{p-s}^q$ experts making correct hypotheses. If $s < p-q$, then there are at least $q+1$ experts making correct hypotheses. The plurality rule assumes that no more than $q$ incorrect hypotheses lead to the same synthetic two-stage least squares coefficient. This assumption is similar in spirit to the plurality assumption used in the invalid IV literature \citep[e.g.,][]{guo2018confidence}. 
We further discuss Assumption \ref{A4} in Section \ref{sec:disA4} of the supplementary material, where we argue that its violation is unlikely. 

In parallel to Observation \ref{observation:2}, we have the following theorem.

\begin{theorem}[Synthetic two-stage regularized regression]\label{thm:l0 plurality}
Suppose that models \eqref{eqn:model1}, \eqref{eqn:model2}, and Conditions \ref{A1}, \ref{A2}, and \ref{A4} hold.
\begin{enumerate}
    \item If \ref{A3} holds, then $\dot{\beta}$ is identifiable via 
     \(    \dot{\beta} = \underset{{\beta}\in \mathbb{R}^{p}}{\argmin}\;\mathbb{E}( Y - \widetilde{X}^\T \beta )^2,
    \)
    subject to $\|\beta\|_0 < p-q$. 
    \item If \ref{A3} fails, then $\dot{\beta}$ is not identifiable, and for all $\widetilde{\beta} \in \underset{{\beta}\in \mathbb{R}^{p}}{\argmin}\;\mathbb{E}( Y - \widetilde{X}^\T \beta )^2$, we have $\|\widetilde{\beta}\|_0 \geq p-q$. 
\end{enumerate}
\end{theorem}

An important feature of Theorem \ref{thm:l0 plurality} is that, given $q$, it is possible to test the sparsity condition \ref{A3} from the observed data. In particular, it shows that under models \eqref{eqn:model1}, \eqref{eqn:model2}, Conditions \ref{A1}, \ref{A2}, and the plurality rule \ref{A4}, the following three statements are equivalent:
\begin{enumerate}
\item[(1)] $\beta$ is identifiable;
\item[(2)] Condition \ref{A3} holds;
\item[(3)] The most sparse least-squares solution to the second-stage regression has an $\ell_0$-norm smaller than $p-q$, i.e.,
\begin{equation}
\label{eqn:nsc}
    \min\limits_{\widetilde{\beta} \in \underset{{\beta}\in \mathbb{R}^{p}}{\argmin}\;\mathbb{E}( Y - \widetilde{X}^\T \beta )^2} \|\widetilde{\beta}\|_0 < p-q.
\end{equation}
\end{enumerate}

It is worth noting that \eqref{eqn:nsc} can be checked from the observed data distribution, so that one may develop a consistent test for Condition \ref{A3} and the identifiability of $\beta$ under models \eqref{eqn:model1}, \eqref{eqn:model2}, and Conditions \ref{A1}, \ref{A2}, and \ref{A4}. See Algorithm \ref{alg:SIV} below for more details.

\section{Estimation via  the synthetic two-stage regularized regression}
\label{sec:esti}
\subsection{Estimation}
Let ${\bf X} \in \mathbb{R}^{n \times p}$ be the design matrix and ${\bf Y} \in \mathbb{R}^{n\times 1}$ denote the observed outcome. Theorem \ref{thm:l0 plurality} suggests the following synthetic two-stage regularized regression for estimating $\beta$:
\begin{equation}
    \label{L0 optimization}
    \begin{split}
    \widehat{\bf X} &= \widehat{\mathbb{E}}({\bf X} \mid \widehat{SIV}),\\
    \widehat{\beta} &= \underset{{\beta}\in \mathbb{R}^p}{\argmin}{\|{\bf Y} - \widehat{\bf X} \beta\|_2^2} \quad \text{subject to } \|\beta\|_0 \leq k,
    \end{split}
\end{equation}
where $k$ is a tuning parameter, $\widehat{\Lambda}$ is an estimator of the loading matrix, and 
 \(
\widehat{SIV} = {\bf X} B_{\widehat{\Lambda}^\perp}.
\)

Several estimators have been proposed to determine the number of latent factors in a factor model. In our simulations and data analysis, we use the estimator developed by \cite{onatski2010determining} to obtain \( \widehat{q} \), as it is applicable in both low- and high-dimensional settings. Likewise, various methods exist for estimating the loading matrix \( \Lambda \). In low-dimensional settings where \( p \) is fixed, we recommend the maximum likelihood estimator of \( \Lambda \), obtained by maximizing the log-likelihood under multivariate normality. In high-dimensional settings, we estimate \( \Lambda \) using principal component analysis (PCA) \citep{bai2003inferential}, which yields a row-consistent estimator for \( \Lambda \), that is, each estimated row \( \widehat{\Lambda}_{i\cdot} \) consistently estimates its true counterpart \( {\Lambda}_{i\cdot} \). This approach does not require the covariance matrix \( \text{Cov}(\epsilon_X) \) to be diagonal.

Finally, we use cross-validation to select the tuning parameter $k$. Algorithm \ref{alg:SIV} summarizes our estimation procedure.

\begin{algorithm}[!ht]
  \caption{\quad The synthetic two-stage regularized regression}
\label{alg:SIV}
\hspace*{0.02in} {\bf Input: }  
 ${\bf X} \in \mathbb{R}^{n\times p}$ (centered), ${\bf Y} \in \mathbb{R}^{n\times 1}$ 
\hspace*{0.02in} 
\begin{algorithmic}[1]  
\State Obtain $\widehat{q}$ from ${\bf X}$ \citep[e.g.,][]{onatski2010determining}. 
\If{$n > p$} 
    obtain $\widehat{\Lambda} \in \mathbb{R}^{p \times q}$ via maximum likelihood estimation, assuming multivariate normality;  
\Else 
    let $\widehat{\lambda}_1 \geq \widehat{\lambda}_2 \geq \ldots \geq \widehat{\lambda}_p$ be the eigenvalues of ${\bf X}^\T {\bf X}/(n-1)$, and let $\widehat{\xi}_1, \widehat{\xi}_2, \ldots, \widehat{\xi}_p$ be the corresponding eigenvectors. Define $\widehat{\Lambda} = (\sqrt{\widehat{\lambda}_1} \widehat{\xi}_1 \; \ldots \; \sqrt{\widehat{\lambda}_q} \widehat{\xi}_q)$.
\EndIf

\State Let $B_{\widehat{\Lambda}^\perp}$ be a semi-orthogonal matrix whose columns are orthogonal to the columns of $\widehat{\Lambda}$. This can be obtained, for example, using the {\tt Null} function from the {\tt MASS} package in {\tt R}.
\State Obtain $\widehat{SIV} = {\bf X} B_{\widehat{\Lambda}^\perp}$.
\State Obtain $\widehat{\bf X} = \widehat{\mathbb{E}}({\bf X}\mid \widehat{SIV})$ via the fitted values from ordinary least squares.
\State Obtain $\widehat{\beta}$ via \eqref{L0 optimization}, where the tuning parameter $k$ is selected via 10-fold cross-validation.
\If{$\;\widehat{q}+\widehat{k}<p$}
    output $\widehat{\beta}$;
\Else 
    $\;\betaz$ is not identifiable.
\EndIf
\end{algorithmic}
\end{algorithm}

\subsection{Theoretical properties}

In this section, we study the theoretical properties of the estimator $\widehat{\beta}$ in Algorithm \ref{alg:SIV}. We consider two paradigms:  
(1) low-dimensional settings, where the dimension of exposure $p$ is fixed; and (2) high-dimensional settings, where $p$ grows with the sample size $n$. For the former, we show that under mild regularity conditions, $\widehat{\beta}$ is $\sqrt{n}$-consistent. For the latter, we show that under mild regularity conditions, $\widehat{\beta}$ achieves a Lasso-type error bound. We also demonstrate variable selection consistency in both scenarios. In our theoretical results, we do not require $\widehat{\Lambda} = \Lambda$; we only need certain norms of $\widehat{\Lambda}$ to be consistent with those of $\Lambda$, which can be achieved by classical estimators.
We first introduce assumptions for the low-dimensional case.

\begin{assumption}(Assumptions for fixed $p$)
\label{ass:Assumptions for fixed p}
\begin{enumerate}[label={B\arabic*}]

    \item \label{B1} All coefficients ${\Lambda}$, $\betaz$, and the function $g(\cdot)$ in models \eqref{eqn:model1} and \eqref{eqn:model2} are fixed and do not change as $n \to \infty$.
    \item \label{B2} $U_i$, $\epsilon_{x,i}$, and $\epsilon_{y,i}$ are independent random draws from the joint distribution of $(U,\epsilon_x,\epsilon_y)$ such that $E(\epsilon_x) = \bm{0}$, $E(U) = \bm{0}$, $\text{Cov}(\epsilon_x) = D$, $\text{Cov}(U) = I_q$, and $(U,\epsilon_x,\epsilon_y)$ are mutually independent. Furthermore, assume that $\text{Var}(\epsilon_y) = \sigma^2$ and $\max_{1 \leq j \leq p}\text{Var}(X_j) = \sigma_x^2$; these parameters are fixed and do not change as $n \to \infty$. 
    \item \label{B3} For the maximum likelihood estimator $\widehat{\Lambda}$, there exists an orthogonal matrix $O \in \mathbb{R}^{q \times q}$ such that $\| \widehat{\Lambda} - \Lambda O \|_2 = O_p(1/\sqrt{n})$.
    \item \label{B4} Let $\Sigma_{\widetilde{X}} = \text{Cov}(\widetilde{X})$. 
    We assume
     \(
    \min_{\theta \in \mathbb{R}^p, \, 0 < \|\theta\|_0 \leq 2s}\frac{\theta^\T \Sigma_{\widetilde{X}} \theta}{\|\theta\|_2^2} > c
    \) 
    for some positive constant $c$.
\end{enumerate}
\end{assumption}

Conditions \ref{B1}--\ref{B2} are standard assumptions for the low-dimensional setting. Given Condition \ref{A2}, Condition \ref{B3} requires that the estimator for factor loadings is root-$n$ consistent. Condition \ref{B4} is the population version of the sparse eigenvalue condition \citep[Assumption 3(b)]{raskutti2011minimax}.

Under these conditions, $\widehat{\beta}$ is root-$n$ consistent and achieves consistency in variable selection.

\begin{theorem}\label{thm:lowd thm} 
Under Conditions \ref{A1}--\ref{A4} and \ref{B1}--\ref{B4}, if the tuning parameter satisfies $\widehat{k} = s$, the following holds:
\begin{enumerate}
    \item[1.] ($\ell_1$-error rate) $\|\widehat{\beta} - \dot{\beta}\|_1 = O_p(n^{-1/2}).$
    \item[2.] (Variable selection consistency) Let $\mathcal{A} = \{j: \dot{\beta}_j = 0\}$ and $\widehat{\mathcal{A}} = \{j: \widehat{\beta}_j = 0\}$. Then $\mathbb{P}(\widehat{\mathcal{A}} = \mathcal{A}) \rightarrow 1$ as $n \rightarrow \infty.$
\end{enumerate}
\end{theorem}

In Theorem \ref{thm:lowd thm}, it is assumed that $\widehat{k} = s$. This is a standard condition in the $\ell_0$-optimization literature \citep[e.g.,][]{raskutti2011minimax,shen2013constrained}.

Next, we consider the high-dimensional case and demonstrate that our estimator exhibits properties similar to those of standard regularized estimators in the high-dimensional statistics literature, including a Lasso-type error bound and consistency in variable selection. We impose the following regularity conditions.

\begin{assumption}(Assumptions for diverging $p$)
\label{ass:Assumptions for diverge p}
\begin{enumerate}[label={C\arabic*}]
    \item \label{C1} $sq^2\log(p)\log(n)/n \rightarrow 0$, $n = O(p)$, and $q + \log(p) \lesssim \sqrt{n}$, where $x \lesssim y$ means there exists a constant $C$ such that $x \leq Cy$.   

    \item \label{C2} The expectation  \(\gamma := \mathbb{E}(Ug(U)) \in \mathbb{R}^q\), the variance  \(\sigma_g^2 = \text{Var}(g(U))\), and the covariance  \(\Gamma := \text{Var}(Ug(U)) \in \mathbb{R}^{q \times q}\) exist. For a matrix  \(M\), let  \(\lambda_{\max}(M)\) and  \(\lambda_{\min}(M)\) denote the maximum and minimum eigenvalues of  \(M\). There exist positive constants  \(C_1\),  \(C_2\), and  \(C_3\) such that 
     \(
    0 < C_1 \leq \min\{ \lambda_{\min}(D), \lambda_{\min}(\Lambda^\T \Lambda/p) \} \leq \max\{ \lambda_{\max}(D), \lambda_{\max}(\Lambda^\T \Lambda/p) \} \leq C_2 < \infty,
    \)
    and 
     \(
    \max\{ \|\gamma\|_2, \;\sigma_g^2, \;\text{Trace}(\Gamma) \} \leq C_3.
    \)
    
    \item \label{C3} Assume the random variables in models \eqref{eqn:model1} and \eqref{eqn:model2} satisfy $E(X) = \bm{0}$, $E(U) = \bm{0}$, and $\text{Cov}(U) = I_q$. We also assume $\epsilon_y$ is independent of $(X,U)$ and $\epsilon_x$ is independent of $U$. Furthermore, assume $\epsilon_y$, $\epsilon_{x,j}$, and $X_j$ are sub-Gaussian random variables with sub-Gaussian parameters $\sigma^2$, $\widetilde{\sigma}_j^2$, and $\sigma_j^2$, respectively. The parameters satisfy $\sigma^2 \leq C_4$, and $C_5 \leq \widetilde{\sigma}_j^2, \sigma_j^2 \leq C_6$ for some constants $C_4, C_5, C_6 > 0$.
    
    \item \label{C4} There exist positive constants $C_7$ and $C_8$ such that $\underset{i \in \mathcal{A}}{\min}|\dot{\beta}_i| \geq n^{C_7-1/2}$ and $s^2(q+1)^2\log{p} \leq n^{2C_7 - C_8}$.
\end{enumerate}
\end{assumption}

Condition \ref{C1} allows the number of exposures $p$ to grow exponentially with the sample size, while the number of latent confounders $q$ grows at a slower polynomial rate. Condition \ref{C2} is a standard assumption in high-dimensional factor analysis \citep{fan2013large,shen2016general} for loading identification. Condition \ref{C3} assumes that the exposures $X_j$ are sub-Gaussian and that the noise level is bounded. Condition \ref{C4} is a standard assumption on minimum signal strength.

\begin{theorem}\label{thm:highd error rate}
Assume that Conditions \ref{A1}--\ref{A4} and \ref{C1}--\ref{C3} hold, and that the tuning parameter satisfies $\widehat{k} = s$. Then:
\begin{enumerate}
    \item[1.] ($\ell_1$-error rate) $\|\widehat{\beta} - \dot{\beta}\|_1 = O_p\left(s(q+1)\sqrt{\frac{\log(p)}{n}}\right).$ 
    
    \item[2.] (Variable selection consistency) Under Condition \ref{C4}, $\mathbb{P}(\widehat{\mathcal{A}} = \mathcal{A}) \rightarrow 1$ as $n \rightarrow \infty.$
\end{enumerate}
\end{theorem}

\begin{remark}
The first part of Theorem \ref{thm:highd error rate} differs from Theorem 1 in \cite{cevid2020spectral}. Their theoretical result relies on the following linear model and decomposition:
\[
Y = X^\T \beta + U^\T \gamma + \epsilon_y = X^\T \beta + X^\T b + (U^\T \gamma - X^\T b) + \epsilon_y,
\]
where $b$ is the best linear predictor of $U$ from $X$. Their result depends on (i) $\|b\|_2 = O\left({1}/{\sqrt{p}}\right)$ and (ii) the term $(U^\T \gamma - X^\T b) + \epsilon_y$ being independent of $X$ under their joint Gaussian assumption. 
In general, these conditions fail to hold under model \eqref{eqn:model2}, {where $g(U)$ is an unknown function.}
\end{remark}

\subsection{Extension to nonlinear settings}

\label{sec:nonlinear_main}
In this section, we extend the SIV method to address scenarios where the treatment  \( X \) has nonlinear effects on the outcome  \( Y \). Revisiting model \eqref{eqn:model2}, we remove the assumption of linearity, allowing both the treatment  \( X \) and the unmeasured confounder  \( U \) to influence the outcome  \( Y \) through nonlinear relationships:
\begin{equation}
Y = f(X;\beta) + g(U) + \epsilon_y \label{eqn:nonlinearmodel}    
\end{equation}
In this model, the treatment influences the outcome through the nonlinear causal function  \(f(\cdot;\beta)\), with  \(\beta \in \mathbb{R}^p\) as the parameter of interest. Our focus is on estimating the parameter  \(\beta\).

The key observation is that, under model \eqref{eqn:model1}, the synthetic instruments  \( \text{SIV} \in \mathbb{R}^{p-q} = B_{\Lambda^\perp}^\T \epsilon_X \) are linear combinations of  \( \epsilon_X \), which are independent of both  \( g(U) \), for any measurable  \( g \), and  \( \epsilon_y \).
Consequently, we have the following vector equation:
\[
\mathbb{E}\{SIV(Y-f(X;\beta))\} = \mathbb{E}\{SIV(g(U)+\epsilon_y)\} = 0,
\]
when  \(\beta\) is set to its true value. Following this, we define the population GMM loss:
\begin{equation}
    \label{eqn:gmmloss}
    G(\beta) = \left\|\mathbb{E}\{SIV (Y-f(X;\beta))\}\right\|_2^2.
\end{equation}

For the function  \(f(X;\beta)\), let  \(\partial f(X;\beta)/\partial \beta\) denote the  \(p \times 1\) column vector  \(({\partial f(X;\beta)}/{\partial \beta_1}
\)  \(, \ldots, {\partial f(X;\beta)}/{\partial \beta_p})^\T\), and let  \(\partial f(X;\beta)/\partial \beta^\T\) denote the  \(1 \times p\) row vector  \(({\partial f(X;\beta)}/{\partial \beta_1}, \ldots, \)
 \({\partial f(X;\beta)}/{\partial \beta_p})\). We make the following assumptions, denoted as D1 and D2, which are generalizations of Conditions \ref{A1} and \ref{A4} in the nonlinear setting.

\begin{enumerate}[label=D\arabic*]
    \item \label{D1} (Invertibility) The matrix  \( \mathbb{E}(X {\partial f(X;\beta)}/{\partial \beta^\T}) \in \mathbb{R}^{p \times p}\) is invertible, and any  \(q \times q\) submatrix of  \(\mathbb{E}^{-1}(X {\partial f(X;\beta)}/{\partial \beta^\T})\Lambda\) is also invertible for any  \(\beta\).
    \item \label{D2} (Nonlinear plurality rule) Let  \(C^*\) be a subset of  \(\{1, 2, \ldots, p\}\) with cardinality  \(q\) and  \(\dot{\beta}_{C^*} \neq 0\), and let the synthetic GMM estimator obtained by assuming  \(\beta_{C^*} = 0\) be 
     \(
    \widetilde{\beta}^{C^*} = \underset{\beta \in \mathbb{R}^p: \beta_{C^*}=0}{\argmin} G(\beta).
    \) 
    The plurality rule assumes that  \(\widetilde{\beta}^{C^*}\) is uniquely defined and that  \(\max\limits_{\beta \in \mathbb{R}^p}|\{C^*: \widetilde{\beta}^{C^*} = \beta\}| \leq q.\)
\end{enumerate}

The following theorem states that  \(\beta\) is identifiable using synthetic GMM in a manner parallel to Theorem \ref{thm:l0 plurality}, but within a nonlinear setting.
\begin{theorem}[Synthetic Generalized Method of Moments]\label{thm:l0 plurality gmm}
Suppose that models \eqref{eqn:model1}, \eqref{eqn:model2}, and Conditions \ref{A2}, \ref{D1}, and \ref{D2} hold.
\begin{enumerate}
    \item If \ref{A3} holds, then  \(\dot{\beta}\) is identifiable via  \(\dot{\beta} = \underset{{\beta} \in \mathbb{R}^{p}}{\argmin}\;G(\beta)\), subject to  \(\|\beta\|_0 < p-q\).
    \item If \ref{A3} fails, then  \(\dot{\beta}\) is not identifiable, and for all  \(\widetilde{\beta} \in \underset{{\beta} \in \mathbb{R}^{p}}{\argmin}\;G(\beta)\), we have  \(\|\widetilde{\beta}\|_0 \geq p-q\).
\end{enumerate}
\end{theorem}

We now discuss the estimation of $\beta$ in finite samples. A natural approach is to replace the expectation in equation~\eqref{eqn:gmmloss} with its empirical counterpart. Specifically, we consider the following loss function:
\begin{equation*}
    G_n(\beta) = \left(\frac{1}{n}\sum_{i=1}^n [SIV_i\{Y_i - f(X_i; \beta)\}]\right)^\T 
    W 
    \left(\frac{1}{n}\sum_{i=1}^n [SIV_i\{Y_i - f(X_i; \beta)\}]\right),
\end{equation*}
where $W \in \mathbb{R}^{(p-q)\times(p-q)}$ is a weight matrix. 

We now discuss the estimation of $\beta$ in finite samples. A natural approach is to replace the expectation in equation~\eqref{eqn:gmmloss} with its empirical counterpart. Specifically, we consider the following loss function:
\begin{equation*}
    G_n(\beta) = \left(\frac{1}{n}\sum_{i=1}^n [SIV_i\{Y_i - f(X_i; \beta)\}]\right)^\T 
    W 
    \left(\frac{1}{n}\sum_{i=1}^n [SIV_i\{Y_i - f(X_i; \beta)\}]\right),
\end{equation*}
where $W \in \mathbb{R}^{(p-q)\times(p-q)}$ is a weight matrix. 
The oracle weight matrix that achieves the highest efficiency in GMM is the inverse of 
$\text{Cov}(\text{SIV})$ \citep[Eq.~17]{burgess2017review}. 
In practice, we estimate $W$ using the inverse of the empirical covariance, 
$\widehat{\text{Cov}}^{-1}(\widehat{\text{SIV}})$.

Finally, let  $\bm{X} \in  \mathbb{R}^{n\times p}$,  $\bm{Y} \in \mathbb{R}^{n\times 1}$,  $f(\bm{X};\beta) \in \mathbb{R}^{n\times 1}$, and  $\bm{SIV} \in \mathbb{R}^{n \times (p-q)}$ denote the relevant random matrices in the finite-sample setting. 
In the low-dimensional setting, where the number of instruments $p - q$ is fixed and smaller than $n$, we consider the following optimization problem:
\begin{equation}
    \label{eqn:nonlinear2SLS}
    \argmin_{\beta \in \mathbb{R}^p } 
    \left\| 
        \bm{SIV}({\bm{SIV}^\T \bm{SIV}})^{-1}{\bm{SIV}^\T (\bm{Y} - f(\bm{X};\beta))} 
    \right\|^2_2 
    \quad \text{subject to } \|\beta\|_0 \leq k.
\end{equation}
The unconstrained optimization in \eqref{eqn:nonlinear2SLS} is also referred to as the nonlinear two-stage least squares estimator in the literature \citep{amemiya1974nonlinear}. 
Under the linear setting where  $f(\bm{X};\beta) =  \bm{X}\beta$, equation~\eqref{eqn:nonlinear2SLS} has a similar form to \eqref{L0 optimization}. 
Optimization~\eqref{eqn:nonlinear2SLS} can be accelerated using a splicing approach, which is designed to efficiently solve the best subset selection problem \citep{doi:10.1073/pnas.2014241117, zhang2023splicing}. 
We establish the theoretical properties of~\eqref{eqn:nonlinear2SLS} in Section~\ref{sec:appthmnonlinear} of the supplementary material, showing that under mild regularity conditions, the proposed estimator achieves both consistency and variable selection consistency in the low-dimensional setting.

\begin{remark}
\label{rem:highdGmm}
In high-dimensional settings where the dimension of $\operatorname{SIV}$, namely $p - q$, exceeds the sample size $n$, the matrix $\operatorname{SIV}^\T \operatorname{SIV}$ is not invertible. 
In this case, one may follow the approach of \citet[][p.~35]{belloni2018high} and choose the weight matrix $W$ as 
\[
W = \operatorname{diag}\!\left( 
    \operatorname{Var}\!\left( 
        \mathbb{E}_n\!\left[ \operatorname{SIV}(Y - f(X; \widetilde{\beta})) \right] 
    \right) 
\right)^{-1},
\]
where $\widetilde{\beta}$ denotes an initial estimator of $\beta$. 
\end{remark}

\section{Simulation studies}
\label{sec:simulation}
In this section, we evaluate the numerical performance of the proposed SIV method and compare it with other methods across various scenarios. First, we assess the performance of the SIV method under a linear outcome model in Section \ref{sec:simu_linear}. 
Next, we evaluate the algorithm in the context of a nonlinear outcome model in Section \ref{sec:sim_nonlinear}.

We provide additional simulation results in the supplementary material. In Section~\ref{sec:simuweakeffect}, we present simulation results for various estimators under dense confounding with many weak effects. In Section~\ref{sec:MS}, we explore an alternative moment selection estimator, inspired by the work of \citet{andrews1999consistent}, and compare it with our proposed estimator. Section~\ref{sec:CI_sup} discusses the construction of confidence intervals for selected causal variables using the \texttt{ivreg} function. Finally, in Section~\ref{sec:countdata}, we extend the SIV algorithm to confounded count data, where the effect of the unmeasured confounder on the outcome cannot be additively separated from the effect of the treatment.


\subsection{Simulation studies with a linear outcome model}
\label{sec:simu_linear}
We begin by evaluating the SIV estimator within a linear outcome model. The model is defined by $f(X;\beta) = X^\T\beta$ and $g(U) = U^\T\gamma$. In our simulations, we let $q = 3$, $s = 5$, and $\beta = (1,1,1,1,1,0,\ldots,0)^\T \in \mathbb{R}^p$. Each element in $\Lambda_{j,k}$ and $\gamma_k$ is independently generated from $\text{Uniform}(-1,1)$ for $j = 1,\ldots,p$ and $k = 1,\ldots,q$. The hidden variables $U_{i,k}$ follow i.i.d. standard normal distributions for $i = 1,\ldots,n$ and $k = 1,\ldots,q$. The random errors are generated as $\epsilon_x \sim \mathbb{N}(0,\sigma_x^2 I_p)$ and $\epsilon_{y} \sim \mathbb{N}(0,\sigma^2)$, where $\sigma_x = 2$ and $\sigma = 5$. We evaluate the performance of our method under the following two settings: (i) low-dimensional cases: $p = 100$ and $n \in \{200,600,1000,\ldots,5000\}$; (ii) high-dimensional cases: $n = 500$ and $p \in \{500,750,1000,\ldots,3000\}$. All simulation results are based on $1000$ Monte Carlo runs. The data-generating mechanism is designed to mimic key features of the real application in Section~\ref{sec:data}; see Section~\ref{sec:comparison_sim&data} of the supplementary material for a comparison between the real and synthetic data.

We compare the following methods in our simulations.

\begin{enumerate}
    \item[1.] (SIV) We implement Algorithm \ref{alg:SIV} and determine $\widehat{q}$ using the method proposed by \cite{onatski2010determining}. A detailed discussion of \cite{onatski2010determining}'s method is provided in Section \ref{rem:Onatski}  of the supplementary material. For cases where $p \leq 30$, we employ a full best subset selection routine to solve the $\ell_0$-optimization problem. 
    When $p > 30$, we utilize the adaptive best subset selection method implemented in the {\tt abess} function in {\tt R}.

    \item[2.] \citep[Lasso,][]{tibshirani1996regression}: We implement the Lasso using the {\tt glmnet} function in {\tt R}, with the tuning parameter selected via 10-fold cross-validation.
    
    \item[3.] \citep[Trim,][]{cevid2020spectral}: We implement \cite{cevid2020spectral}'s method using the code available from \url{https://github.com/zijguo/Doubly-Debiased-Lasso}, with an update detailed in Section \ref{sec:update_supp} of the supplementary material.
    
    \item[4.] \citep[Null,][]{miao2023identifying}: For the low-dimensional settings, we implement \cite{miao2023identifying}'s method using the code available from \url{https://www.tandfonline.com/doi/suppl/10.1080/01621459.2021.2023551}. 
    For the high-dimensional settings, their method cannot be applied directly because $\xi$ in their procedure cannot be solved by ordinary least squares. Therefore, we replace ordinary least squares with the Lasso, with the tuning parameter selected by 10-fold cross-validation.
    
    \item[5.] (IV-Lasso) Motivated by a reviewer’s suggestion, we consider the following two-step ``IV-Lasso'' procedure.  
 {First, we apply the Lasso as a screening step to identify candidate causal predictors of $Y$. Specifically, we solve
\[
\widetilde{\beta} = \underset{\beta \in \mathbb{R}^p}{\arg\min} \left\{ \|{\bf Y} - \widehat{{\bf X}} \beta\|_2^2 + \lambda \|\beta\|_1 \right\},
\]
where $\widehat{\bf X} = \widehat{\mathbb{E}}({\bf X} \mid \widehat{\text{SIV}})$. 
Let $\widehat{\mathcal{A}} = \{j : \widetilde{\beta}_j \neq 0\}$ denote the set of variables selected by the Lasso.}
     In the second step, we fit a linear model of  \(Y\) on  \(\widehat{X}_{\widehat{\mathcal{A}}}\) using ordinary least squares, yielding estimates  \(\widehat{\beta}^{\;\text{IV-Lasso}}_{\widehat{\mathcal{A}}}\). We set  \(\widehat{\beta}^{\;\text{IV-Lasso}}_{\widehat{\mathcal{A}}^c} = 0\).
\end{enumerate}

\begin{figure}[!htbp]
    \centering{
    \begin{subfigure}{0.49\textwidth}
    \includegraphics[scale = 0.33]{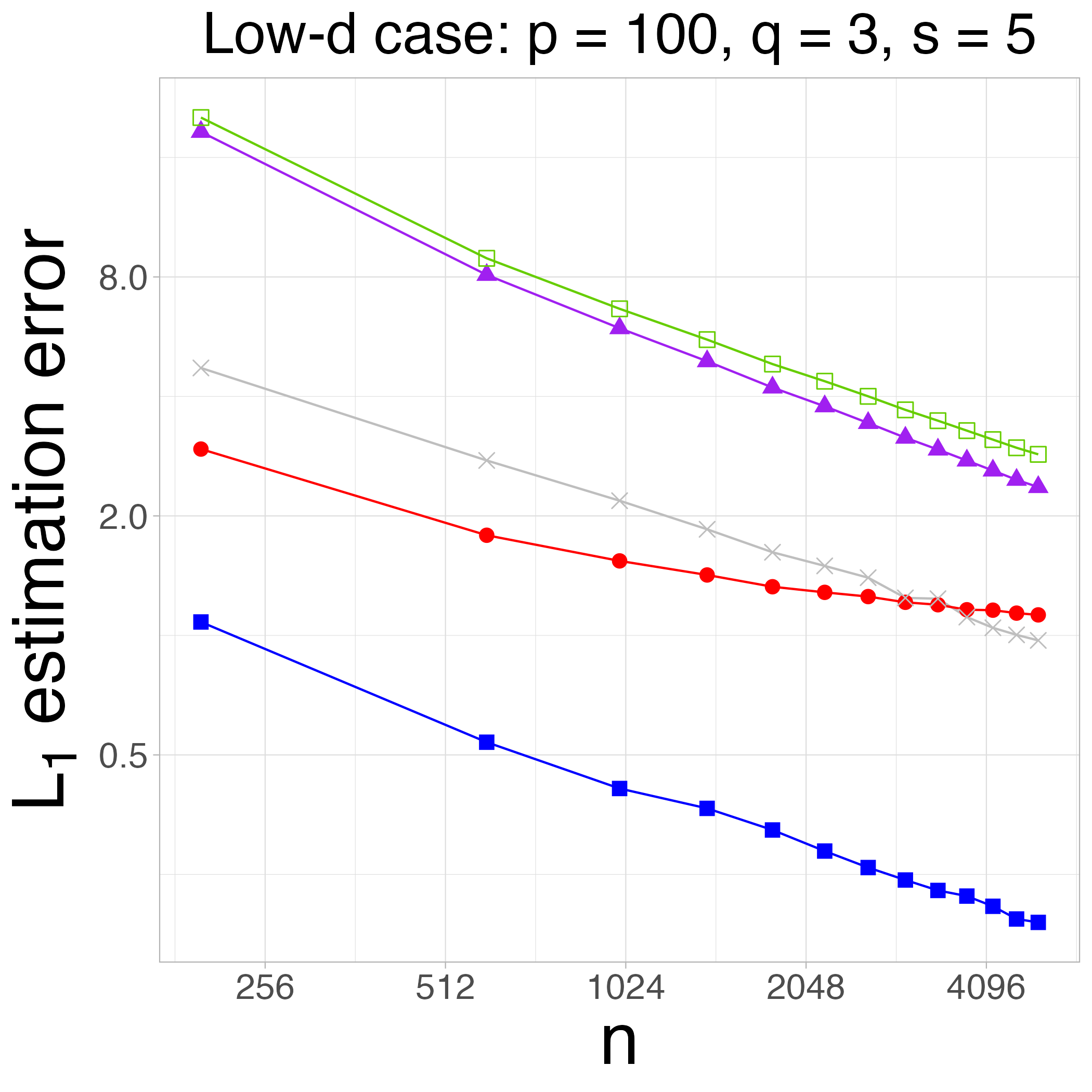}
  \centering
    \caption{$p = 100$, and $n$ varies from $200$ to $5000$.}
     \label{fig:fixp}
  \end{subfigure}}
  \hfill
  \centering{
  \begin{subfigure}{0.49\textwidth}
  \centering
    \includegraphics[scale = 0.33]{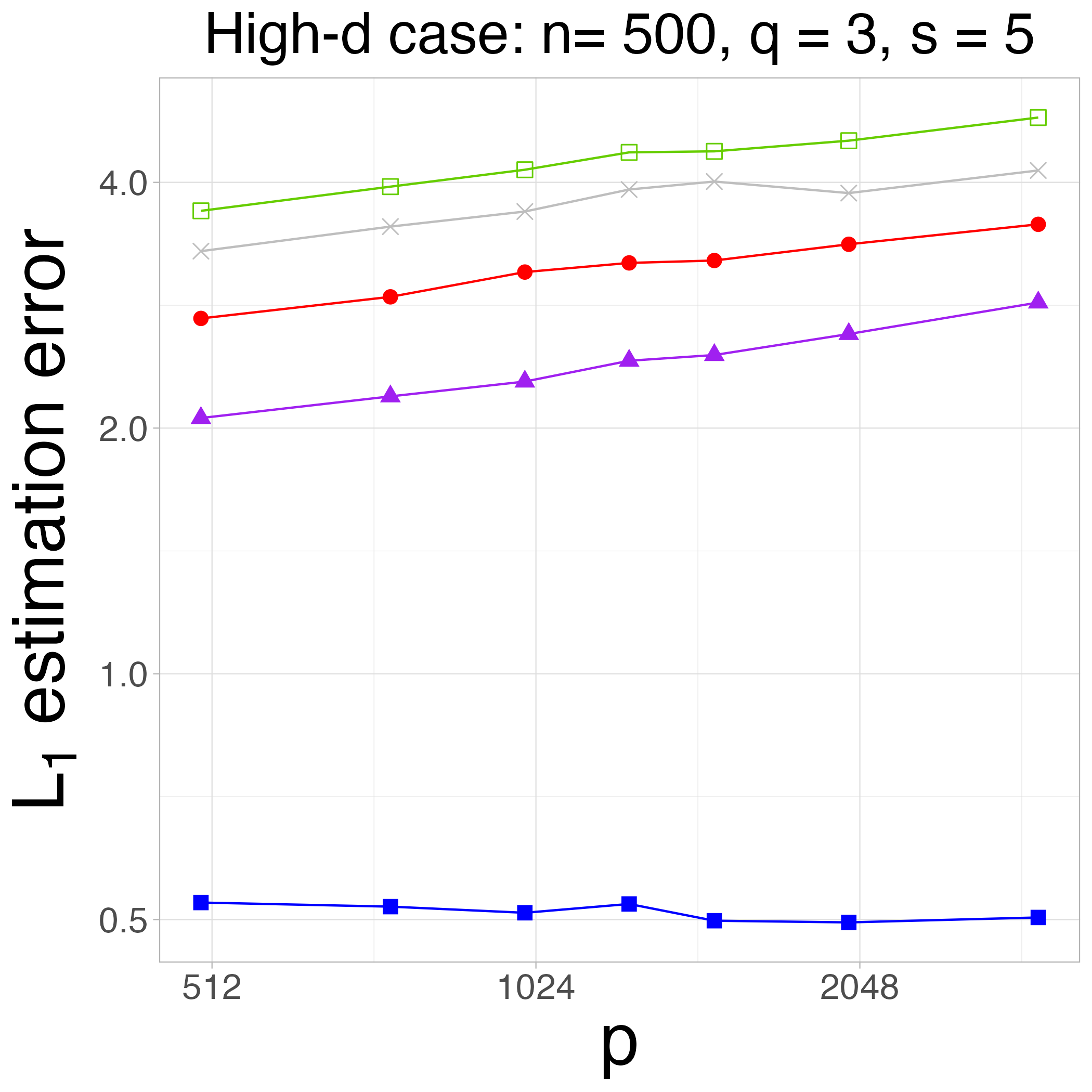}
    \caption{$n = 500$, and $p$ varies from $500$ to $3000$.}
      \label{fig:fixn}
  \end{subfigure}
  \hspace*{\fill}
  }
    \caption{Estimation errors $||\widehat{\beta} - \beta||_1$ for SIV ($\blacksquare$, blue), Lasso ($\CIRCLE$, red), Trim ($\blacktriangle$, purple), Null ($\square$, green), and IV-Lasso($\times$, grey) based on $1000$ Monte Carlo runs. 
    }
      \label{fig:Estimation errors}
\end{figure}

\begin{figure}[!htbp]
    \centering{
    \begin{subfigure}{0.49\textwidth}
    \includegraphics[scale = 0.33]{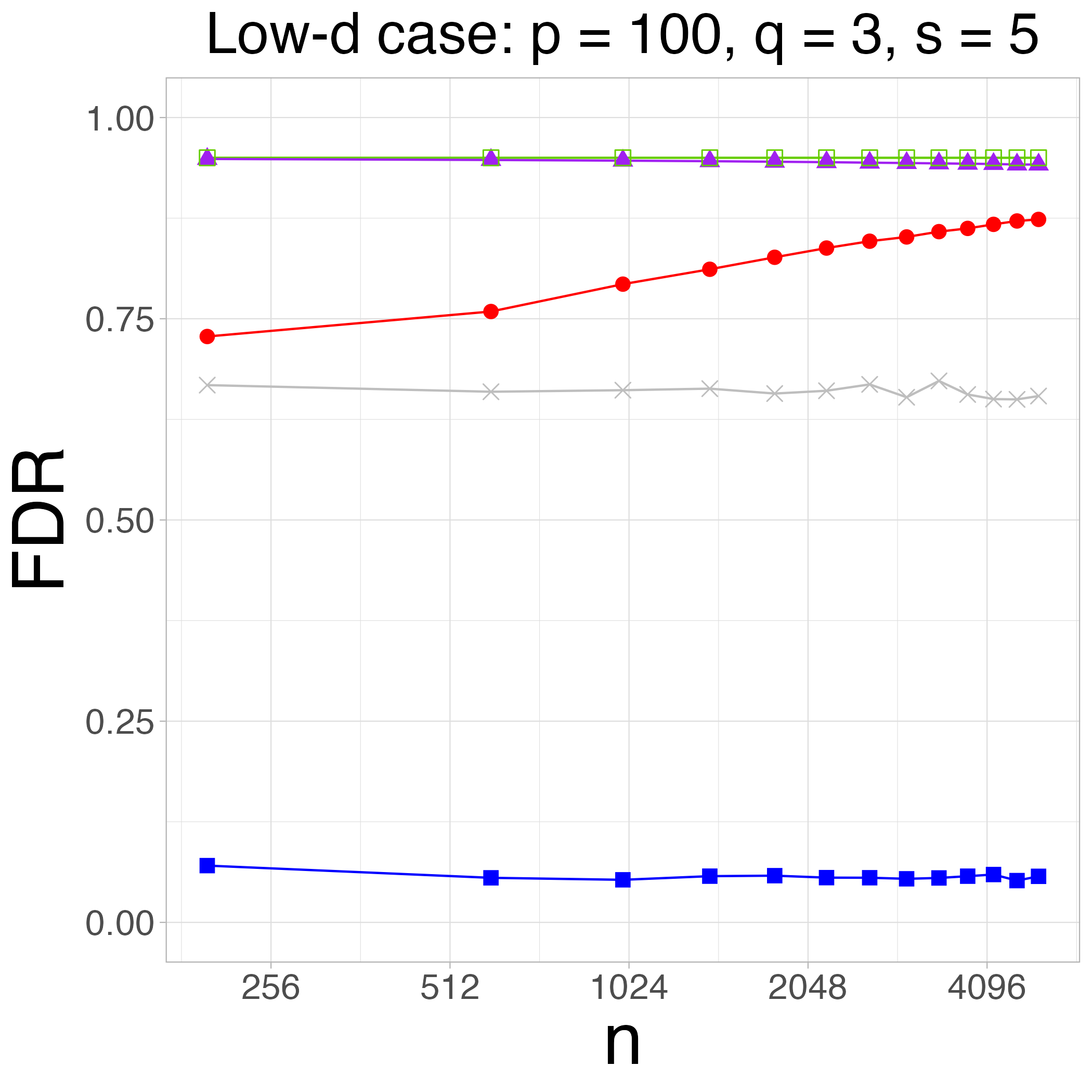}
  \centering
    \caption{$p = 100$, and $n$ varies from $200$ to $5000$.}
     \label{fig:fixpsparsity}
  \end{subfigure}}
  \hfill
  \centering{
  \begin{subfigure}{0.49\textwidth}
  \centering
    \includegraphics[scale = 0.33]{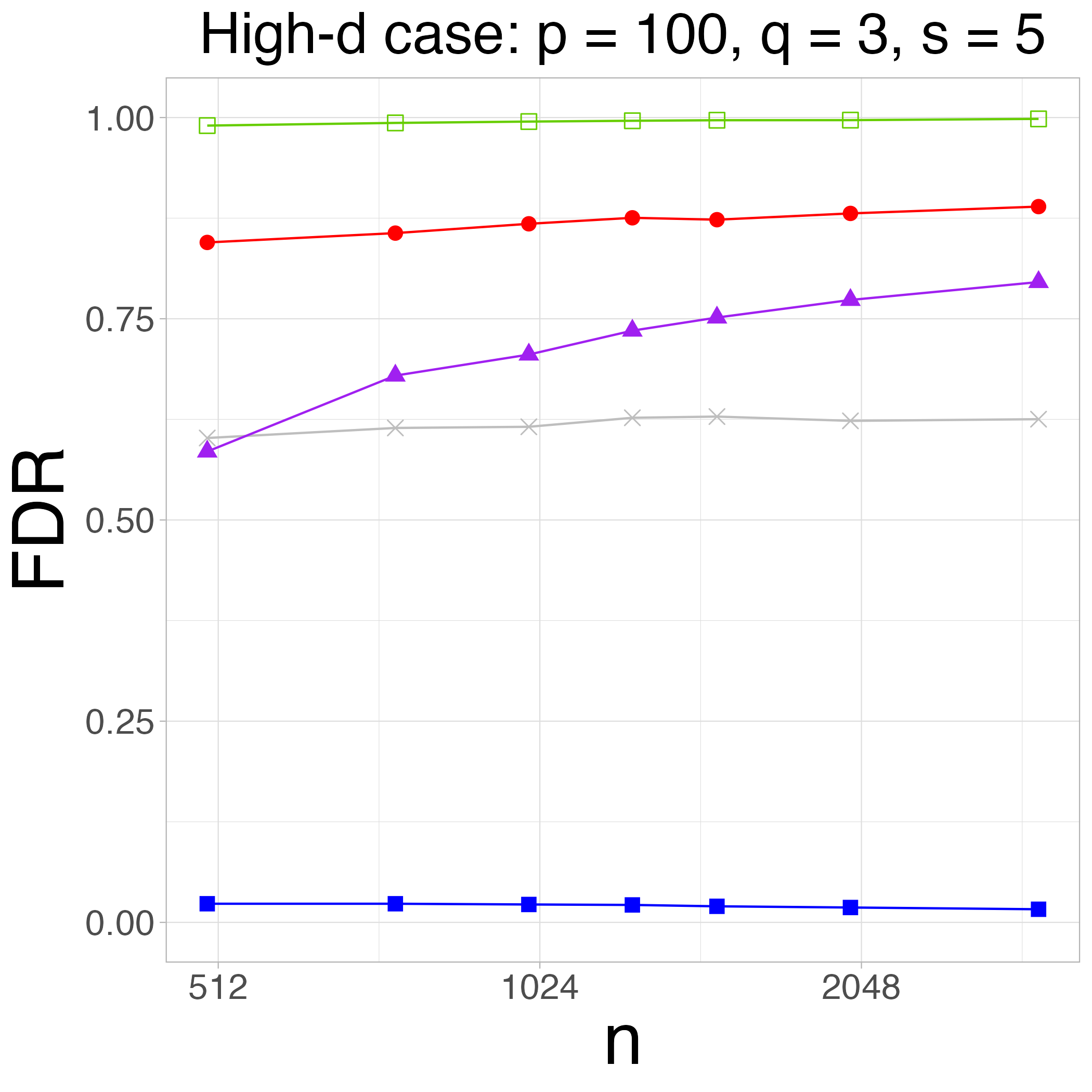}
    \caption{$n = 500$, and $p$ varies from $500$ to $3000$.}
      \label{fig:fixnsparsity}
  \end{subfigure}
  \hspace*{\fill}
  }
    \caption{False discovery rate(FDR) for SIV ($\blacksquare$, blue), Lasso ($\CIRCLE$, red), Trim ($\blacktriangle$, purple), Null ($\square$, green), and IV-Lasso($\times$, grey) based on $1000$ Monte Carlo runs. }
    \label{fig:FDR}
\end{figure}


We present the $\ell_1$-estimation errors of the five methods in Figures \ref{fig:Estimation errors}. In low-dimensional cases, as illustrated in Figure \ref{fig:fixp}, the bias of the Lasso estimator stabilizes as the sample size grows. The Lasso estimator does not account for unmeasured confounding and is therefore not expected to perform well. The Lasso, Trim, and Null methods exhibit considerable $\ell_1$-estimation bias relative to other methods. Compared to IV-Lasso, our SIV estimator demonstrates superior performance. This advantage arises from the $\ell_0$-optimization employed in the SIV method, which enables more accurate variable selection, as illustrated in Figure \ref{fig:fixpsparsity} later.

To ensure a fair comparison across all methods, we use cross-validation to select $\lambda$. However, it is well known that cross-validation can lead to overfitting and an increased false discovery rate. In practice, researchers may adopt the ``one-standard-error" rule, which selects the largest $\lambda$ such that the cross-validation error remains within one standard error of its minimum \citep{hastie2009elements,kang2016instrumental}. In the simulation settings we have considered, we find that the ``one-standard-error rule'' allows the IV-Lasso method to perform as well as the SIV method. See Section~\ref{sec:ivlasso1se} of the supplementary material for details on how the ``one-standard-error” rule improves the empirical performance of IV-Lasso.

For the high-dimensional settings, we can see from Figure \ref{fig:fixn} that our SIV method consistently outperforms the comparison methods. As discussed in Section \ref{sec:intro}, the true correlations between $X$ and $Y$ are non-sparse. This explains the large estimation errors of the Lasso method. The Null method exhibits even larger biases than the Lasso method. The Trim method, designed specifically for high-dimensional settings, outperforms both the Lasso and Null methods. However, our estimator still shows a much smaller bias than the Trim estimator. In Section~\ref{sec:trim_sup} of the supplementary material, we provide additional discussion, simulations, and comments on why the Trim estimator performs less favorably compared to our estimator.

We also observe from Figure \ref{fig:fixn} that the IV-Lasso estimator underperforms compared to the naive Lasso estimator in high-dimensional settings. This underperformance occurs because the naive Lasso method applies  \(\ell_1\)-penalization, which drives the coefficients of incorrectly selected  \(\widehat{\beta}_i\) towards zero. In contrast, the second step of IV-Lasso negates this shrinkage, causing the coefficients of incorrectly selected  \(\widehat{\beta}_i\) to diverge from zero, thereby increasing estimation bias.

Since the underlying $\betaz$ is sparse, we also report the performance of variable selection for all the methods in Figure \ref{fig:FDR}. All these methods correctly classify the true causes of the outcome as active exposures, that is, $\widehat{\mathcal{A}} \supseteq \mathcal{A}$. Thus, we only report the average false discovery rates, denoted as $\#\{\widehat{\mathcal{A}}\setminus \mathcal{A}\}/\#\widehat{\mathcal{A}}$, across 1000 Monte Carlo runs. It is evident that our proposal achieves the lowest false discovery rate among all the methods in both the low- and high-dimensional settings. 

We further evaluate the performance of our proposed algorithm in settings with non-diagonal  \( \text{Cov}(\epsilon_x) \). We generate $\epsilon_{x,i} \sim \mathbb{N}(0,D)$. 
For the low-dimensional setting, we randomly select 20 pairs of  \( i, j \in \{1, 2, \ldots, p\} \) and set  \( D_{i,j} = D_{j,i} = 1 \). The list of pairs is provided in Section \ref{sec:detail:nondiagonal} of the supplementary material. In high-dimensional settings, we set  \( D_{i,j} = 4 \times 0.3^{|i-j|} \). All other aspects of the data-generating mechanism remain unchanged. In the low-dimensional scenario, we use the Robust Principal Component Analysis method \citep{candes2011robust} to estimate the low-dimensional structure of the covariance matrix and factor loadings. For the high-dimensional scenario, we directly use the Principal Component Analysis method to estimate factor loadings. These are implemented using the {\tt R} packages {\tt rpca} and {\tt prcomp}, respectively.

\begin{figure}[!htbp]
\centering{
    \begin{subfigure}{0.49\textwidth}
    \includegraphics[scale = 0.33]{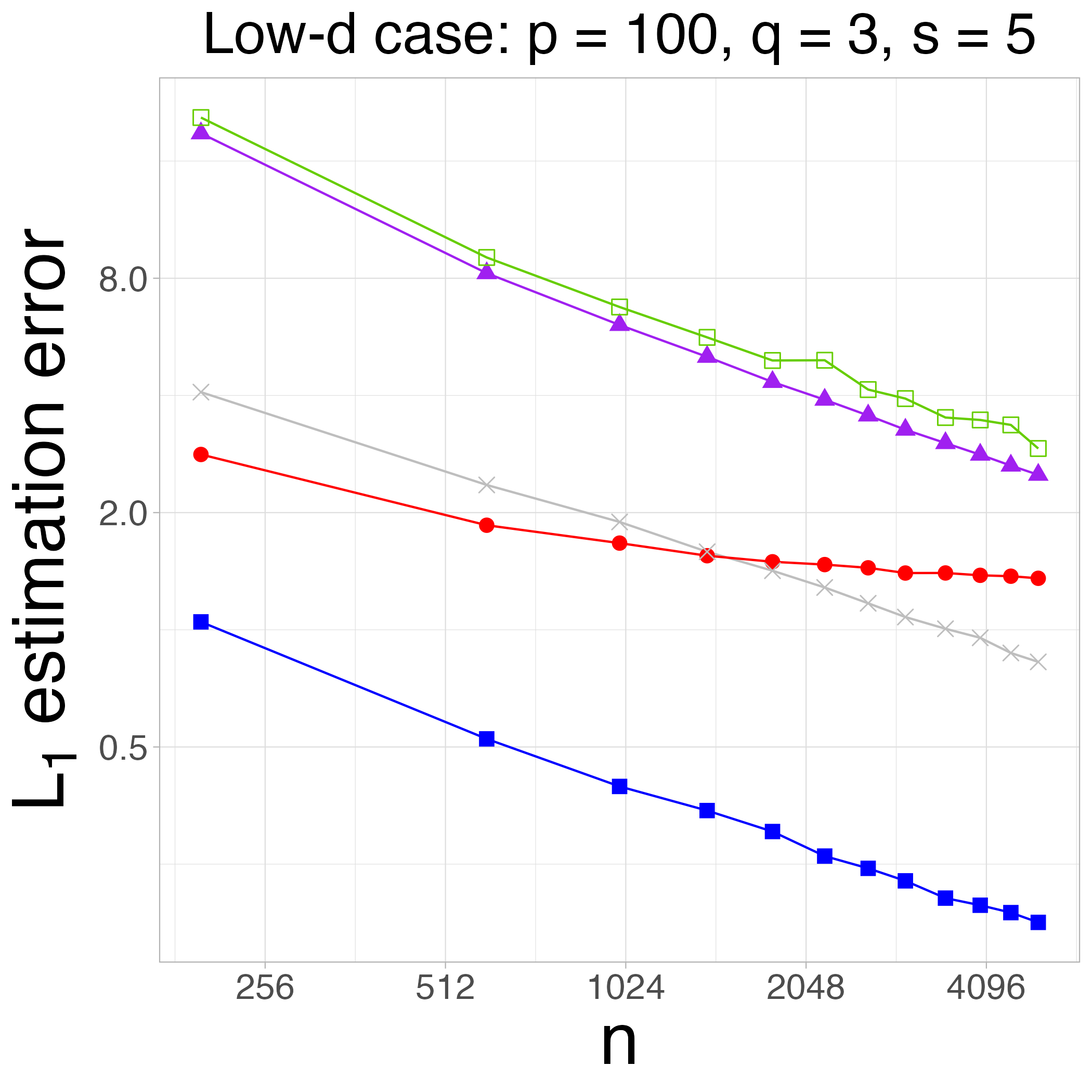}
  \centering
    \caption{$p = 100$, and $n$ varies from $200$ to $5000$.}
     \label{fig:fixp_correlated}
  \end{subfigure}}
  \hfill
  \centering{
  \begin{subfigure}{0.49\textwidth}
  \centering
    \includegraphics[scale = 0.33]{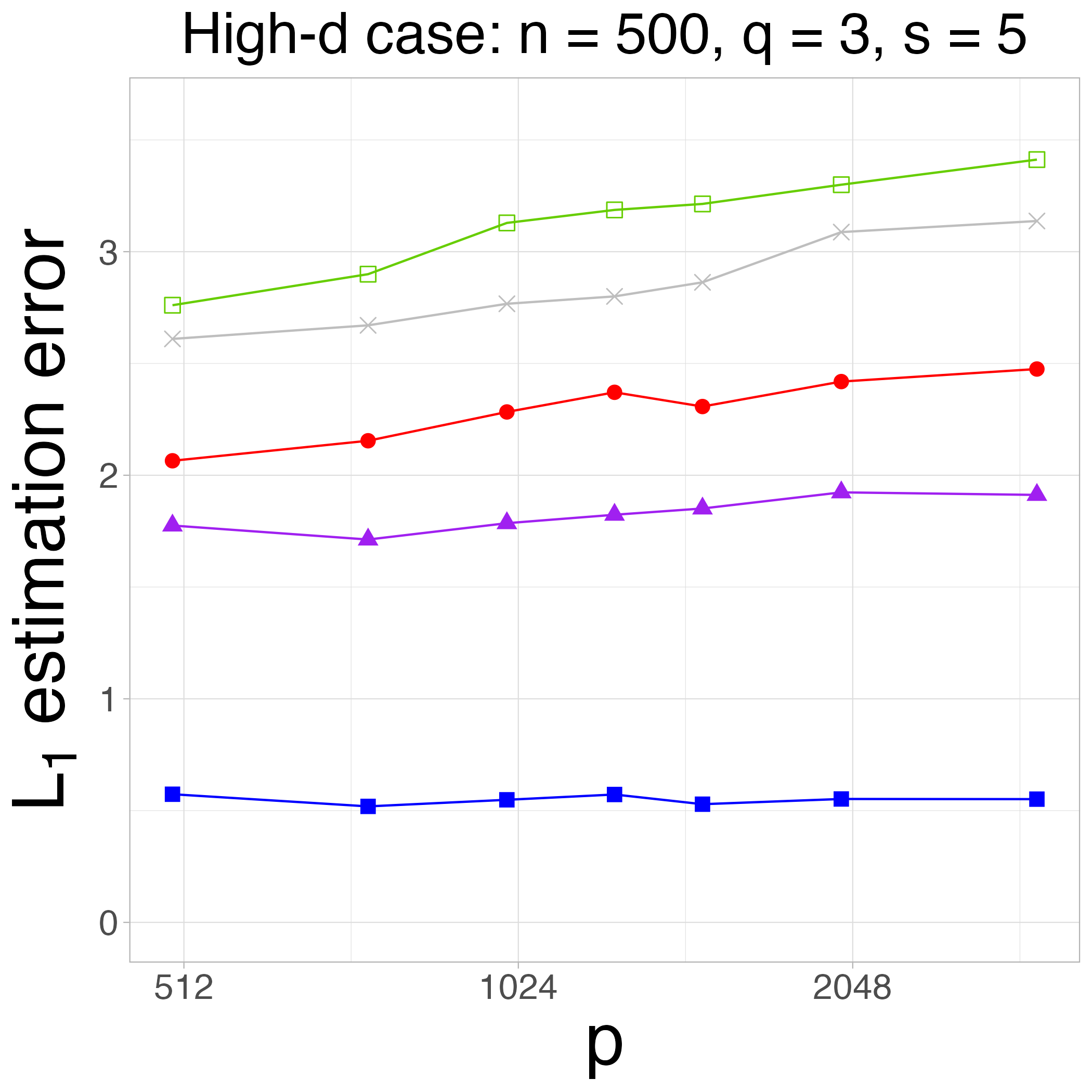}
    \caption{$n = 500$, and $p$ varies from $500$ to $3000$.}
      \label{fig:fixn_correlated}
  \end{subfigure}
  \hspace*{\fill}
  }
    \caption{Estimation errors $||\widehat{\beta} - \beta||_1$ with non-diagonal $D$ for SIV ($\blacksquare$, blue), Lasso ($\CIRCLE$, red), Trim ($\blacktriangle$, purple), Null ($\square$, green), and IV-Lasso($\times$, grey) based on $1000$ Monte Carlo runs. 
    \label{fig:estimation_nondiagonal}
    }
\end{figure}

\begin{figure}[!htbp]
    \centering{
    \begin{subfigure}{0.49\textwidth}
    \includegraphics[scale = 0.33]{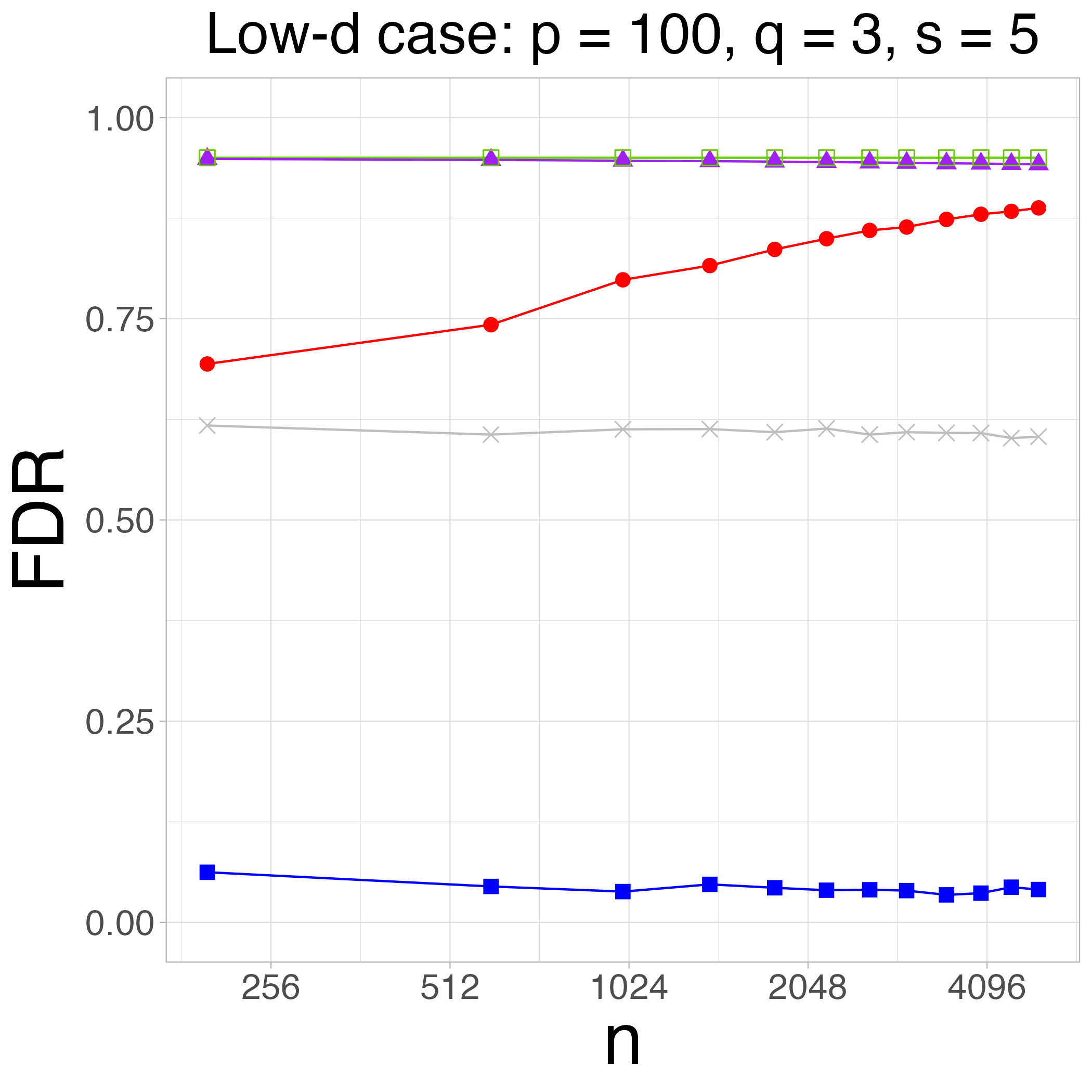}
  \centering
    \caption{$p = 100$, and $n$ varies from $200$ to $5000$.}
     \label{fig:fixpsparsity_correlated}
  \end{subfigure}}
  \hfill
  \centering{
  \begin{subfigure}{0.49\textwidth}
  \centering
    \includegraphics[scale = 0.33]{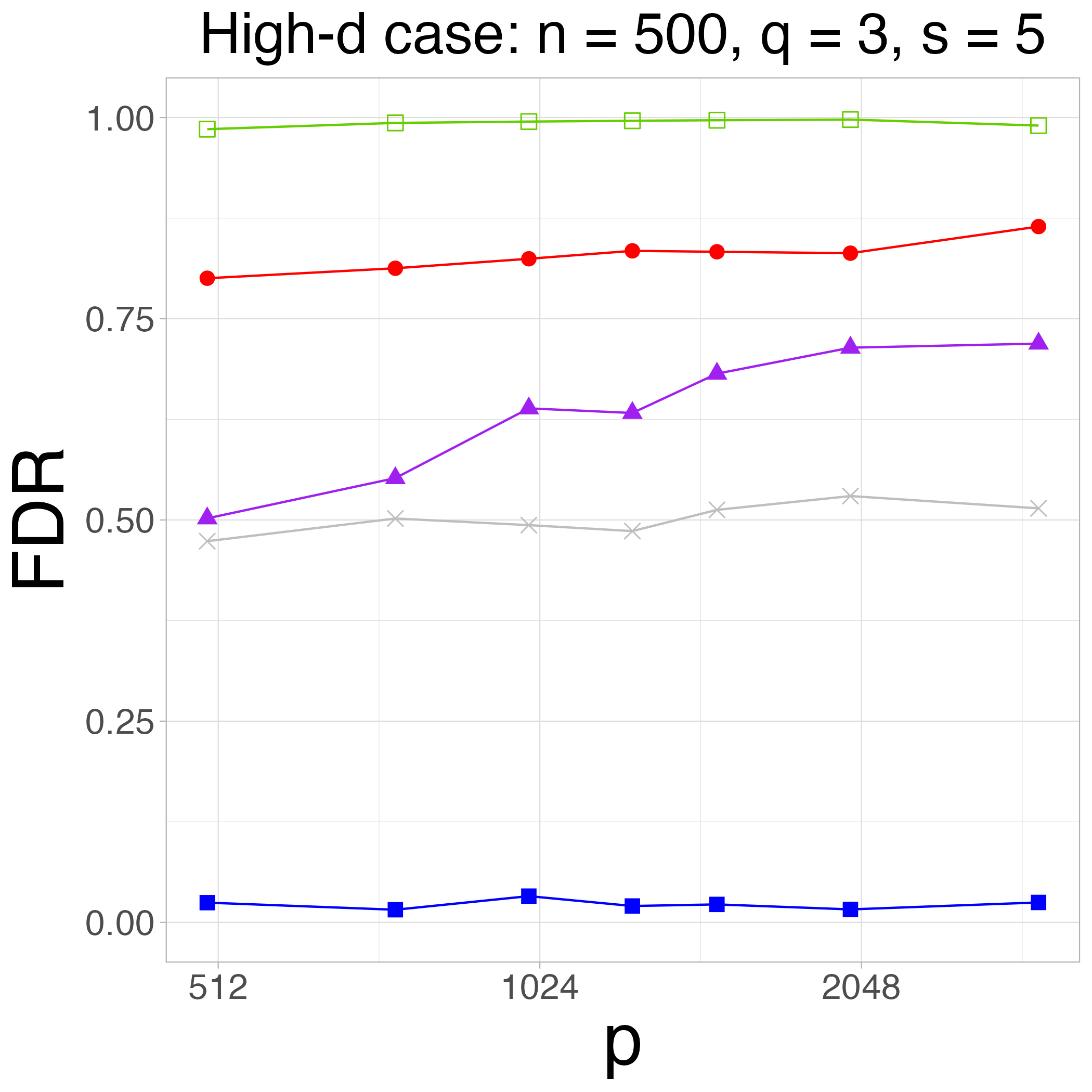}
    \caption{$n = 500$, and $p$ varies from $500$ to $3000$.}
      \label{fig:fixnsparsity_correlated}
  \end{subfigure}
  \hspace*{\fill}
  }
    \caption{False discovery rate (FDR) with non-diagonal $D$ for SIV ($\blacksquare$, blue), Lasso ($\CIRCLE$, red), Trim ($\blacktriangle$, purple), Null ($\square$, green), and IV-Lasso($\times$, grey) based on $1000$ Monte Carlo runs. }
    \label{fig:FDR_nondiagonal}
\end{figure}

The simulation results, presented in Figures  \ref{fig:estimation_nondiagonal} and \ref{fig:FDR_nondiagonal},  illustrate the $\ell_1$-estimation errors and false discovery rates for various methods. These methods exhibit similar trends in $\ell_1$-error performance as observed in the previous setting with uncorrelated errors. The Trim method shows a slight improvement, achieving lower $\ell_1$-errors and false discovery rates compared to the previous simulation results. The performance of the SIV method remains consistent with its performance when $D$ is a diagonal matrix. Additionally, our method continues to outperform the other comparison methods.

\subsection{Simulation studies with non-linear outcome models}
\label{sec:sim_nonlinear}
We then evaluate the performance of our proposed estimator \eqref{eqn:nonlinear2SLS} with nonlinear outcome models. We set  \(q = 2\),  \(s = 2\), and  \(p = 10\). We consider two different settings for $f(X)$. In the first setting,  \(f(X;\beta) = \sum^{10}_{j=1} X_j^3\beta_j\) with  \(\beta = (0.3, 0.3, 0, 0, \ldots, 0)^\T \in \mathbb{R}^{10}\) and  \(g(U) = U_1^3\gamma_1 + U_2^3\gamma_2\). In the second setting,  \(f(X;\beta) = \exp(X^\T \beta)\) and  \(g(U) = (U^3)^{\T}\gamma\). Each element in  \(\Lambda_{j,k}\) and  \(\gamma_k\) is independently generated from  \(\mathbb{N}(0,1)\) for  \(j = 1, \ldots, p\) and  \(k = 1, \ldots, q\). The hidden variables  \(U_{i,k}\) follow i.i.d. standard normal distributions for  \(i = 1, \ldots, n\) and  \(k = 1, \ldots, q\). The random errors are generated as  \(\epsilon_x \sim \mathbb{N}(0, \sigma^2_x I_p)\) and  \(\epsilon_y \sim \mathbb{N}(0, \sigma^2)\), where  \(\sigma_x = 2\) and  \(\sigma = 1\). We evaluate the performance of our estimators with  \(n \in \{1000, \ldots, 5000\}\). All simulation results are based on 1000 Monte Carlo runs. 

For comparison, we did not implement the other methods in Section \ref{sec:simu_linear}, as they are not designed for nonlinear outcome models. Instead, we consider a popular method for addressing endogeneity in high-dimensional settings \citep[e.g.][]{wang2019blessings,ouyang2023high,fan2024latent}, which first estimates the unmeasured confounders $U$ and then directly adjusts for its estimate $\widehat{U}$ in the regression modeling. Specifically, we compare the proposed method with the following two variations of the so-called U-hat method, with the tuning parameter  \(k\) selected using 10-fold cross-validation in all cases:
\begin{enumerate}
   \item[1.] (SIV): We obtain $\widehat{\beta}$ from \eqref{eqn:nonlinear2SLS}. The $\ell_0$-optimization is accelerated using the splicing technique \citep{zhang2023splicing}.
   
   \item[2.] (U-hat1): First, we obtain  \(\widehat{\bm{U}} \in \mathbb{R}^{n \times q}\) using the equation  \(\widehat{\bm{U}} = \bm{X} \widehat{\text{Cov}}(X)^{-1} \widehat{\Lambda}\). Next, we obtain  \(\widehat{\beta}\) by solving the following optimization problem:
   \[
   (\widehat{\beta},\widehat{\gamma}) = \underset{\beta \in \mathbb{R}^p, \gamma \in \mathbb{R}^q}{\argmin} \|\bm{Y} - f(\bm{X}; \beta) - \widehat{\bm{U}} \gamma\|_2^2 \quad \text{subject to} \quad \|\beta\|_0 \leq k.
   \]
   
   \item[3.] (U-hat2): We first obtain  \(\widehat{\bm{U}} \in \mathbb{R}^{n \times q}\) similarly. We then obtain  \(\widehat{\beta}\) by solving the following optimization problem:
   \[
   (\widehat{\beta},\widehat{\gamma}) = \underset{\beta \in \mathbb{R}^p, \gamma \in \mathbb{R}^q}{\argmin} \|\bm{Y} - f(\bm{X}; \beta) - \widehat{\bm{U}}^3 \gamma\|_2^2 \quad \text{subject to} \quad \|\beta\|_0 \leq k,
   \]
   where  \(\widehat{\bm{U}}^3 \in \mathbb{R}^{n \times q}\) is defined as  \(\{\widehat{\bm{U}}^3\}_{i,j} = \{\widehat{\bm{U}}_{i,j}\}^3\). Note that this method assumes knowledge of the specific form of $g(U)$, which is generally not available in practice.
\end{enumerate}

\begin{figure}[!htbp]
    \centering{
    \begin{subfigure}{0.49\textwidth}
    \includegraphics[scale = 0.35]{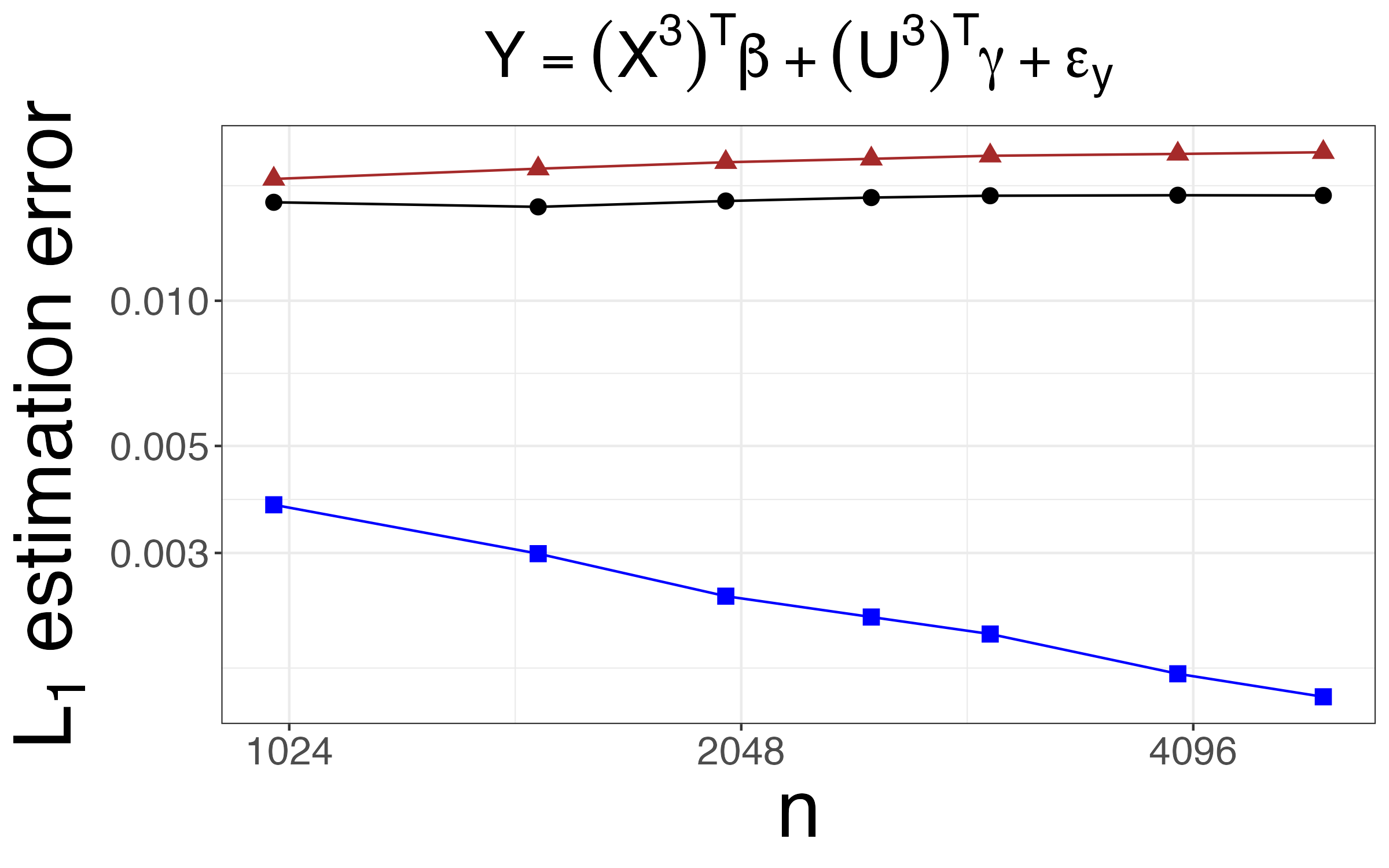}
  \centering
    \caption{Non-linear setting 1.}
     \label{fig:nonlinearsim1}
  \end{subfigure}}
  \hfill
  \centering{
  \begin{subfigure}{0.49\textwidth}
  \centering
    \includegraphics[scale = 0.35]{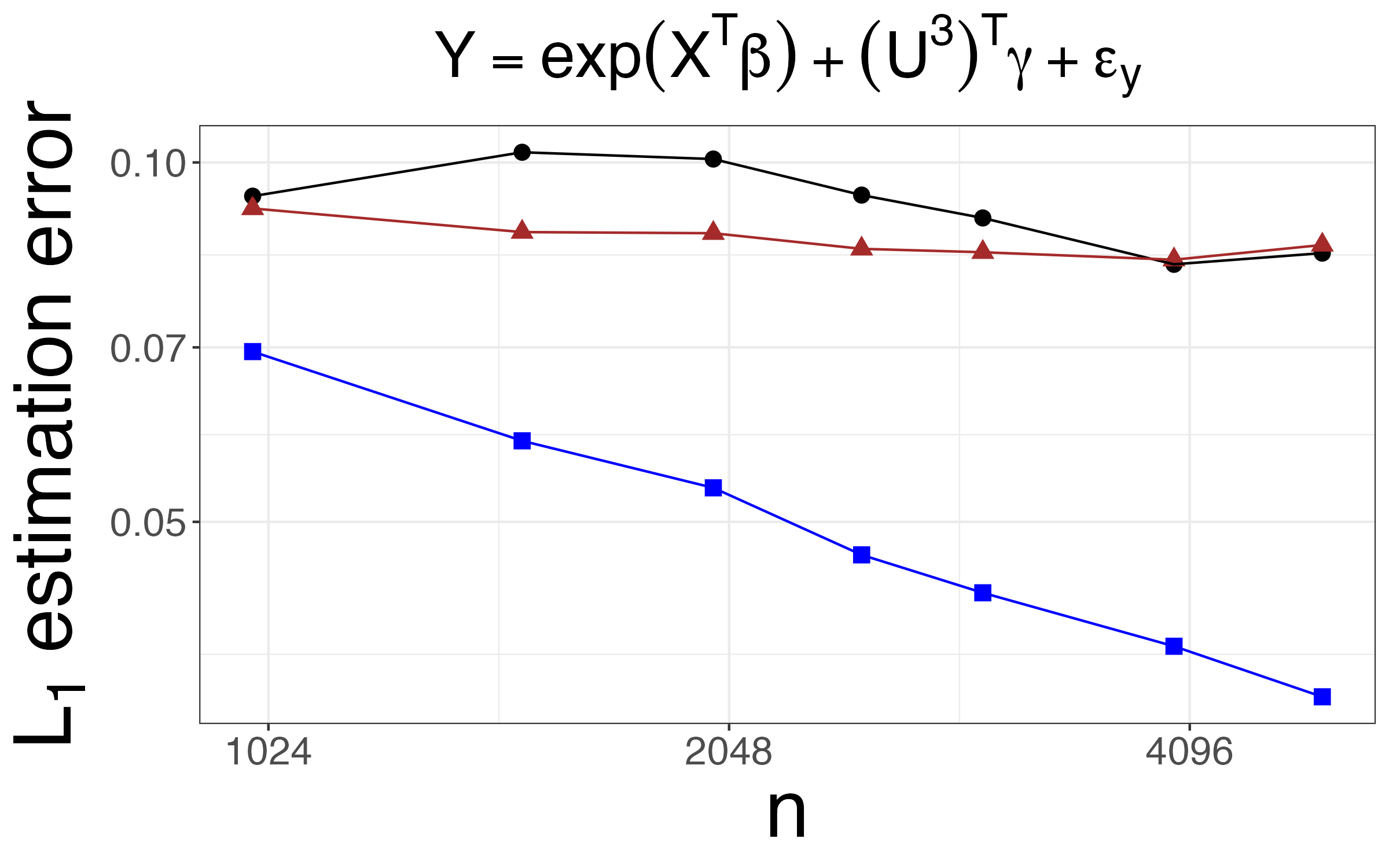}
    \caption{Non-linear setting 2.}
      \label{fig:nonlinearsim2}
  \end{subfigure}
  \hspace*{\fill}
  }
    \caption{Simulation results for nonlinear models with $p = 10$ and $n = 1000, 1500, \dots, 5000$. The methods compared are SIV ($\blacksquare$, blue), U-hat1 ($\CIRCLE$, black), and U-hat2 ($\blacktriangle$, brown). }
    \label{fig:nonlinear}
\end{figure}

Figure~\ref{fig:nonlinear} displays the $\ell_1$-estimation errors for all estimators. The results indicate that U-hat1 and U-hat2 perform similarly, with their biases stabilizing as the sample size increases. In contrast, the bias of the proposed estimator decreases with larger sample sizes, demonstrating its consistency in estimating the causal parameter $\beta$ under nonlinear outcome models. In Section~\ref{sec:nonlinear,nondigonal} of the supplementary material, we further consider a setting with a nonlinear outcome model and a non-diagonal $\text{Cov}(\epsilon_x)$. The results suggest that the SIV method remains consistent in this complex setting.

\begin{remark}
    Under the linear models in Section \ref{sec:simu_linear}, results from the U-hat (including U-hat1 and U-hat2) and SIV methods coincide numerically. While the \( U \)-hat method has been proposed in the literature \citep[e.g.,][]{wang2019blessings}, its validity has been widely challenged \citep[e.g.,][]{ogburn2019comment,grimmer2023naive}. From this perspective, our results may be viewed as a justification for the \( U \)-hat method in the special and unrealistic case of the linear outcome model \eqref{eqn:model-lineary}. 
    
In general, our approach differs fundamentally from the U-hat methods. The U-hat1 method yields consistent estimators of $\beta$ only when the treatment--outcome relationship is linear. As shown in Section~\ref{sec:Uhat} of the supplementary material, U-hat1 fails under the more general model  
\[
Y = f(X; \beta) + g(U) + \epsilon_y,
\]
where $f(X; \beta)$ may be nonlinear. In particular, we derive a necessary condition (Equation~\eqref{eqn:identifiabilityequation_supp}) that must hold for U-hat1 to consistently estimate $\beta$, but this condition is typically violated in nonlinear outcome models. The U-hat2 method, on the other hand, relies on modeling the latent confounder--outcome relationship, introducing additional assumptions that are implausible in practice and do not resolve the underlying identification challenge.

In contrast, our approach constructs instrumental variables, which are agnostic to the form of unmeasured confounding, even if the causal relationship is nonlinear. Because of this, we believe it is a more robust approach in this setting. To our knowledge, we are the first to apply the IV framework to this problem, offering a method that accommodates nonlinear causal effects and arbitrary dependence on unmeasured confounders \( U \).
See Section~\ref{sec:Uhat} of the supplementary material for further discussion and a comparison between the proposed method and the U-hat methods.
\end{remark}

\section{Real data application}
\label{sec:data}
To further illustrate the proposed synthetic instrumental variable method, we reanalyzed a mouse obesity dataset described by \cite{wang2006genetic}. The study involved a cross of 334 mice between the C3H strain and the susceptible C57BL/6J (B6) strain on an ApoE-null background, which were fed a Western diet for 16 weeks. The dataset includes genotype data on 1,327 SNPs, gene expression profiles of 23,388 liver tissue genes, and clinical information such as body weights. \cite{lin2015regularization} previously analyzed this dataset using regularized methods for high-dimensional instrumental variable regression, treating the SNPs as potential instruments, the gene expressions as treatments, and identifying 17 genes likely to affect mouse body weight. \cite{gleason2021robust} also discussed controversies surrounding the use of SNPs as instruments for estimating the effects of gene expression. \cite{miao2023identifying} applied their method to estimate the causal effects associated with these 17 genes. However, their approach cannot be used to estimate effects associated with the full set of 23,388 genes, as it only accommodates low-dimensional exposures. In our analysis, we use the same dataset to identify the genes that influence mouse body weight and to estimate the magnitude of these effects. Notably, our method does not depend on genotype data or other instrumental variables and can handle high-dimensional exposures.

We followed the procedure described in \cite{lin2015regularization} to preprocess the dataset. Genes with a missing rate greater than 0.1 were removed, and the remaining missing gene expression data were imputed using nearest neighbor averaging \citep{troyanskaya2001missing}. We also removed genes that could not be mapped to the Mouse Genome Database (MGD) and those with a standard deviation of gene expression levels less than $0.1$. A marginal linear regression model was fitted between the mice's body weight and sex to subtract the estimated sex effect from the body weight, adjusting for the effect of sex. The fitted residual was used as the outcome $Y$, and the gene expression levels were centered and standardized as multiple treatments. After data cleaning and merging gene expression and clinical data, the final dataset comprised $p = 2819$ genes from $n = 306$ mice ($154$ female and $152$ male).

The estimator proposed by \cite{onatski2010determining} suggests that there are three unobserved latent factors, so we applied our method with $\widehat{q}=3$, using a linear outcome model as the working model. Five genes were found to affect mouse body weight. Under the plurality rule \ref{A4}, we conclude that the causal effects are identifiable as $\widehat{q} + \widehat{s} = 8 \ll p = 2819$. Specifically, our analysis suggests that increasing the gene expression levels of $Igfbp2$, $Rab27a$, $Dct$, $Ankhd1$, and $Gck$ by one standard deviation leads to changes of $-1.98([-2.69,-1.27])$, $1.88([1.26,2.51])$, $1.43([0.86, 2.02])$, $-1.33([-1.90,-0.77])$, and $1.17([0.69,1.66])$ grams in mouse body weight, respectively. The empirical confidence intervals were constructed using the \texttt{ivreg} function; see Section \ref{sec:CI_sup} for more details.

We compared our approach with six methods: (i) the Lasso method; 
(ii) the two-stage regularization (2SR) method \citep{lin2015regularization}, which leverages SNPs as high-dimensional instrumental variables to estimate the causal effect; (iii) the auxiliary variable method \citep{miao2023identifying}, which focuses on the 17 genes identified by \cite{lin2015regularization} as having non-zero effects and uses the five SNPs selected by \cite{lin2015regularization} as auxiliary variables; (iv) the null variable method \citep{miao2023identifying}, which assumes that more than half of the 17 genes have zero effects on mouse body weight; (v) the Trim method \citep{cevid2020spectral}; and (vi) the IV-Lasso method, as detailed in Section \ref{sec:simulation}. Detailed results for these comparison methods are included in Section \ref{sec:comparison} of the supplementary material. The number of active genes found by these methods, defined as genes with non-zero effects on mouse body weight, is 87, 17, 4, 2, 4, and 14, respectively. Consistent with the simulation results, the Lasso method identifies many more active exposures than our method, and its selected set includes all genes identified by our approach. Compared to the other methods, both the 2SR and auxiliary variable methods rely on additional SNP information.

All methods, except the null variable method, identify the expression of the $Igfbp2$ gene (insulin-like growth factor binding protein 2) as a cause of obesity, which is known to prevent obesity and protect against insulin resistance \citep{wheatcroft2007igf}. Additionally, we identify four other genes potentially linked to obesity. The $Rab27a$ gene (Ras-related protein Rab-27A) is involved in insulin granule docking in pancreatic $\beta$ cells, and Rab27a-mutated mice show glucose intolerance after a glucose load \citep{kasai2005rab27a}, suggesting its positive effect on body weight. The $Dct$ gene (Dopachrome Tautomerase) has been associated with obesity and glucose intolerance \citep{kim2015macrophage}, with overexpression observed in the visceral adipose tissue of morbidly obese patients \citep{randhawa2009evidence}. The $Gck$ gene (Glucokinase) plays a key role in blood glucose recognition, and its overexpression is linked to insulin resistance \citep{randhawa2009evidence}, which may explain its impact on body weight.

\section{Discussion}
\label{sec:discussion}
In this paper, we study how to identify and estimate causal effects with a multi-dimensional treatment in the presence of unmeasured confounding. Our key assumption is a sparse causation assumption, which in many contexts serves as an appealing alternative to the widely adopted sparse association assumption. We develop a synthetic instrument approach to identify and estimate causal effects without the need to collect additional exogenous information, such as instrumental variables. Our estimation procedure can be formulated as an $\ell_0$-optimization problem and can therefore be solved efficiently using off-the-shelf packages.

A distinctive feature of our framework is that it allows the use of Algorithm \ref{alg:SIV} to consistently test the sparsity condition \ref{A3} under the other model assumptions. In practice, however, we observe that this test can be unstable in finite samples, particularly in boundary cases where $p \approx s + q$. Developing more stable tests for the sparsity condition \ref{A3} is an interesting avenue for future research.

We have focused on a linear treatment model \eqref{eqn:model1}. In a general nonlinear treatment model, where the relationship between treatment $X$ and the unmeasured factor $U$ is nonlinear, nonlinear factor analysis could be used to fit $X = m(U; \Lambda) + \epsilon$. However, identifying a function $h(\cdot)$ such that $h(X)$ is independent of $U$ remains a significant challenge. Extending this framework to accommodate nonlinear treatment models is left for future research.

We have also focused on the identification and estimation problems. Assuming that the $\ell_0$-penalization procedure \eqref{L0 optimization} accurately selects the true non-zero causal effects, standard M-estimation theory can be used to construct pointwise confidence intervals. However, constructing uniformly valid confidence intervals for the causal parameters remains a challenge, as statistical inference after model selection is typically not uniform \citep{leeb2005model}. One promising approach is to build on a uniformly valid inference method for the standard $\ell_0$-penalization procedure, which, to the best of our knowledge, remains an open problem in the statistical literature.

\section*{Acknowledgements}
We thank Xin Bing, Zhichao Jiang, Wang Miao, Thomas Richardson, James Robins, Dominik Rothenhäusler, and Xiaochuan Shi for their helpful discussions and constructive comments. We also extend our gratitude to the Editor, the Associate Editor, and the anonymous referees for their valuable and thoughtful input, which has significantly improved the quality of this manuscript.

\section*{Supplementary material}
\label{SM}
The supplementary material  contains further discussions of the assumptions and additional simulation results. It also includes more examples and proofs of all the theorems and lemmas.  
\bibliography{ref,ref_revision2}

\clearpage
\appendix

\renewcommand{\thelemma}{S.\arabic{lemma}}
\renewcommand{\thetheorem}{S.\arabic{theorem}}
\renewcommand{\theproposition}{S.\arabic{proposition}}
\renewcommand{\theremark}{S.\arabic{remark}}
\renewcommand{\thealgorithm}{S\arabic{algorithm}}

\setcounter{section}{0}
\renewcommand{\thesection}{S.\arabic{section}}

\setcounter{equation}{0}
\renewcommand{\theequation}{S\arabic{equation}}

\setcounter{figure}{0}
\renewcommand{\thefigure}{S\arabic{figure}}

\setcounter{table}{0}
\renewcommand{\thetable}{S\arabic{table}}

  \title{\bf \Large{Web-based supporting materials for ``The synthetic instrument: From sparse association to sparse causation''}}
   \date{}
   \author{}
  \maketitle

\begin{center}
   \vspace{-10pt}
  \author{\large Dingke Tang$^{1}$, Dehan Kong$^{2}$, and Linbo Wang$^{2}$\footnote{Address for correspondence: Linbo Wang,      Department of Statistical Sciences, University of Toronto, 700 University Avenue, 9th Floor,
Toronto, ON, Canada, M5G 1Z5 \\
Email: linbo.wang@utoronto.ca} \\
  \vspace{10pt}
  $^{1}$Department of Mathematics and Statistics, University of Ottawa, Ottawa, Ontario, Canada.\\
  $^{2}$ Department of Statistical Sciences, University of Toronto, Toronto, Ontario, Canada}\\ \vspace{20pt}

\end{center}

\setcounter{equation}{0}
\setcounter{figure}{0}
\setcounter{table}{0}
\makeatletter
\renewcommand{\theequation}{S\arabic{equation}}
\renewcommand{\thefigure}{S\arabic{figure}}
\renewcommand{\thetable}{S\arabic{table}}
\setcounter{section}{0}


The supplementary material is organized as follows: Section \ref{sec:identification_supp} presents additional discussion for our results. Section \ref{sec:lemmas} contains the lemmas used to prove the theorems. Section \ref{sec:lemmaproof} includes the proofs of the lemmas stated in Section \ref{sec:lemmas}. In Section \ref{sec:thmproof}, we prove the main theorems stated in the paper. Section \ref{sec:comparison} presents the real data results for all comparison methods. Section \ref{sec:additional_simulation} provides additional simulation results. Finally, we discuss the U-hat1 method in Section \ref{sec:Uhat}.


\section*{Notation}
We use $\widehat{\beta}$ to denote the solution to the $\ell_0$ optimization problem \eqref{L0 optimization}, and $\dot{\beta}$ to denote the true value of $\beta$ in model \eqref{eqn:model2}.
Let $\mathcal{A} = \{ j \mid \dot{\beta}_j \neq 0 \}$ with $|\mathcal{A}| = s$.  
Let ${\bf X} \in \mathbb{R}^{n \times p}$ be the design matrix of multiple causes, and let $\widehat{\bf X} = \widehat{\mathbb{E}}({\bf X} \mid \widehat{\mathrm{SIV}})$ denote the projected design matrix.  

Define $\Sigma_X = \mathbb{E}(X X^\top)$, $D = \mathrm{Cov}(\epsilon_x)$, $\widehat{\Sigma}_X = {\bf X}^\top {\bf X}/(n-1)$, and $\widehat{\Sigma}_{\widehat{X}} = \widehat{\bf X}^\top \widehat{\bf X}/(n-1)$.  
For two positive sequences $a_n$ and $b_n$, we write $a_n \lesssim b_n$ if there exists a constant $C > 0$ such that $a_n \leq C b_n$ for all $n$; $a_n \asymp b_n$ if both $a_n \lesssim b_n$ and $b_n \lesssim a_n$; and $a_n \ll b_n$ if $\limsup_{n \to \infty} a_n/b_n = 0$.  

For a matrix $A$, we use $A_{\cdot,j}$ and $A_{i,\cdot}$ to denote its $j$th column and $i$th row, respectively. For an index set $J$, let $A_{J,\cdot}$ and $A_{\cdot,J}$ denote the submatrices of $A$ containing only rows or columns in $J$, while $A_{-J,\cdot}$ and $A_{\cdot,-J}$ denote the submatrices obtained by deleting the corresponding rows or columns.  
We use $\|A\|_F$, $\|A\|_2$, and $\|A\|_\infty$ to denote the Frobenius norm, spectral norm, and element-wise maximum norm of $A$, respectively.  
We write $\mu_i(A)$ for the $i$th singular value of $A$, and $\lambda_i(A)$ for its $i$th eigenvalue. The maximum and minimum eigenvalues of $A$ are denoted by $\lambda_{\max}(A)$ and $\lambda_{\min}(A)$, respectively.  

Throughout the appendix, we assume that $q$ is known; the estimation of $q$ is discussed in Section~\ref{rem:Onatski}.  

In the high-dimensional setting where $p$ is allowed to diverge, consider the singular value decomposition (SVD) and principal component analysis (PCA) of ${\bf X}$ and ${\bf X}^\top {\bf X}/(n-1)$, respectively:
\begin{equation}\label{eqn:singular,pca}
\begin{split}
    {\bf X} &= \sqrt{n-1}\, (\widehat\eta_1\;\widehat\eta_2\;\ldots\;\widehat\eta_p)
        \,\mathrm{diag}(\sqrt{\widehat\lambda}_1,\ldots,\sqrt{\widehat\lambda}_p)
        ({\widehat\xi}_1\;\ldots\;{\widehat\xi}_p)^\top,\\
    \frac{{\bf X}^\top {\bf X}}{n-1} &= 
        ({\widehat\xi}_1\;\ldots\;{\widehat\xi}_p)\,
        \mathrm{diag}(\widehat\lambda_1,\ldots,\widehat\lambda_p)\,
        ({\widehat\xi}_1\;\ldots\;{\widehat\xi}_p)^\top.
\end{split}
\end{equation}
Here $\widehat\lambda_1 \geq \widehat\lambda_2 \geq \cdots \geq \widehat\lambda_k > 0 = \widehat\lambda_{k+1} = \cdots = \widehat\lambda_p$, with $k = \mathrm{Rank}({\bf X})$.  
Define $\widehat{\Lambda} = (\sqrt{\widehat\lambda}_1 \widehat\xi_1\;\ldots\;\sqrt{\widehat\lambda}_q \widehat\xi_q)$,  
${B}_{\widehat\Lambda^\perp} = (\widehat\xi_{q+1}\;\widehat\xi_{q+2}\;\ldots\;\widehat\xi_p)$,  
and $H = (\widehat\eta_1\;\widehat\eta_2\;\ldots\;\widehat\eta_q) \in \mathbb{R}^{n \times p}$ such that $H^\top H = I_q$.

\section{Discussion on Assumptions}
\label{sec:identification_supp}

\subsection{Discussion on condition A1}
\label{sec:discussA1}

\subsubsection*{Implications of Assumption A1 when $q=1$} 

If $\mathrm{Cov}(\epsilon_x)$ is diagonal, then Assumption A1 implies that any $q \times q$ submatrix of $\Lambda$ is invertible. In particular, when $q=1$, this means that no element of $\Lambda$ can be zero. However, this need not hold when $\mathrm{Cov}(\epsilon_x)$ is not diagonal. To see this, let $D := \mathrm{Cov}(\epsilon_x)$. By the Woodbury matrix identity, we have
\begin{equation*}
   \begin{split}
\mathrm{Cov}(X)^{-1}\Lambda &= (D + \Lambda^\top \Lambda)^{-1}\Lambda\\
&= \{D^{-1} - D^{-1}\Lambda(I_q + \Lambda^\top D^{-1}\Lambda)^{-1}\Lambda^\top D^{-1}\}\Lambda  \\
&= D^{-1}\Lambda - D^{-1}\Lambda(I_q + \Lambda^\top D^{-1}\Lambda)^{-1}\Lambda^\top D^{-1}\Lambda \\
&= D^{-1}\Lambda - D^{-1}\Lambda(I_q + \Lambda^\top D^{-1}\Lambda)^{-1}(\Lambda^\top D^{-1}\Lambda + I_q - I_q) \\
&= D^{-1}\Lambda( I_q  + \Lambda^\top D^{-1}\Lambda  )^{-1}.
\end{split}
\end{equation*}
If $D$ is diagonal and Assumption A1 holds—so that any $q \times q$ submatrix of $\mathrm{Cov}(X)^{-1}\Lambda \in \mathbb{R}^{p \times q}$ is invertible—then any $q \times q$ submatrix of $\Lambda$ must also be invertible. For $q = 1$, this implies that no element of $\Lambda$ is zero. 

Nevertheless, even if some $\Lambda_j$ are zero, {\bf the causal parameter \(\beta\) remains identifiable} under analogous conditions, and {\bf the SIV method can still be used to identify and estimate the parameter of interest}.

We now discuss how to identify $\beta$ under these weaker conditions. Since $\Lambda$ is identifiable up to a rotation, we can still identify the set $\{j: \Lambda_j = 0\}$. Define $\mathcal{A} = \{j:\; \beta_j \neq 0\}$ and $\mathcal{B} = \{j:\; \Lambda_j \neq 0\}$. Let $p = \dim(X)$, $p_0 = \#\mathcal{B}$, $s = \|\beta\|_0$, and $s_0 = \#(\mathcal{A}\cap \mathcal{B})$. Because $\Lambda_{\mathcal{B}^c} = 0$, there is no confounder that can affect $X_{\mathcal{B}^c}$ and $Y$, so $\beta_{\mathcal{B}^c}$ can be obtained directly via regression of $Y$ on $X_{\mathcal{B}^c}$. Identification of $\beta_{\mathcal{B}}$ then follows from Theorem \ref{thm:l0 plurality}. Specifically, $\beta_{\mathcal{B}}$ is identifiable if $s_0 \leq (p_0-1)/2$ under the majority rule, or if $s_0 < (p_0-1)$ under the plurality rule. These conditions parallel Assumptions A3 and A3' in the main text.

Next, we illustrate how to use the SIV method to estimate $\beta$ when some $\Lambda_j = 0$. Consider a scenario suggested by a reviewer: we have five treatments $(X_1, \ldots, X_5)$. Among them, $X_1$, $X_4$, and $X_5$ affect $Y$, while $X_4$ and $X_5$ are not confounded by $U$. This setting is shown in Figure \ref{fig:0lambda_supp}.

\begin{figure}
  \centering  
  \scalebox{0.75}{
  \begin{tikzpicture}[->,>=stealth',shorten >=1pt,auto,node distance=2.3cm,
     semithick, scale=0.50]
     pre/.style={-,>=stealth,semithick,blue,ultra thick,line width = 1.5pt}]
     \tikzstyle{every state}=[fill=none,draw=black,text=black]
    \node[shade] (U) at (0, 2){$U$};
     \node[] (X2) at (0,-4) {$X_2$};
     \node[] (X1) at (-4,-4) {$X_1$};
     \node[] (X3) at (4,-4) {$X_3$};
     \node[] (X4) at (8, -4) {$X_4$};
     \node[] (X5) at (12,-4) {$X_5$};
     \node[] (Y) at  (0,-10) {{$Y$}};
    \path  (U) edge [left] node {$\Lambda_1$} (X1);
    \path (U) edge [left] node {$\Lambda_2$} (X2);
    \path (U) edge node {$\Lambda_3$} (X3);
    \path (X1) edge [right] node {$\dot\beta_1$} (Y)
          (X4) edge [left] node {$\dot\beta_4$} (Y)
          (X5) edge node {$\dot\beta_5$} (Y);
    \draw (0.8,2) arc(90:-90:13cm and 6cm);
      \node[] (m) [right = 0.5cm of X5] {${\gamma}$};
    \end{tikzpicture}
    }
    \caption{Graphical illustration for unconfounded treatments.}
\label{fig:0lambda_supp}
\end{figure}

Suppose $\Lambda_1 = \Lambda_2=1$, $\Lambda_3 = 2$, and $\Lambda_4 = \Lambda_5 = 0$. Then
\[
\Lambda = \begin{pmatrix}
1\\
1\\
2\\
0\\
0
\end{pmatrix},\quad  
B_{\Lambda^\perp} = \begin{pmatrix}
\frac{1}{\sqrt{2}}  & \frac{1}{\sqrt{3}}  & 0 &0 \\
-\frac{1}{\sqrt{2}} & \frac{1}{\sqrt{3}}  & 0 &0 \\
0  & -\frac{1}{\sqrt{3}} & 0 &0 \\
0  & 0  & 1 &0 \\
0  & 0  & 0 &1 
\end{pmatrix},\quad 
SIV = B_{\Lambda^\perp}^\top X = \begin{pmatrix}
        \frac{X_1- X_2}{\sqrt{2}}\\
        \frac{X_1+X_2-X_3}{\sqrt{3}}\\
        X_4 \\
        X_5   
\end{pmatrix}.
\]

In this setting, variables $\{X_i\}$ with $i \in \{1,2,3\}$ are uncorrelated with those $\{X_j\}$ with $j \in \{4,5\}$. Linear regression of $X$ on $SIV$ yields $\widehat{X} = (\widehat{X}_1, \ldots, \widehat{X}_5)^\top$, where $\widehat{X}_i = \mathbb{E}(X_i \mid (X_1 - X_2)/\sqrt{2}, (X_1+X_2 - X_3)/\sqrt{3})$ for $i \in \{1,2,3\}$, while $\widehat{X}_4 = X_4$ and $\widehat{X}_5 = X_5$. Note that in this case, the two unconfounded treatments are themselves SIVs, and $(\widehat{X}_1,\widehat{X}_2,\widehat{X}_3)$ remain uncorrelated with $(\widehat{X}_4,\widehat{X}_5)$.

To show that regressing $Y$ on $\widehat{X}$ is equivalent to running two separate regressions—one of $Y$ on $(\widehat{X}_1, \widehat{X}_2, \widehat{X}_3)$ with a sparsity constraint, and another of $Y$ on $(X_4, X_5)$—consider
\begin{equation}
    \label{eqn:lambda0e1_supp}
        \widehat{\beta} = \underset{\beta \in \mathbb{R}^{5},  \|\beta\|_0 \leq k}{\arg\min} \; \mathbb{E}\left( Y - \sum_{j=1}^5 \widehat{X}_j \beta_j \right)^2. 
\end{equation}
Let $\dot\beta = (\dot\beta_1, 0, 0, \dot\beta_4, \dot\beta_5)^\top$, $Y_1 = \dot\beta_1 X_1 + U\gamma + \epsilon_y$, and $Y_2 = \dot\beta_4 X_4 + \dot\beta_5 X_5$, so $Y = Y_1 + Y_2$. Then
\begin{equation}
\label{eqn:lambda0e2_supp}
    \begin{split}
        \mathbb{E}\Big(Y - \sum_{j=1}^5 \widehat{X}_j \beta_j\Big)^2 
        &= \mathbb{E}\Big\{(Y_1 - \sum_{j=1}^3 \widehat{X}_j \beta_j) + (Y_2 - \sum_{j=4}^5 \widehat{X}_j \beta_j)\Big\}^2 \\
        &= \mathbb{E}(Y_1 - \sum_{j=1}^3 \widehat{X}_j \beta_j)^2 
         + \mathbb{E}(Y_2 - \sum_{j=4}^5 X_j \beta_j)^2,
    \end{split}
\end{equation}
where the last equality holds because the two terms are uncorrelated. Thus,
\[
\widehat{\beta} = \underset{\beta \in \mathbb{R}^{5},  \|\beta\|_0 \leq k}{\arg\min} 
\Big\{\mathbb{E}(Y_1 - \sum_{j=1}^3 \widehat{X}_j \beta_j)^2 
+ \mathbb{E}(Y_2 - \sum_{j=4}^5 \widehat{X}_j \beta_j)^2\Big\}.
\]

Minimizing over $\beta_{4,5} \in \mathbb{R}^2$ gives $(\widehat{\beta}_4, \widehat{\beta}_5) = (\dot\beta_4, \dot\beta_5)$, while minimizing over $\beta_{1,2,3} \in \mathbb{R}^3$ with $\|\beta_{1,2,3}\|_0 \leq 1$ yields $(\widehat{\beta}_1, \widehat{\beta}_2, \widehat{\beta}_3) = (\dot\beta_1, 0, 0)$ by Theorem 1. Hence, our algorithm successfully identifies and estimates $\beta$ even when some $\Lambda_j = 0$. This argument extends directly to the more general case where $p > 3$, $q = 1$, and some $\Lambda_j = 0$.

\subsubsection*{Discussion of Assumption A1 when \( q \geq 2 \)}

If \( q \geq 2 \) and \(\text{Cov}(\epsilon_x)\) is a diagonal matrix, Assumption A1 implies that any \( q \times q \) submatrix of \( \Lambda \) should be invertible. We now discuss cases where Assumption A1 is violated when \( q \geq 2 \):

(i.) If all \( q \times q \) submatrices of \( \Lambda \) are not invertible, we may find \(\tilde{U}\) with \(\text{dim}(\tilde{U}) = q_0 < q\) to quantify the unmeasured confounding, thus simplifying the analysis to the \( q_0 < q \) scenario. Here, \(\text{Rank}(\Lambda) = q_0 < q\), and the matrix can be decomposed using a rank factorization as \( \Lambda = \widetilde{\Lambda} \cdot C \), where \( \widetilde{\Lambda} \) is a \( p \times q_0 \) matrix with full column rank, and \( C \) is a \( q_0 \times q \) matrix with full row rank. We define \( \widetilde{U} = C U \), indicating that the treatments \( X \) are confounded by \( \widetilde{U} \in \mathbb{R}^{q_0} \).
\[
X = \Lambda U + \epsilon_x = \widetilde{\Lambda} \widetilde{U} + \epsilon_x
\]

(ii.) When a row \( \Lambda_i \in \mathbb{R}^q \) is zero, our algorithm remains effective. In such cases, a column in \( B_{\Lambda^\perp} \) corresponds to \( e_i \), where \( e_i \in \mathbb{R}^p \) is the unit vector with its \( i \)th element as 1 and all other elements as 0. Consequently, a variable in SIV will be \( X_i \) itself, and \( \widehat{X}_i \) will also be \( X_i \). Regression of \( Y \) on \( \widehat{X} \) will yield the causal parameter \( \beta \), similar to the scenario when \( q=1 \).

(iii.) In cases where some \( q \times q \) submatrices of \( \Lambda \in \mathbb{R}^{p \times q} \) are not invertible, we argue that such scenarios are rare. It would require identifying two groups of treatments, \(\{X_{i1}, X_{i2}, \ldots, X_{iq}\}\) and \(\{X_{j1}, X_{j2}, \ldots, X_{jq}\}\), where the effects from \( U \) to \(\{X_{i1}, X_{i2}, \ldots, X_{iq}\}\) are linearly dependent, while those to \(\{X_{j1}, X_{j2}, \ldots, X_{jq}\}\) are linearly independent. In practice, finding confounders with such a structure is unlikely.

\subsection{Discussion on Condition \ref{A4}}
\label{sec:disA4}
We now discuss the plurality rule. First, consider the simplest case where \( q = 1 \). Define the adjusted loading matrix \(\widetilde{\Lambda} = \text{Cov}^{-1}(X) \Lambda \in \mathbb{R}^{p \times 1}\). The plurality rule is violated only if there exist two indices \(1 \leq i, j \leq p\) such that
\begin{align}
        \tag{a}\label{eqn:a} &\beta_i \neq 0, \;\; \beta_j \neq 0,\\
        \tag{b}\label{eqn:b} &\frac{\beta_i}{\widetilde{\Lambda}_i} = \frac{\beta_j}{\widetilde{\Lambda}_j}.
\end{align}
Equation \eqref{eqn:b} implies that the causal effects of \( X_i \) and \( X_j \) on the outcome \( Y \) are \textbf{exactly proportional} to their corresponding adjusted factor loadings, which is unlikely to occur in practice.

We now discuss the more general setting where \( q \geq 2 \). Define the adjusted loading matrix \(\widetilde{\Lambda} = \text{Cov}^{-1}(X) \Lambda \in \mathbb{R}^{p \times q}\). Let \(C^*_{(1)}, \ldots, C^*_{(q+1)}\) be subsets of \(\{1, 2, \ldots, p\}\) with cardinality \( q \) and \(\dot{\beta}_{C^*_{(i)}} \neq 0\). The plurality rule is violated only if the following equation holds:
\begin{equation}
    \tag{c}
    \label{eqn:c}
    \widetilde{\Lambda}^{-1}_{\{C^*_{(1)}, \cdot\}} \beta_{C^*_{(1)}} = \ldots = \widetilde{\Lambda}^{-1}_{\{C^*_{(q+1)}, \cdot\}} \beta_{C^*_{(q+1)}}.
\end{equation}
It is unlikely to find \( q+1 \) different subsets \( C^*_{(1)}, \ldots, C^*_{(q+1)} \) such that equation \eqref{eqn:c} holds.

These conditions are similar in spirit to the faithfulness assumption commonly assumed in the causal discovery literature \citep{pearl2009causality_sup}; we refer interested readers to \cite{uhler2013geometry_sup} for more discussions related to this topic.

We now present an example where the plurality rule is violated. 
\begin{example}
Assume \( q = 1 \). \( (X, Y) \) are generated via the equations \( X = \Lambda U + \epsilon_x \) and \( Y = X^\top \dot{\beta} + U \gamma + \epsilon_y \), with parameters \( \Lambda = (1, 1, \ldots, 1)^\top \in \mathbb{R}^{p \times 1} \), \( \dot{\beta}_1 = \dot{\beta}_2 = \ldots = \dot{\beta}_s = 1 \), \( \dot{\beta}_{s+1} = \ldots = \dot{\beta}_p = 0 \), \(\gamma = 1\), and random variables \(U, \epsilon_y \sim \mathbb{N}(0,1)\), \( \epsilon_x \sim \mathbb{N}(0, I_p)\).

We examine how the identifiability of \( \dot{\beta} \) varies with different values of \( s \). Let \( SIV = B_{\Lambda^\perp}^\top X \) and \( \widetilde{X} = \mathbb{E}(X \mid SIV) \). We have the following result for regression \( Y \sim \widetilde{X} \):
\begin{equation*}
    \begin{split}
    &\argmin_{\beta \in \mathbb{R}^p, \; \|\beta\|_0 < p-1} \mathbb{E}(Y - \widetilde{X} \beta)^2 = \{\dot{\beta}, \;  1_p - \dot{\beta} \}.\\
    \end{split}
\end{equation*}
\end{example}

We now provide the proof of this example. From Lemma \ref{lem:l0 structure}, we know that 
$$
\argmin_{\beta \in \mathbb{R}^p} \mathbb{E}(Y-\widetilde{X}\beta)^2 = \{\dot \beta + \Sigma_X^{-1}\Lambda\alpha \mid \alpha \in \mathbb{R}\}.
$$

In this example, $\Sigma_X^{-1}\Lambda = (I_p + \Lambda \Lambda^\top)^{-1} \Lambda = \Lambda/(1+p)$. Adding the sparsity constraint, we have 
\begin{equation*}
    \argmin_{\beta \in \mathbb{R}^p,\; ||\beta||_0 < p-1} \mathbb{E}(Y-\widetilde{X}\beta)^2 
    =\{\dot\beta + \Lambda\alpha \mid \alpha\in \mathbb{R},\;||\dot\beta + \Lambda\alpha||_0 < p-1 \} 
    =  \{\dot \beta,\;  1_p -\dot\beta  \},\\
\end{equation*}
So far, we have proven the equations in the above example. This example demonstrates that \( s < p - q \) does not guarantee the identifiability of \(\beta\), as the regression yields two possible solutions: \(\dot{\beta}\) and \( 1_p - \dot{\beta} \). In scenarios where the plurality rule is violated, as shown in this example, reduced-rank regression with the constraint \( s < p - q \) cannot uniquely determine \(\beta\).

\subsection{Weak instruments and identification}
\label{sec:weakivandidentification}

Instead of the setting described in Section~\ref{sec:discussA1} and Figure~\ref{fig:0lambda_supp}, one reviewer pointed out a challenging scenario in which some variables are unconfounded and have no effect on the outcome. Specifically, consider the case where \( p = 5 \), \( q = 1 \), and \( s = 3 \), with only \( X_1 \), \( X_2 \), and \( X_3 \) affecting \( Y \), and loadings \( \Lambda_4 = \Lambda_5 = 0 \). In this setting, the parameter of interest is not identifiable. We now discuss the performance of our estimator under this challenging case.

It is still possible to test whether the parameter is identifiable in this scenario. We describe an approach based on assessing the {\bf uniqueness} of the solution to the relevant optimization problem, which serves as a proxy for the identifiability of \( \beta \).

\subsection*{Setting, Observation, and Algorithm}

Consider the scenario you suggested, where the latent confounder $U$ affects treatments $X_1$, $X_2$, and $X_3$, and these variables also have nonzero causal effects on the outcome. Figure~\ref{fig:0lambda_sup_new} provides a graphical illustration of this setting.

\begin{figure}[ht]
  \centering  
  \scalebox{0.7}{
  \begin{tikzpicture}[->,>=stealth',shorten >=1pt,auto,node distance=2.3cm,
     semithick, scale=0.50]
     pre/.style={-,>=stealth,semithick,blue,ultra thick,line width = 1.5pt}]
     \tikzstyle{every state}=[fill=none,draw=black,text=black]
    \node[shade] (U) at (0, 2){$U$};
     \node[] (X2) at (0,-4) {$X_2$};
     \node[] (X1) at (-4,-4) {$X_1$};
     \node[] (X3) at (4,-4) {$X_3$};
     \node[] (X4) at (8, -4) {$X_4$};
     \node[] (X5) at (12,-4) {$X_5$};
     \node[] (Y) at  (0,-10) {{$Y$}};
    \path  (U) edge [left] node {$\Lambda_1$} (X1);
    \path (U) edge [left] node {$\Lambda_2$} (X2);
    \path (U) edge node {$\Lambda_3$} (X3);
    \path 
      (X1) edge [right] node {$ {\dot\beta_1}$} (Y)
      (X2) edge [right] node {$ {\dot\beta_2}$} (Y)
      (X3) edge node {$ {\dot\beta_3}$} (Y);
    \draw (0.8,2) arc(90:-90:13cm and 6cm);
    \node[] (m) [right = 0.5cm of X5] {${\gamma}$};
    \end{tikzpicture}
    }
    \caption{Graphical illustration of the setting described in Section~\ref{sec:weakivandidentification}.}
\label{fig:0lambda_sup_new}
\end{figure}

When applying our algorithm, we obtain $\widehat{s} = \|\widehat{\beta}\|_0 = 2$. Hence the sparsity condition $\widehat{s} + \widehat{q} = 3 < 4 = 5 - 1 = p - q$ is satisfied. A direct application of the sparsity check in equation~\eqref{eqn:nsc} could therefore lead to the erroneous conclusion that $\beta$ is identifiable. Instead, motivated by Theorem~\ref{thm:l0 plurality}, we conclude that $\beta$ is not identifiable by observing that the solution set
\begin{equation}
    \label{eqn:solution}
    \left\{\widehat{\beta}: \widehat{\beta} \in \underset{\beta \in \mathbb{R}^5, \, \|\beta\|_0=2}{\argmin} \|Y-\widehat{X}\beta\|_2^2\right\}
\end{equation}
is not unique. This non-uniqueness arises both at the population level and in finite-sample numerical settings. 

Specifically, consider the following minimizers:
\begin{align*}
    &\widehat{\beta}^{(1,2)} = \underset{\beta_1 \neq 0,\; \beta_2 \neq 0,\;\beta_3=\beta_4=\beta_5=0}{\argmin} \|Y-\widehat{X}\beta\|_2^2,\\
    &\widehat{\beta}^{(1,3)} = \underset{\beta_1 \neq 0,\; \beta_3 \neq 0,\;\beta_2=\beta_4=\beta_5=0}{\argmin} \|Y-\widehat{X}\beta\|_2^2,\\
    &\widehat{\beta}^{(2,3)} = \underset{\beta_2 \neq 0,\; \beta_3 \neq 0,\;\beta_1=\beta_4=\beta_5=0}{\argmin} \|Y-\widehat{X}\beta\|_2^2.
\end{align*}
We observe that $\widehat{\beta}^{(1,2)}$, $\widehat{\beta}^{(1,3)}$, and $\widehat{\beta}^{(2,3)}$ all belong to the solution set \eqref{eqn:solution}, confirming its non-uniqueness. Based on this observation, we develop Algorithm~\ref{alg:identifiability_sup} to formally test the identifiability of $\beta$. We also perform simulation studies, which demonstrate that this test performs well in finite samples.

\begin{algorithm}[!ht]
\caption{\quad Testing Identifiability of Causal Effects via Synthetic Instruments and Uniqueness}
\label{alg:identifiability_sup}
  \hspace*{0.02in} {\bf Input: }  
  ${\bf X} \in \mathbb{R}^{n\times p}$ (centered), ${\bf Y} \in \mathbb{R}^{n\times 1}$  
  \hspace*{0.02in} 
  \begin{algorithmic}[1]  
    \State Obtain $\widehat{\bm X}$ using Algorithm~\ref{alg:SIV} in the manuscript;
    \State Obtain $\widehat{\beta}$ and $\widehat{s}$ by solving \eqref{L0 optimization} in the manuscript with cross-validation;
    \State Set Identifiability = True;
    \State Define the collection of index sets $M=\{\mathcal{A}\mid \mathcal{A}\subset \{1,2,3,\dots,p\},|\mathcal{A}|=\widehat{s}\}$;
    \State For $\mathcal{A}\in M$, define the optimizer: 
    $$\widetilde{\beta}^{(\mathcal{A})}:= \argmin_{\beta \in \mathbb{R}^p: \beta_{\mathcal{A}^c}=0}\|\bm Y-\widehat{\bm X}\beta\|_2^2.$$

    \If {  
      $\exists\; \widetilde{\beta}^{(\mathcal{A})}$, where $\mathcal{A}\in M$, such that 
    $$
   \| \widetilde{\beta}^{(\mathcal{A})}-\widehat{\beta}\|_0 \neq 0, \quad \text{and}\quad\|\bm Y-\widehat{\bm X}\; \widetilde{\beta}^{(\mathcal{A})}\|_2^2 - \|\bm Y-\widehat{\bm X}\;\widehat{\beta}\|_2^2 \leq tol
    $$} 
       Identifiability = False;
    \ElsIf{ $\widehat{s}+\widehat{q} \geq p $ } 
      \State Identifiability = False;
    \EndIf
    \State \Return Identifiability.
  \end{algorithmic}
\end{algorithm}

\subsection*{Simulation and Results}

We now evaluate the performance of Algorithm~\ref{alg:identifiability_sup} in a finite-sample setting. We conducted a simulation study with 1,000 repetitions, each involving $p = 5$ predictors. We considered scenarios in which $s = 1, 2,$ or $3$ of the predictors were causal. The data included a single unobserved confounder ($q = 1$), specified by a loading matrix $\Lambda = (3, 1, 2, 0, 0)^\top$ and an effect size of $\gamma = 1$. The true coefficients for the relevant predictors were $\beta_1=\cdots=\beta_s = 1$. In each repetition, we generated the unobserved confounder $U \sim \mathcal{N}(0, 1)$, followed by the treatments $X = U \Lambda + \epsilon_X$, where $\epsilon_{X} \sim \mathcal{N}(0, I_5)$. The outcome variable was then generated as $Y = X\beta + U\gamma + \epsilon_Y$, where $\epsilon_Y \sim \mathcal{N}(0, 1)$. 

\begin{figure}[ht]
    \centering
    \includegraphics[width=0.6\linewidth]{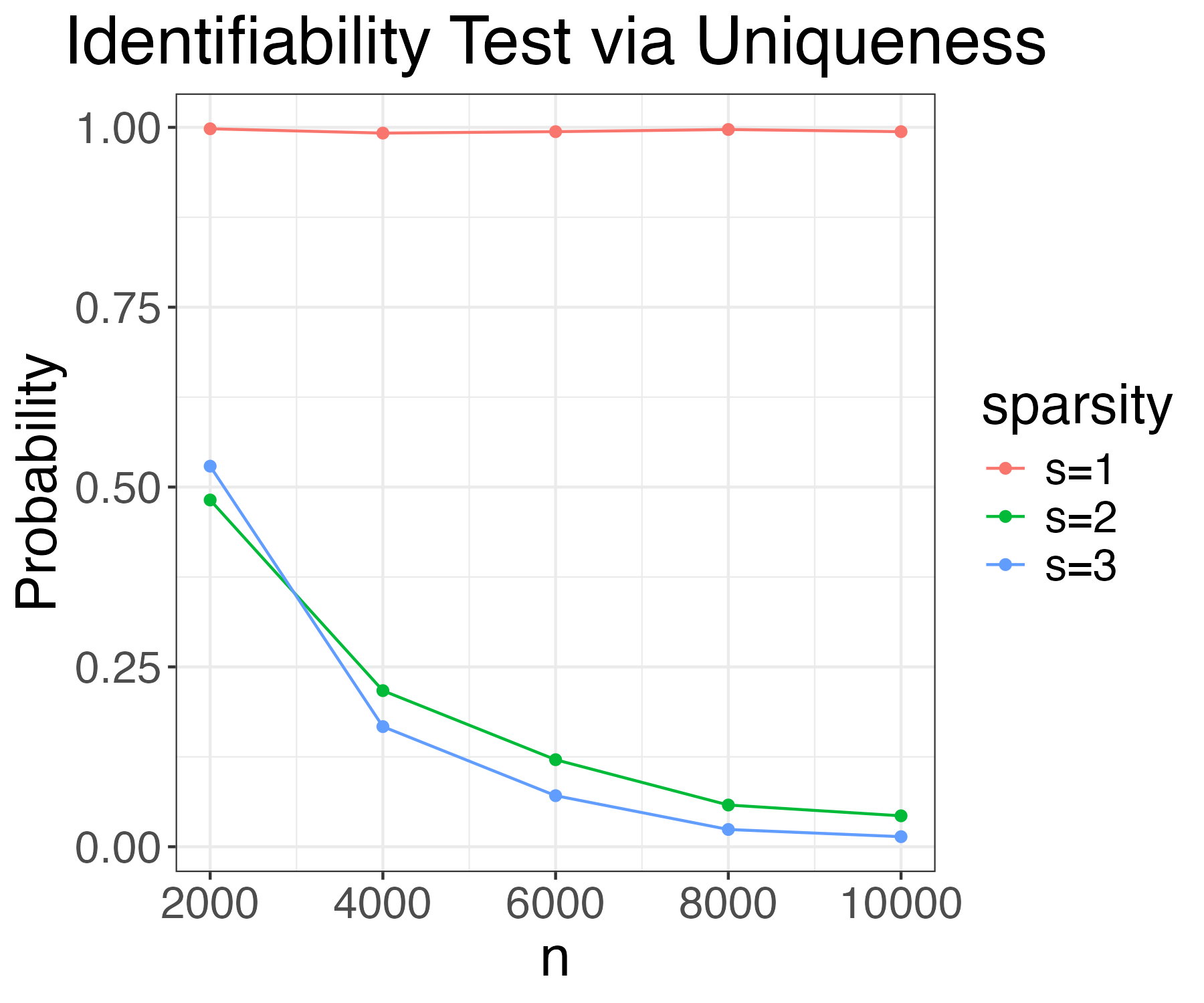}
    \caption{The probability that Algorithm~\ref{alg:identifiability_sup} concludes that the model is identifiable for different $s$ and $n$. Results are based on 1,000 Monte Carlo runs.}
    \label{fig:identifiability_sup}
\end{figure}

We applied our algorithm to determine whether the parameter $\beta$ is identifiable. The parameters described above were fixed, and the sample size varied from 2,000 to 10,000. To account for numerical precision, we set $tol = 10^{-5}$. The simulation results are presented in Figure~\ref{fig:identifiability_sup}. As shown in the figure, when $s=1$ and the parameter is identifiable, the algorithm correctly recognizes identifiability with probability close to 1 across all settings. When $s = 2$ or $3$, where the parameter is not identifiable, the algorithm correctly detects non-identifiability with high probability as the sample size increases. These results confirm that our algorithm performs well in finite samples.

\subsection{Determining $q$}
\label{rem:Onatski}

In practice, when the number of unmeasured confounders is unknown, it is necessary to conduct a test to determine this quantity. Our treatment model utilizes a factor model with \( q \) latent factors, suggesting that techniques from factor analysis can be employed to identify the number of latent factors, and consequently, the number of unmeasured confounders. We adopt the estimator introduced by \cite{onatski2010determining_sup}, which determines the number of unobserved confounders based on the maximum eigengap. The proposed criterion for the eigenvalue difference is as follows:
\begin{equation*}
\widehat{q} = \max \{ i \leq r_{\max} : \widehat{\lambda}_i - \widehat{\lambda}_{i+1} \geq t_0 \},
\end{equation*}
where \( t_0 \) is a given threshold, \( r_{\max} \) is the prespecified maximum number of factors, and \( \widehat{\lambda}_1, \widehat{\lambda}_2, \ldots \) are the eigenvalues of the matrix \(\bm{X^\T X}/n\) in decreasing order. We select \( t_0 \) using the method provided in their paper and set \( r_{\max} = 10 \). We directly apply their algorithm in our simulations, which performs well in both high- and low-dimensional settings.

\section{Lemmas}
\label{sec:lemmas}

\subsection{Lemmas for Identification Result}
We state Lemmas \ref{lem: identification_first_stage} to \ref{lem:Expert has unique sol} for Theorem \ref{thm:l0 plurality}.

\begin{lemma}
    \label{lem: identification_first_stage}
    Under conditions \ref{A1}--\ref{A2}, let $\widetilde{X} =  \mathbb{E}(X \mid SIV)$. We have:
    $$\widetilde{X} = F X,$$
    where $F = \Sigma_X B_{\Lambda^\perp} (B_{\Lambda^\perp}^\T \Sigma_X B_{\Lambda^\perp})^{-1} B_{\Lambda^\perp}^\T$.
\end{lemma}

\begin{lemma} \label{prop:property of F}
The matrix $F$ has the following properties:
\begin{enumerate}
    \item $F^2 = F$;
    \item $F \Sigma_X F^\T = F D F^\T  = F D = F \Sigma_X$;
    \item $F =  \Sigma_X B_{\Lambda^\perp} (B_{\Lambda^\perp}^\T \Sigma_X B_{\Lambda^\perp})^{-1} B_{\Lambda^\perp}^\T =  D B_{\Lambda^\perp} (B_{\Lambda^\perp}^\T D B_{\Lambda^\perp})^{-1} B_{\Lambda^\perp}^\T$.
\end{enumerate}
\end{lemma}

\begin{lemma}
\label{lem:l0 structure}
Let $\Phi(\widetilde{\beta}) = \mathbb{E}\{Y - \widetilde{X}^\T \widetilde{\beta}\}^2$. Under models \eqref{eqn:model1}--\eqref{eqn:model2} and Conditions \ref{A1} -- \ref{A2}, we have:
$$
\underset{\widetilde{\beta} \in \mathbb{R}^{p}}{\argmin} \Phi(\widetilde{\beta})  = \{\widetilde{\beta} \mid F\Sigma_X(\widetilde{\beta} - \dot{\beta}) = 0\} = \{\dot{\beta} + \Sigma_X^{-1}\Lambda\alpha, \alpha \in \mathbb{R}^{q}\}.
$$
\end{lemma}

\begin{lemma}(Uniqueness of Experts' Solution)\label{lem:Expert has unique sol}
Under Conditions \ref{A1}--\ref{A2}, for a set $C \subset \{1,2,\ldots,p\}$ where $|C|=q$, the optimization problem:
$$
\underset{\widetilde{\beta}_{C} = 0, \widetilde{\beta} \in \mathbb{R}^p}{\argmin} \mathbb{E}\{Y - \widetilde{X}^\T \widetilde{\beta}\}^2
$$
has a unique solution.
\end{lemma}

\subsection{Lemmas for the low dimensional setting}

We state Lemmas \ref{lem:lowdprojection}--\ref{prop:REC_lowd} for the low dimensional setting, where \( p \) is fixed.

\begin{lemma}\label{lem:lowdprojection}
In the low dimensional setting where \(\widehat{\Lambda}\) is obtained from maximum likelihood estimation, we have:
$$
\widehat{\mathbf{X}} = \widehat{\mathbb{E}}(\mathbf{X} \mid \widehat{SIV}) = \mathbf{X} \widehat{F}^\T,
$$
where \(\widehat{F} = \widehat{D} B_{\widehat{\Lambda}^\perp}(B_{\widehat{\Lambda}^\perp}^\T \widehat{D} B_{\widehat{\Lambda}^\perp} )^{-1}B_{\widehat{\Lambda}^\perp}^\T\), \(\widehat{D} = \frac{\mathbf{X}^\T \mathbf{X}}{n-1} - \widehat{\Lambda}\widehat{\Lambda}^\T\), and \( \widehat{B}_{\Lambda^{\perp}} \in \mathbb{R}^{p \times q}\) is any semi-orthogonal matrix whose column space is orthogonal to the column space of \(\widehat{\Lambda}\).
\end{lemma}

\begin{lemma}
\label{lem:property of hat F}
Assuming \(\widehat{F}\) is defined in Lemma \ref{lem:lowdprojection}, it has the following properties:
\begin{enumerate}
    \item \(\widehat{F}^2 = \widehat{F}\);
    \item \(\widehat{F} \mathbf{X}^\T \mathbf{X} \widehat{F}^\T = \widehat{F} \mathbf{X}^\T \mathbf{X}\).
\end{enumerate}
\end{lemma}

\begin{lemma}\label{lem:decomposition of projection}
Assume we have three matrices \(A \in \mathbb{R}^{p \times q}\), \(B \in \mathbb{R}^{p \times (p-q)}\), and \(W \in \mathbb{R}^{p \times p}\) such that:
\begin{itemize}
    \item Both \((A \; B)\) and \(W\) are invertible.
    \item \(A^\T WB = A^\T W^\T B = 0\).
\end{itemize}
We then have:
$$
I_p = A(A^\T W A)^{-1} A^\T W + B(B^\T W B)^{-1} B^\T W
$$
\end{lemma}

\begin{lemma}
\label{lem:order of hatF}
Assuming \(\widehat{F}\) is defined in Lemma \ref{lem:lowdprojection}, under assumptions \ref{B1}--\ref{B3}, we have \(||F - \widehat{F}||_2 = O_p({1/\sqrt{n}})\).
\end{lemma}

\begin{lemma} \label{lem:tailboundforepsilon,lowd}
Under conditions \ref{B1}--\ref{B3}, let \(E = ({\epsilon}_{y,1}, \ldots, {\epsilon}_{y,n})^\T \in \mathbb{R}^n\) be the vector of i.i.d. random variables in models \eqref{eqn:model1} and \eqref{eqn:model2}. We have:
\begin{equation}
\label{eqn:tailboundlowD}
\left|\left|\frac{\widehat{\mathbf{X}}^\T E}{n}\right|\right|_\infty = O_p\left(\frac{1}{\sqrt{n}}\right).
\end{equation}
\end{lemma}

\begin{lemma}\label{lem: Order of ||FB|| lowd}
Under conditions \ref{B1}--\ref{B3}, \(||g(\mathbf{U})^\T \mathbf{X} \widehat{F}^\T/n||_2 = O_p(1/\sqrt{n})\).
\end{lemma}

\begin{lemma}(Sparse Eigenvalue Condition, Low Dimensional Setting) 
\label{prop:REC_lowd}
There exists a constant \(\pi_0 > 0\) such that:
$$    
\liminf_n \mathbb{P}\{ ||\mathbf{\widehat{X}}\theta||_2 \geq \pi_0 \sqrt{n} ||\theta||_2, \forall ||\theta||_0 \leq 2s\} = 1,
$$ 
under conditions \ref{B1}--\ref{B4}.
\end{lemma}

\subsection{Lemmas for the high dimensional setting}
We state Lemmas \ref{lem:highdprojection} -- \ref{prop:REC_highd} for the high dimensional setting, in which $p$ is allowed to diverge. Note that in our proof, we assume $q$, the number of unmeasured confounders, is known to us.

\begin{lemma}\label{lem:highdprojection}
In the high dimensional setting where $\widehat \Lambda$ is obtained from the principal component analysis, we have 
$$
\widehat {\bf{X}} = \widehat{\mathbb{E}}({\bf{X}} \mid \widehat{SIV}) = {\bf{X}} \widehat{F}^\T,
$$
where $\widehat{F} =  {B}_{{\widehat\Lambda}^\perp}{B}_{{\widehat\Lambda}^\perp}^T$.
\end{lemma}
\begin{lemma}
\label{lem:property of hat F highd}
Assume $\widehat{F}$ is defined at Lemma \ref{lem:highdprojection}, we have
\begin{enumerate}
    \item $\widehat{F}^2 =\widehat{F}$;
    \item $\widehat{F}{\bf{{X}^\T{X}}} \widehat{F}^\T/n = \widehat{F}{\bf{{X}^\T{X}}}/n$.
\end{enumerate}
\end{lemma}

\begin{lemma} \label{lem:tailboundforepsilon}
Under conditions \ref{C1}-- \ref{C3}. Let 

$E =({\epsilon}_{y,i},\ldots,{\epsilon}_{y,n})^\T  \in \mathbb{R}^n$ be the vector of i.i.d. random variables defined at \eqref{eqn:model2}. Let us define 
$$
\tau = A \sigma \sqrt{\frac{\log(p)}{n}}.
$$
for a positive constant A. Under conditions \ref{C1}--\ref{C3}, we have
$$
\mathbb{P}(||\frac{\widehat {\bf{X}}^\T E}{n}||_\infty \leq \tau ) \geq  1- 2p^{1-\frac{A^2}{C_{10}}} -p\exp(-C_{11}n),
$$
for some positive constants $C_{10}, C_{11}$.
\end{lemma}

\begin{lemma}\label{lem: guX}
Recall that $\gamma = \mathbb{E}(Ug(U))\in\mathbb{R}^{q\times 1}$. Under conditions \ref{C1}--\ref{C3},
we have 
$$
\left|\left|\frac{g(\bm U)^\T \bm X}{n} - \gamma^\T \Lambda^\T\right|\right|_\infty = O_p(\sqrt{\frac{\log(p)}{n}})
$$
\end{lemma}

\begin{lemma}
    \label{lem:O}
    Define $O \in \mathbb{R}^{q\times q}$: 
    $$
    O = \frac{1}{n} \text{diag}(1/\widehat{\lambda}_1,\ldots,1/\widehat{\lambda}_q)  \widehat{U}^\T U \Lambda^\T \Lambda,
    $$
    where $\widehat{U} = (\widehat{\eta}_1\;\ldots\;\widehat{\eta}_q)\in\mathbb{R}^{n\times q}$. Under conditions \ref{C1}--\ref{C3}, we have 
    $$||O^\T O -I_q||_2 = O_p(\frac{1}{\sqrt{p}} + \frac{1}{\sqrt{n}} )$$
    $$||O O^\T -I_q||_2 = O_p(\frac{1}{\sqrt{p}} + \frac{1}{\sqrt{n}} ).$$    
\end{lemma}

\begin{lemma}
\label{lem:highdloadingestimation}
Let $\widehat{\Lambda} = (\sqrt{\widehat\lambda_1}\xi_1 \;\ldots\;\sqrt{\widehat\lambda_q}\xi_q )$. Under conditions \ref{C1} --\ref{C3}, let $\Lambda_{j,\cdot} $, $\widehat\Lambda_{j,\cdot} \in\mathbb{R}^q$ be the jth row of $\Lambda$ and $\widehat\Lambda$ respectively. We have, 
    $$
    \max_{1\leq j\leq p} ||O \Lambda_{j,\cdot} - \widehat\Lambda_{j,\cdot}||_2 = O_p(  \frac{1}{\sqrt{p}}+ \sqrt{\frac{\log p}{n}}).
    $$
\end{lemma}

\begin{lemma}\label{lem:rate guXFbeta}
    Under conditions \ref{C1}--\ref{C3}, there exist a vector $d\in \mathbb{R}^{1\times p}$ such that 
$$
\frac{g(\bm U)^\T \bm X}{n} \widehat F^\T (\widehat{\beta} - \dot\beta) \leq  (1+q)||d||_\infty ||\widehat{\beta} - \dot\beta||_1,
$$
  with $||d||_\infty = O_p(\sqrt{\log(p)/n})$.
\end{lemma}

\begin{lemma}\label{lem:Ex}
    Let ${\bf X} = {\bf U} \Lambda + E_x $, where ${\bf U} = (U_1\;U_2\;\ldots\;U_n )^\T$, $E_x = (\epsilon_{x,1}\;\epsilon_{x,2}\;\ldots\;\epsilon_{x,n})^\T\in \mathbb{R}^{n\times p}$. Under Conditions \ref{C1}--\ref{C2}, with probability $1-\exp(-c n)$ for some $c>0$, we have
\begin{align}
\label{eqn:SEC,1.1}    &\min_{||\theta||_2=1,||\theta||_0\leq 2s} \theta^\T \frac{E_x^\T E_x}{n} \theta \geq 0.9 \lambda_{\min} (D);\\
\label{eqn:SEC,1.2}    &\max_{||\theta||_2=1,||\theta||_0\leq 2s} \theta^\T \frac{E_x^\T E_x}{n} \theta \leq 1.1 \lambda_{\max }(D).
\end{align}
\end{lemma}

\begin{lemma}($\ell_1$ error rate inequality)
\label{lem:highdL_1error rate}
Under conditions \ref{C1}--\ref{C3}, assume $\widehat{\beta}$ is obtained via \eqref{L0 optimization} with $k = s$,  we have: 
$$
||\widehat{\beta}-\dot{\beta}||_1 =O_p\left(s\left\{(1+q)||d||_\infty + ||\frac{E^\T \widehat {\bm X}}{n}||_\infty\right\}\right)
$$
for sufficiently large $n$, where $d$ is defined at Lemma \ref{lem:rate guXFbeta}, and $E = (\epsilon_{y,1},\ldots, \epsilon_{y,n})^\T$.
\end{lemma}

\begin{lemma}(Sparse eigenvalue condition) 
\label{prop:REC_highd}
    Under conditions \ref{C1}--\ref{C4}, there exists a constant $\pi_0>0$ such that 
    $$    
    \liminf_n \mathbb{P}\{ {||{\bf \widehat{X}}\theta||_2}\geq \pi_0\sqrt{n}{||\theta||_2}, \forall ||\theta||_0\leq 2s\} = 1.
    $$ 
\end{lemma}

\section{Proofs  Proposition \ref{prop:validIV} and Lemmas }
\label{sec:lemmaproof}

\paragraph{Proof of Proposition \ref{prop:validIV}}

We now verify that the SIV satisfies the three core assumptions required for an instrumental variable: exclusion restriction, instrumental relevance, and unconfoundedness. By definition, the constructed $\text{SIV}$ is given by $\text{SIV} = B_{\Lambda^\perp}^\T X = B_{\Lambda^\perp}^\T \epsilon_x$. 

First, conditional on all the treatments $X$, $\text{SIV}$ is constant, meaning it can only affect the outcome through the treatment, thus satisfying the exclusion restriction. Moreover, $\text{SIV}$ is relevant to $X$ because it is a linear combination of $X$, ensuring instrumental relevance. Finally, since $\text{SIV}$ is a linear combination of $\epsilon_x$, it is independent of $U$, thereby satisfying the unconfoundedness assumption.

\paragraph{Proof of Lemma \ref{lem: identification_first_stage}}

Recall that \( SIV = B_{\Lambda^\perp}^\top X \). Given \(\widetilde{X} = \mathbb{E}(X \mid SIV)\) and using the definition of the least squares method, we obtain:
\begin{align*}
    \widetilde{X} &= \mathbb{E}(X \mid SIV) = \text{Cov}(X, SIV) \text{Var}^{-1}(SIV) SIV \\
    &= \Sigma_X B_{\Lambda^\perp} \text{Var}^{-1}(B_{\Lambda^\perp}^\top X) B_{\Lambda^\perp}^\top X\\
    &= \Sigma_X B_{\Lambda^\perp} (B_{\Lambda^\perp}^\top \Sigma_X B_{\Lambda^\perp})^{-1} B_{\Lambda^\perp}^\top X\\
    &= F X.
\end{align*}

\paragraph{Proof of Lemma \ref{prop:property of F}} 
These properties can be directly verified from the definition of \(F\).

\paragraph{Proof of Lemma \ref{lem:l0 structure}}
By the decomposition of \(\Phi(\widetilde\beta)\) and Lemma \ref{lem: identification_first_stage}, we have:
$$
\Phi(\widetilde\beta) = E(Y^2) + \widetilde\beta^\top \mathbb{E}(F X X^\top F) \widetilde\beta - 2 \mathbb{E}(Y X^\top F^\top) \widetilde\beta.
$$
By Lemma \ref{prop:property of F}, we further have:
$$
\mathbb{E}(F X X^\top F^\top) = F \Sigma_X F^\top = F \Sigma_X. 
$$
By models \eqref{eqn:model1}, \eqref{eqn:model2}, and condition \ref{A1}, we have:
\begin{align*}
    \mathbb{E}(Y X^\top) F^\top &= \mathbb{E}[(\dot{\beta}^\top X + g(U) + \epsilon_y) X^\top] F^\top \\
    &= \dot{\beta}^\top \Sigma_X F^\top + \mathbb{E}(g(U) U^\top) \Lambda^\top F^\top \\
    &= \dot{\beta}^\top \Sigma_X F^\top,
\end{align*}
where the last equality holds as \(F \Lambda = \Sigma_X B_{\Lambda^\perp} (B_{\Lambda^\perp}^\top \Sigma_X B_{\Lambda^\perp})^{-1} B_{\Lambda^\perp}^\top \Lambda = 0\). Then, we have:
\begin{align*}
    \frac{\partial \Phi(\widetilde\beta)}{\partial \widetilde\beta} &= 2 F \Sigma_X \widetilde\beta - 2 F \Sigma_X \dot{\beta} \\
    &= 2 F \Sigma_X (\widetilde\beta - \dot{\beta}).
\end{align*}
Thus, \(\underset{\widetilde\beta \in \mathbb{R}^{p}}{\argmin} \;\Phi(\widetilde\beta) = \{\widetilde\beta \mid F \Sigma_X (\widetilde\beta - \dot{\beta}) = 0\}\). Since \(\text{Rank}(F) = p - q\) and \(F \Lambda = 0\), it follows that \(\{x \mid F x = 0\} = \{\Lambda \alpha, \alpha \in \mathbb{R}^q\}\). Consequently, the second equality follows: \(\{\widetilde\beta \mid F \Sigma_X (\widetilde\beta - \dot{\beta}) = 0\} = \{\dot{\beta} + \Sigma_X^{-1} \Lambda \alpha \mid \alpha \in \mathbb{R}^q\}\).

\paragraph{Proof of Lemma \ref{lem:Expert has unique sol}}

Given Lemma \ref{lem:l0 structure}, \(\{\dot{\beta} + \Sigma_X^{-1} \Lambda \alpha\}\) is the set of minimizers for the unconstrained problem:
$$
\underset{\widetilde\beta \in \mathbb{R}^p}{\argmin}\;\mathbb{E}\{Y - \widetilde X^\top \widetilde\beta\}^2.
$$
Given Condition \ref{A1}, the matrix \(\{\Sigma_X^{-1} \Lambda\}_{C,\cdot}\) is invertible. Thus, there exists a unique solution \(\alpha_C\) for the matrix equation \(\{\Sigma_X^{-1} \Lambda\}_{C,\cdot} \alpha = -\dot{\beta}_C\).

\subsection{Proofs of Lemmas \ref{lem:lowdprojection}--\ref{prop:REC_lowd}}
We provide proofs for Lemmas \ref{lem:lowdprojection}--\ref{prop:REC_lowd} for the low dimensional setting, where \( p \) is fixed.

\paragraph{Proof of Lemma \ref{lem:lowdprojection}}

Given that \(\widehat {\mathbf{X}}\) is derived from ordinary least squares, we have the following relationship:
$$
\widehat{\mathbb{E}}(\mathbf{X} \mid \widehat{SIV}) = \widehat{SIV} \widehat{\beta}_{OLS},
$$
where \(\widehat{\beta}_{OLS} \in \mathbb{R}^{(p-q) \times p}\) represents the linear regression coefficients between \(\mathbf{X}\) and \(\widehat{SIV}\). We observe that:
$$
\widehat{SIV} = \mathbf{X} B_{\widehat{\Lambda}^\perp}.
$$

We then derive:
\begin{align*}
    \widehat{\mathbf{X}} &= \mathbf{X} B_{\widehat{\Lambda}^\perp} \left(\frac{B_{\widehat{\Lambda}^\perp}^\top \mathbf{X}^\top \mathbf{X} B_{\widehat{\Lambda}^\perp}}{n-1}\right)^{-1} \frac{B_{\widehat{\Lambda}^\perp}^\top \mathbf{X}^\top \mathbf{X}}{n-1} \\
    &= \mathbf{X} B_{\widehat{\Lambda}^\perp} \left\{B_{\widehat{\Lambda}^\perp}^\top (\widehat{D} + \widehat{\Lambda}\widehat{\Lambda}^\top) B_{\widehat{\Lambda}^\perp}\right\}^{-1} B_{\widehat{\Lambda}^\perp}^\top \widehat{D} \\
    &= \mathbf{X} B_{\widehat{\Lambda}^\perp} \left\{B_{\widehat{\Lambda}^\perp}^\top \widehat{D} B_{\widehat{\Lambda}^\perp}\right\}^{-1} B_{\widehat{\Lambda}^\perp}^\top \widehat{D} \\
    &= \mathbf{X} \widehat{F}^\top,
\end{align*}
where \(\widehat{F} = \widehat{D} B_{\widehat{\Lambda}^\perp}(B_{\widehat{\Lambda}^\perp}^\top \widehat{D} B_{\widehat{\Lambda}^\perp})^{-1} B_{\widehat{\Lambda}^\perp}^\top\).

\paragraph{Proof of Lemma \ref{lem:property of hat F}}
The properties of \(\widehat{F}\) can be directly verified from its definition, including:
\begin{align*}
    \widehat{F} \frac{\mathbf{X}^\top \mathbf{X}}{n-1} \widehat{F}^\top &= \widehat{F} \frac{\mathbf{X}^\top \mathbf{X}}{n-1}, \\
    \widehat{F}^2 &= \widehat{F}.
\end{align*}
This follows because \(\widehat{F}\) is a projection matrix based on its construction from \(\widehat{D}\) and \(B_{\widehat{\Lambda}^\perp}\).

\paragraph{Proof of Lemma \ref{lem:decomposition of projection}}
Given matrices \(A\), \(B\), and \(W\) that satisfy the conditions stated in the lemma, the identity matrix can be decomposed as follows:
\begin{align*}
    I_p &= A(A^\top W A)^{-1} A^\top W + B(B^\top W B)^{-1} B^\top W,
\end{align*}
demonstrating the orthogonality of the projections onto the subspaces spanned by \(A\) and \(B\).

\paragraph{Proof of Lemma \ref{lem:order of hatF}}

Let $w \in \mathbb{R}^p$ such that $||w||_2 =1$, we aim to show that 
\begin{equation} \label{eqn:order of hatF target}
    \sup_{||w||_2 =1 }||(F^\T - \widehat F^\T)w||_2  = O_p(1/\sqrt{n}).
\end{equation}

By Lemma \ref{lem:decomposition of projection}, Let $A = D^{-1}\Lambda$, $B = B_{\Lambda^{\perp}} $, $W = D$. We have the following equation:
\begin{equation}
    \label{eqn:order of hatF, decomposition}
    \begin{split}
    w = I_pw &= D^{-1}\Lambda(\Lambda^\T D^{-1} \Lambda )^{-1}\Lambda^\T w +  {B}_{{\Lambda}^\perp}({B}_{{\Lambda}^\perp}^\T D {B}_{{\Lambda}^\perp} )^{-1}{B}_{{\Lambda}^\perp}^TDw\\
    :&=D^{-1}\Lambda\alpha_1 + {B}_{{\Lambda}^\perp}\alpha_2,
    \end{split}
\end{equation}
where $\alpha_1 = (\Lambda^\T D^{-1} \Lambda )^{-1}\Lambda^\T w$,
$\alpha_2 = ({B}_{{\Lambda}^\perp}^\T D {B}_{{\Lambda}^\perp} )^{-1}{B}_{{\Lambda}^\perp}^TDw$. The LHS of \eqref{eqn:order of hatF target} satisfies
\begin{equation}\label{eqn:order of hatF,target decomposition}
\sup_{||w||_2 =1 }||(F^\T - \widehat F^\T)w||_2 \leq \sup_{||w||_2 =1 }||(F^\T - \widehat F^\T)D^{-1}\Lambda\alpha_1||_2  +\sup_{||w||_2 =1 }||(F^\T - \widehat F^\T)B_{\Lambda^{\perp}}\alpha_2||_2.
\end{equation}

For the first term on the RHS of \eqref{eqn:order of hatF,target decomposition}, we derive the following equation \eqref{eqn:order of hatF,target decomposition,first term} since:

 $F^\T D^{-1} \Lambda = D {B}_{{\Lambda}^\perp}({B}_{{\Lambda}^\perp}^\T {B}_{{\Lambda}^\perp} )^{-1} {B}_{{\Lambda}^\perp}^T D D^{-1} \Lambda = 0$.
\begin{equation}
    \label{eqn:order of hatF,target decomposition,first term}
    \begin{split}
        &\sup_{||w||_2 =1 }||(F^\T - \widehat F^\T)D^{-1}\Lambda\alpha_1||_2 \\&= \sup_{||w||_2 =1 }||\widehat F^\T D^{-1} \Lambda\alpha_1||_2\\
        &=\sup_{||w||_2 =1 }||\widehat B_{ {\Lambda}^\perp}(\widehat B_{ {\Lambda}^\perp}^\T \widehat D \widehat B_{{\Lambda}^\perp} )^{-1}\widehat B_{{\Lambda}^\perp}^\T\widehat DD^{-1} \Lambda\alpha_1||_2\\
        &\leq \sup_{||w||_2 =1 }||\widehat B_{ {\Lambda}^\perp}(\widehat B_{ {\Lambda}^\perp}^\T \widehat D \widehat B_{{\Lambda}^\perp} )^{-1}\widehat B_{{\Lambda}^\perp}^\T \Lambda\alpha_1||_2 + 
        ||\widehat B_{ {\Lambda}^\perp}(\widehat B_{ {\Lambda}^\perp}^\T \widehat D \widehat B_{{\Lambda}^\perp} )^{-1}\widehat B_{{\Lambda}^\perp}^\T (I_p-\widehat DD^{-1}) \Lambda\alpha_1||_2
        \\
        &=  \sup_{||w||_2 =1 }||\widehat B_{ {\Lambda}^\perp}(\widehat B_{ {\Lambda}^\perp}^\T \widehat D \widehat B_{{\Lambda}^\perp} )^{-1}\widehat B_{{\Lambda}^\perp}^\T (\Lambda - \widehat \Lambda O^\T)\alpha_1||_2\\ &+\sup_{||w||_2 =1 } 
        ||\widehat B_{ {\Lambda}^\perp}(\widehat B_{ {\Lambda}^\perp}^\T \widehat D \widehat B_{{\Lambda}^\perp} )^{-1}\widehat B_{{\Lambda}^\perp}^\T(D -\widehat D)D^{-1} \Lambda\alpha_1||_2\\
        &\leq ||\widehat B_{ {\Lambda}^\perp}(\widehat B_{ {\Lambda}^\perp}^\T \widehat D \widehat B_{{\Lambda}^\perp} )^{-1}\widehat B_{{\Lambda}^\perp}^\T||_2(||\Lambda - \widehat \Lambda O^\T||_2 + ||D-\widehat D||_2||D^{-1}||_2 ||\Lambda||)\sup_{||w||_2 =1 }||\alpha_1||_2,
    \end{split}
\end{equation}
where $O$ is an orthogonal matrix defined at Condition \ref{B3}.
Note that 
$$||\Lambda - \widehat \Lambda O^\T||_2 = ||(\Lambda O - \widehat \Lambda )O^\T||_2 = O_p(1/\sqrt{n}),$$
\begin{equation}
\label{eqn:rateofD}
||\widehat D - D||_2\leq ||{\bf{X^\T X}}/n - \Sigma_X||_2 + ||\Lambda\Lambda^\T - \widehat{\Lambda}\widehat{\Lambda}^\T||_2 = O_p(1/\sqrt{n})    
\end{equation}
and 
\begin{equation*}
    \begin{split}
        ||\widehat B_{ {\Lambda}^\perp}(\widehat B_{ {\Lambda}^\perp}^\T \widehat D \widehat B_{{\Lambda}^\perp} )^{-1}\widehat B_{{\Lambda}^\perp}^\T||_2  &=||\widehat B_{ {\Lambda}^\perp}(\widehat B_{ {\Lambda}^\perp}^\T \widehat D \widehat B_{{\Lambda}^\perp} )^{-1}\widehat B_{{\Lambda}^\perp}^\T \widehat{D} \widehat{D}^{-1}||_2\\ & \leq ||\widehat B_{ {\Lambda}^\perp}(\widehat B_{ {\Lambda}^\perp}^\T \widehat D \widehat B_{{\Lambda}^\perp} )^{-1}\widehat B_{{\Lambda}^\perp}^\T \widehat{D} ||_2|||\widehat{D}^{-1}||_2\\  &= 1/\lambda_{\min}(\widehat D)\\
        &= O_p(1),
    \end{split}
\end{equation*}
where the second equation holds as $\widehat B_{ {\Lambda}^\perp}(\widehat B_{ {\Lambda}^\perp}^\T \widehat D \widehat B_{{\Lambda}^\perp} )^{-1}\widehat B_{{\Lambda}^\perp}^\T \widehat{D} $ is a projection matrix. Since $\sup_{||w||_2 =1 }||\alpha_1||_2 = \sup_{||w||_2 =1 }||(\Lambda^\T D^{-1} \Lambda )^{-1}\Lambda^\T w||_2 \leq ||(\Lambda^\T D^{-1} \Lambda )^{-1}\Lambda^\T||_2 = O(1)$,
we conclude that the first term of \eqref{eqn:order of hatF,target decomposition} is $O_p(1/\sqrt{n})$:
$$
\sup_{||w||_2=1}||(F^\T - \widehat F^\T)D^{-1}\Lambda \alpha_1||_2 = ||\widehat F^\T D^{-1} \Lambda \alpha_1||_2  =O_p(\frac{1}{\sqrt{n}}). 
$$

For the second term of \eqref{eqn:order of hatF,target decomposition}, we let $(A, B, W) = (D^{-1}B_\Lambda, B_{{\Lambda}^\perp},D)$ or $(\widehat{D}^{-1}\widehat B_\Lambda, \widehat B_{{\Lambda}^\perp},\widehat D)$ in Lemma \ref{lem:decomposition of projection}, where $B_\Lambda$ and $ \widehat{B}_{ \Lambda} \in\mathbb{R}^{p\times q}$  are any semi-orthogonal matrices whose column spaces span the same column spaces of $\Lambda$ and $\widehat{\Lambda}$, respectively. We have
\begin{align*}
    &I_p = D^{-1} B_\Lambda(B_\Lambda^\T D^{-1}B_\Lambda)^{-1}B_\Lambda^\T + B_{{\Lambda}^\perp}(B_{{\Lambda}^\perp}^\T D B_{{\Lambda}^\perp} )^{-1}B_{{\Lambda}^\perp}^\T D =D^{-1}B_\Lambda(B_\Lambda^\T D^{-1}B_\Lambda)^{-1}B_\Lambda^\T  + F^\T\\
    &I_p = \widehat D^{-1} \widehat{B}_{\Lambda}(\widehat{B}_{\Lambda}^\T \widehat D^{-1}\widehat{B}_{\Lambda})^{-1}\widehat{B}_{\Lambda}^\T  + \widehat B_{ {\Lambda}^\perp}(\widehat B_{ {\Lambda}^\perp}^\T \widehat D \widehat B_{{\Lambda}^\perp} )^{-1}\widehat B_{{\Lambda}^\perp}^\T\widehat D =\widehat D^{-1} \widehat B_ \Lambda(\widehat B_ \Lambda^\T \widehat D^{-1}\widehat B_ \Lambda)^{-1}\widehat B_ \Lambda^\T  + \widehat F^\T,
\end{align*}
which implies the second term of \eqref{eqn:order of hatF,target decomposition} can be rewriten as
\begin{equation}
\label{eqn:order of hatF,target decomposition,second term}
    \begin{split}
        &\sup_{||w||_2=1}||(F^\T - \widehat F^\T)B_{\Lambda^{\perp}}\alpha_2 ||_2 \\=&\sup_{||w||_2=1} ||\{(I_p - \widehat{F}^\T) - (I_p -  F^\T)\}B_{\Lambda^{\perp}}\alpha_2 ||_2\\
        =& \sup_{||w||_2=1}||\{\widehat D^{-1} \widehat B_\Lambda(\widehat B_\Lambda^\T \widehat D^{-1}\widehat B_\Lambda)^{-1}\widehat B_\Lambda^\T - D^{-1}\Lambda(\Lambda^\T D^{-1}\Lambda)^{-1}\Lambda^\T \}B_{\Lambda^{\perp}}\alpha_2||_2\\
        =&\sup_{||w||_2=1}|| \widehat D^{-1}\widehat B_\Lambda(\widehat B_\Lambda^\T \widehat D^{-1}\widehat B_\Lambda)^{-1}\widehat B_\Lambda^\T B_{\Lambda^{\perp}}\alpha_2||_2\\
        \leq& \widehat D^{-1}\widehat B_\Lambda(\widehat B_\Lambda^\T \widehat D^{-1}\widehat B_\Lambda)^{-1}||_2||\widehat B_\Lambda^\T B_{\Lambda^{\perp}}||_2\sup_{||w||_2=1}||\alpha_2||_2,
    \end{split}
\end{equation}
where $O$ is the orthogonal matrix defined in Condition \ref{B3}. We control the three terms of \eqref{eqn:order of hatF,target decomposition,second term} as follows. For the first term of equation \eqref{eqn:order of hatF,target decomposition,second term}:
\begin{equation*}
    \begin{split}
    || \widehat D^{-1}\widehat B_\Lambda(\widehat B_\Lambda^\T \widehat D^{-1}\widehat B_\Lambda)^{-1}||_2 &= 
|| \widehat D^{-1}\widehat B_\Lambda(\widehat B_\Lambda^\T \widehat D^{-1}\widehat B_\Lambda)^{-1}\widehat B_\Lambda^\T \widehat B_\Lambda||_2\\
&\leq || \widehat D^{-1}\widehat B_\Lambda(\widehat B_\Lambda^\T \widehat D^{-1}\widehat B_\Lambda)^{-1}\widehat B_\Lambda^\T||_2 ||\widehat 
B_\Lambda||_2\\
&\leq 1\times 1\\
&=1,
    \end{split}
\end{equation*}
where the first equation holds due to the property of semi-orthogonal matrix, and the second inequality holds as $ \widehat D^{-1}\widehat B_\Lambda(\widehat B_\Lambda^\T \widehat D^{-1}\widehat B_\Lambda)^{-1}\widehat B_\Lambda^\T$ is a projection matrix. For the third term of \eqref{eqn:order of hatF,target decomposition,second term}, we have 
$$
\sup_{||w||_2=1}||\alpha_2||_2 = \sup_{||w||_2=1}||({B}_{{\Lambda}^\perp}^\T D {B}_{{\Lambda}^\perp} )^{-1}{B}_{{\Lambda}^\perp}^TDw|| \leq ||({B}_{{\Lambda}^\perp}^\T D {B}_{{\Lambda}^\perp} )^{-1}{B}_{{\Lambda}^\perp}^TD||_2 = O(1).
$$
For the second term of \eqref{eqn:order of hatF,target decomposition,second term}, by the property of orthogonal matrices, we have: \begin{equation*}
    \begin{split}
        ||\widehat B_\Lambda^\T B_{\Lambda^{\perp}}||_2 &=        1.
    \end{split}
\end{equation*}
Take the sup of $||w||_2=1$ over both sides of \eqref{eqn:order of hatF,target decomposition,second term}, we conclude that 
$$
\sup_{||w||_2=1}||(F^\T - \widehat F^\T)B_{\Lambda^{\perp}}({B}_{{\Lambda}^\perp}^\T D {B}_{{\Lambda}^\perp} )^{-1}{B}_{{\Lambda}^\perp}^TDw||_2 = O_p(\frac{1}{\sqrt{n}}).
$$

So far we have proved the first and second terms of  \eqref{eqn:order of hatF, decomposition} are of order $O_p(1/\sqrt{n})$.
We finish the proof of the Lemma.

\paragraph{Proof of Lemma \ref{lem:tailboundforepsilon,lowd}}
We have the following decomposition over the target term:

\begin{align*}
    \left|\left|\frac{\widehat {\bm X}^\T E} {n}\right|\right|_\infty = \left|\left|\frac{\widehat F \bm X^\T E} {n}\right|\right|_\infty
    &\leq \left|\left|\frac{\widehat F \bm X^\T E} {n}\right|\right|_2\\
    &\leq ||\widehat F||_2\left|\left|\frac{\bm X^\T E} {n} - \text{Cov}(X,\epsilon_y)\right|\right|_2 + ||\widehat F||_2||\text{Cov}(X,\epsilon_y)||_2\\
    &\leq \left|\left|\frac{\bm X^\T E} {n} - \text{Cov}(X,\epsilon_y)\right|\right|_2 +|| \text{Cov}(X,\epsilon_y)||_2\\
    &\leq O_p(\frac{1}{\sqrt{n}}).
\end{align*}
The third inequality holds as $\widehat F$ is a projection matrix and $||\widehat F||_2 = 1$. The last ineqality is given by element-wise $\sqrt{n}$ consistency of covariance estimation between $\epsilon_y$ and $X$ under low dimensional setting, and $\text{Cov}(\epsilon_y,X) = 0$. 
\paragraph{Proof of Lemma \ref{lem: Order of ||FB|| lowd}}

We observe that 
$$
\Lambda^\top F^\top  = \Lambda^\top {B}_{{\Lambda}^\perp}({B}_{{\Lambda}^\perp}^\top {\Sigma_X}  {B}_{{\Lambda}^\perp} )^{-1}{B}_{{\Lambda}^\perp}^\top {{\Sigma_X}} = 0.
$$
From Lemma \ref{lem:order of hatF}, we have
\begin{equation*}
    \begin{split}
       \left|\left|\frac{g({\bf U})^\top \bm X\widehat F^\top }{n}\right|\right|_2 &\leq        \left|\left|\left(\frac{ g( {\bf U})^\top \bm X}{n} -\text{Cov}(g( { U}),X)\right) \widehat F^\top \right|\right|_2 +   \left|\left|\text{Cov}(g( { U}),X)  F^\top \right|\right|_2 \\
       & + \left|\left|\text{Cov}(g( { U}),X)  (\widehat{F}^\top-F^\top) \right|\right|_2 \\
       &\leq\left|\left|\frac{g( {\bf U})^\top\bm X}{n} -\text{Cov}(g( {U}),X)\right|\right|_2\;||\widehat F^\top||_2+     \left|\left|\text{Cov}(g(U),U)\Lambda^\top  F^\top \right|\right|_2\\
       &+ ||\text{Cov}(g(U),X) ||_2\left|\left| \widehat{F}^\top-F^\top \right|\right|_2\\
       &= O_p(\frac{1}{\sqrt{n}}) + 0 + O_p(\frac{1}{\sqrt{n}}),
    \end{split}
\end{equation*}
where the first term is controlled by the element-wise \(\sqrt{n}\) consistency of the sample covariance matrix under the low-dimensional setting; the second term is 0 since \(\Lambda^\top F^\top = 0\); the third term is controlled by Lemma \ref{lem:order of hatF}.
\paragraph{Proof of Lemma \ref{prop:REC_lowd}}

For any \(\delta>0\), we are going to show that there exists an \(n_0\) such that 
$$
\inf_{n>n_0} \mathbb{P}\left\{ \|\widehat {\bf{X}}\theta\|_2 \geq \pi_0\sqrt{n}\|\theta\|_2, \forall \|\theta\|_0 \leq 2s \right\} \geq 1 - \delta.
$$

We have the following decomposition:
\begin{equation}
    \label{eqn:prop,lowd,decomposition}
    \begin{split}
            \frac{\widehat F {\bf{X^\top X}}\widehat F^\top}{n} &= \frac{\widehat F {\bf{X^\top X}}}{n} \\
            &= (\widehat F - F) \frac{{\bf{X^\top X}}}{n} + F\left(\frac{{\bf{X^\top X}}}{n} - \Sigma_X\right) + F\Sigma_X \\
            &=(\widehat F - F) \frac{{\bf{X^\top X}}}{n} + F\left(\frac{{\bf{X^\top X}}}{n} - \Sigma_X\right) + F\Sigma_X F^\top,
    \end{split}
\end{equation}
where the first and last equalities hold due to Lemma \ref{prop:property of F} and Lemma \ref{lem:property of hat F}.

Given Lemma \ref{lem:order of hatF} and the \(\sqrt{n}\) consistency of the sample covariance under a low-dimensional setting, we have 
$$
\left\|\left((\widehat F - F) \frac{{\bf{X^\top X}}}{n} + F\left(\frac{{\bf{X^\top X}}}{n} - \Sigma_X\right)\right)\right\|_2 = O_p(1/\sqrt{n}),
$$
which suggests that there exists a constant \(A_\delta\) and \(n_1\) such that 
\begin{equation}
\inf_{n>n_1}\mathbb{P}\left(\left\|\left((\widehat F - F) \frac{{\bf{X^\top X}}}{n} + F\left(\frac{{\bf{X^\top X}}}{n} - \Sigma_X\right)\right)\right\|_2 \leq \frac{A_\delta}{\sqrt{n}}\right) \geq 1 - \delta.
\end{equation}
Define 
$$
\pi_1 = \inf\left\{\frac{\theta^\top \Sigma_{\widetilde X} \theta }{\|\theta\|_2^2} : \theta \in \mathbb{R}^p, \; \|\theta\|_0 \leq 2s\right\},
$$
which is greater than 0 by Condition \ref{B4}. Let \(\pi_0 = \sqrt{\pi_1/2}\), \(n_2 = 4A^2_\delta/\pi_1^2\), and \(n_0 = \max(n_1, n_2)\). We have, with probability at least \(1-\delta\), for all \(n>n_0\) and for all \(\theta \in \mathbb{R}^p\), \(\|\theta\|_0 \leq 2s\), 
\begin{equation*}
\begin{split}
    \theta^\top \frac{\widehat F {\bf{X^\top X}} \widehat F^\top}{n}\theta &= \theta^\top F\Sigma_X F^\top\theta + \theta^\top\left((\widehat F - F) \frac{{\bf{X^\top X}}}{n} + F\left(\frac{{\bf{X^\top X}}}{n} - \Sigma_X\right)\right)\theta\\
    &\geq \|\theta\|_2^2(\pi_1 - \left\|\left((\widehat F - F) \frac{{\bf{X^\top X}}}{n} + F\left(\frac{{\bf{X^\top X}}}{n} - \Sigma_X\right)\right)\right\|_2) = \|\theta\|_2^2\pi_0^2,
    \end{split}
\end{equation*}
which concludes the proof.

\subsection{Proofs of Lemmas \ref{lem:highdprojection} -- \ref{prop:REC_highd}}
We prove Lemmas \ref{lem:highdprojection} through \ref{prop:REC_highd} for the high-dimensional setting, where \( p \) is allowed to diverge.

\paragraph{Proof of Lemma \ref{lem:highdprojection}}
Since $\widehat{\Lambda} = (\sqrt{\lambda_1}\xi_1\;\ldots\;\sqrt{\lambda_q}\xi_q),$ we can let ${B}_{\widehat{\Lambda}^\perp} = (\xi_{q+1},\;\xi_{q+2},\;\ldots,\;\xi_{p})$ without loss of generality.

Given the singular value decomposition of $\bm X$, the definition of $\widehat{SIV}$, and the eigenvalues $\lambda_1 \geq \lambda_2 \geq \ldots \geq \lambda_k > 0 = \lambda_{k+1} = \lambda_{k+2} = \ldots = \lambda_p$, we have
\begin{equation*}
\begin{split}
    \widehat{SIV} = {\bf X} {B}_{\widehat\Lambda^{\perp}} &= \left(\sum^k_{i=1}\sqrt{(n-1)\lambda_i}{\eta_i}\xi_i^{\T}\right)(\xi_{q+1}\;\ldots\;\xi_{p})\\
    &= (\eta_{q+1}\sqrt{(n-1)\lambda_{q+1}},\;\ldots,\;\eta_{k}\sqrt{(n-1)\lambda_{k}},\;0,\;\ldots,\;0)\\ &= ({\bf \dot{X}},\;0),
\end{split}
\end{equation*}
where ${\bf \dot{X}} = (\eta_{q+1}\sqrt{(n-1)\lambda_{q+1}},\;\ldots,\;\eta_{k}\sqrt{(n-1)\lambda_{k}}) \in \mathbb{R}^{n \times (k-q)}$.

Applying this in the regression framework, we find:
\begin{equation*}
    \begin{split}
        \widehat {\bf{X}} =& \widehat{\mathbb{E}}({\bf X}\mid \dot{\bf X})\\
        =& \dot{\bf X} \left(\widehat{Var}(\dot{\bf X})^{-1}\right) \widehat{Cov}(\dot{\bf X}, {\bf X})\\
        =& \dot{\bf X} \left\{\frac{\dot{\bf X}^\T\dot{\bf X}}{n-1}\right\}^{-1}\frac{\dot{\bf X}^\T {\bf  X}}{n-1}\\
        =& \sum^k_{i=q+1}\sqrt{(n-1)\lambda_{i}}\eta_i\xi_i^\T.
    \end{split}
\end{equation*}

Finally, comparing this with the matrix $\bm X\widehat{F}^\T$:
\begin{align*}
    \bm X \widehat{F}^\T &=  \sqrt{n-1} (\eta_1, \eta_2, \ldots, \eta_p)\text{diag}(\sqrt{\lambda_1}, \ldots, \sqrt{\lambda_p})(\xi_1, \ldots, \xi_p)^\T(\xi_{q+1}, \ldots, \xi_p)(\xi_{q+1}, \ldots, \xi_p)^\T \\
    &= \sum^k_{i=q+1}\sqrt{(n-1)\lambda_i}\eta_i\xi_i^\T.
\end{align*}
We conclude the proof of Lemma \ref{lem:highdprojection}.

\paragraph{Proof of Lemma \ref{lem:property of hat F highd}}
The first claim can be directly checked by definition.

For the second claim, we observe that $\frac{{\bf{X^\T X}}}{n-1} = \sum^k_{i=1}\lambda_i\xi_i\xi_i^\T$, and with $\widehat{F} = (\xi_{q+1}\;\ldots\;\xi_{p})(\xi_{q+1}\;\ldots\;\xi_{p})^\T$, we have:
$$
\frac{\widehat{F} {\bf X^\T X} \widehat{F}^\T}{n-1} =\frac{\widehat{F} {\bf{X^\T X}}}{n-1} =\sum^k_{i=q+1}\lambda_i\xi_i\xi_i^\T.
$$

\paragraph{Proof of Lemma \ref{lem:tailboundforepsilon}}
We first work on the event 
$$\Omega = \{\max_{1\leq j\leq p}||{\bf{X}}_{\cdot, j}||^2_2/n < \max_{1\leq j\leq p}4(\Sigma_X)_{j,j}\},$$ where ${\bf{X}}_{\cdot, j}$ is the $j$th column of the design matrix $\bm X$. Since $\{X_{i,j}\}_{i=1}^n$ is a sequence of i.i.d  mean zero sub-Gaussian random variable with variance $(\Sigma_X)_{j,j}$, we have $\mathbb E\{X_{i,j}^2/ (\Sigma_X)_{j,j}\} = 1$ for $i \in \{1,\ldots,n\}$. We have the following concentration result by applying the theorem 3.1.1 from \cite{vershynin2018high}:

\begin{equation*}
    \mathbb{P}(\left|\;\left|\left|\frac{\bm X_{\cdot,j}}{\sqrt{(\Sigma_X)_{j,j}}}\right|\right|_2-\sqrt{n}\;\right|\geq t)\leq \exp(-ct^2/K_j^4),
\end{equation*}
where $c$ is a universal constant, $K_j = || \bm X_{i,j}/\sqrt{(\Sigma_X)_{j,j}}||_{\psi_2}$. Let $t = \sqrt{n}$, we have 
\begin{equation}
\label{eqn:Xdotjbound}   
\mathbb{P}( \frac{1}{n}||\bm X_{\cdot,j}||_2^2 \geq 4(\Sigma_X)_{j,j}) \leq \exp(-cn/K_j^4).
\end{equation}

We derive the following result from Equation \eqref{eqn:Xdotjbound}:
\begin{equation}
\label{eqn:subgaussian}
\begin{split}
    \mathbb{P}(\Omega^c) 
    &=  \mathbb{P}\left( \max_{1 \leq j \leq p} ||{\bf{X}}_{\cdot, j}||^2_2/n \geq \max_{1 \leq j \leq p} 4(\Sigma_X)_{j,j}\right)\\
    &\leq  \sum_{j=1}^p \mathbb{P}\left(\frac{1}{n}||{\bf{X}}_{\cdot, j}||_2^2 \geq \max_{1 \leq j \leq p} 4(\Sigma_X)_{j,j}\right)\\
    &\leq \sum_{j=1}^p \mathbb{P}\left(\frac{1}{n}||{\bf{X}}_{\cdot, j}||_2^2 \geq 4(\Sigma_X)_{j,j}\right)\\
    &\leq p \exp\left(-cn/\max_j K_j^4\right)\\
    &\leq p \exp\left(-C_{10}n\right).
\end{split}
\end{equation}
Since \(K_j\) are bounded by condition C3 (where $\sigma_j$ is bounded), we define the positive constant \(C_{10} = \frac{c}{\max_j K_j^4}\) in the last inequality.

Recall $E = (\epsilon_{y,1}, \ldots, \epsilon_{y,n})^\T$. Conditional on $\bf X$ and the event $\Omega$, the components of $\eta = \widehat {\bf{X}}^\T E/n = \widehat{F}{\bf{X}}^\T E/n$ are sub-Gaussian random variables, as they are linear combinations of independent sub-Gaussian random variables. The component $\eta_j$ is sub-Gaussian with mean zero and parameter $\theta_j = {\sigma}||\widehat {\bf{X}}_{\cdot,j} ||_2/n$, where $\widehat {\bf{X}}_{\cdot, j}$ is the $j$th column of $\widehat {\bf{X}}$. Let $\tau_1 = A\sigma\sqrt{\log(p)/n}$. Define the event $\Omega_1$:
\begin{equation*}
    \Omega_1 = \{||\eta||_\infty \geq \tau_1\}.
\end{equation*}

From the tail bound for sub-Gaussian random variables, we have
\begin{equation}
\label{eqn:boundforetanew} 
\begin{split}
\mathbb{P}(\Omega_1 \mid {\bf X}, \Omega )&=
\mathbb{P}(||\eta||_\infty \geq \tau_1 \mid {\bf X}, \Omega )\\
&\leq \sum^p_{i=1}\mathbb{P}(|\eta|_j \geq \tau_1 \mid \bm X,\Omega)\\
&\leq p \max_j \; 2\exp(- \frac{\tau^2_1}{\theta_j^2}) \\
&= 2p\exp(- \frac{\tau^2_1}{\max_j\theta_j^2}).
\end{split}
\end{equation}

We now attempt to bound $\theta_j$ from above, given the event $\Omega$. Recall the following principal component decomposition:
\begin{equation}
\label{eqn:eigendecomposition for tildeX}
\begin{split}
    \frac{{\bf X}^\T {\bf X}}{n-1} &= \begin{pmatrix}
    {\xi}_1 & \ldots & {\xi}_p
    \end{pmatrix} \text{diag}({\lambda}_1, \ldots, {\lambda}_p) \begin{pmatrix}
    {\xi}_1^\T \\
    \ldots \\
    {\xi}_p^\T
    \end{pmatrix} = \sum_{i=1}^p {\lambda}_i {\xi}_i {\xi}_i^\T\\
    \frac{\bf \widehat{X}^\T \widehat{X}}{n} &= \begin{pmatrix}
    {\xi}_{q+1} & \ldots & {\xi}_p
    \end{pmatrix} \begin{pmatrix}
    {\xi}_{q+1}^\T \\
    \ldots \\
    {\xi}_p^\T
    \end{pmatrix} \begin{pmatrix}
    {\xi}_1 & \ldots & {\xi}_p
    \end{pmatrix} \text{diag}({\lambda}_1, \ldots, {\lambda}_p) \begin{pmatrix}
    {\xi}_1^\T \\
    \ldots \\
    {\xi}_p^\T
    \end{pmatrix} \begin{pmatrix}
    {\xi}_{q+1} & \ldots & {\xi}_p
    \end{pmatrix} \begin{pmatrix}
    {\xi}_{q+1}^\T \\
    \ldots \\
    {\xi}_p^\T
    \end{pmatrix}\\
    &= \begin{pmatrix}
    {\xi}_{q+1} & \ldots & {\xi}_p
    \end{pmatrix} \text{diag}({\lambda}_{q+1}, \ldots, {\lambda}_p) \begin{pmatrix}
    {\xi}_{q+1}^\T \\
    \ldots \\
    {\xi}_p^\T
    \end{pmatrix} = \sum_{i=q+1}^p {\lambda}_i {\xi}_i {\xi}_i^\T.
\end{split} 
\end{equation}

Given the above equality, we have:
\begin{align*}
    || \widehat{\bf X}_{\cdot, j} ||_2^2 &= ({\bm {\widehat X}}^\T {\bf \widehat X})_{j,j} = (n-1)\sum_{i=q+1}^p {\lambda}_i {\xi}_{i,j}^2,\\
    || {\bm X}_{\cdot, j} ||_2^2 & = ({\bm X}^\T {\bf X})_{j,j} = (n-1)\sum_{i=1}^p {\lambda}_i {\xi}_{i,j}^2,
\end{align*}
from which we can infer that $||\widehat {\bm {X}}_{\cdot, j}||^2_2 \leq ||{\bm {X}}_{\cdot, j}||_2^2$. This implies
$$
\theta_j^2  = {\sigma}^2||\widehat {\bf{X}}_{\cdot, j} ||^2_2/n^2 \leq \frac{{\sigma}^2}{n}4(\Sigma_X)_{j,j} \leq \frac{4{\sigma}^2C_6^2}{n},
$$
under event $\Omega$, where $C_6$ is defined in Condition \ref{C3}. So we have
$$
\exp(-\frac{\tau^2_1}{\max_j\theta_j^2})\leq \exp(-\frac{A^2\log(p)/n}{4C_6^2/n}) = p^{-\frac{A^2}{4C_6^2}}.
$$
Define the positive constant $C_{11}=4C_6^2$. Together with \eqref{eqn:boundforetanew}, under the event $\Omega$, we have 
$$\mathbb{P}(\Omega_1 \mid \bm X, \Omega  ) \leq 2p\exp(- \frac{\tau_1^2}{\max_j\theta_j^2}) \leq 2p^{1-\frac{A^2}{C_{11}}}.$$
Take the expectation over the conditional distribution of $X\mid \Omega$, we have
\begin{equation}
\label{eqn:tailboundaprt1}
\begin{split}
\mathbb{P}(\Omega_1 \mid  \Omega  ) \leq 2p\exp(- \frac{\tau_1^2}{\max_j\theta_j^2}) \leq 2p^{1-\frac{A^2}{C_{10}}}.
\end{split}
\end{equation}

Given equations \eqref{eqn:subgaussian}, and \eqref{eqn:boundforetanew}, we have the following probability statement

\begin{equation}
    \begin{split}
        \mathbb{P}(||\frac{\widehat {\bf{X}}^\T E}{n}||_\infty \geq \tau )  &\leq \mathbb{P}(\Omega_1)\\
        &\leq \mathbb{P}(\Omega_1\mid \Omega)P(\Omega) + \mathbb{P}(\Omega_1\mid \Omega^c)P(\Omega^c)\\
        &\leq \mathbb{P}(\Omega_1\mid \Omega)  + P(\Omega^c)\\
        &\leq2p^{1-\frac{A^2}{C_{11}}}  + p\exp(-C_{10}n),
    \end{split}
\end{equation}
which concludes the proof.

\paragraph{Proof of Lemma \ref{lem: guX}}
Recall $\bm{X} = \bm{U} \Lambda^\T + E_x$, where $\bm{U} = (U_1\; U_2\; \ldots\; U_n)^\T \in \mathbb{R}^{n \times q}$, and $E_x = (\epsilon_{x,1}\; \epsilon_{x,2}\; \ldots\; \epsilon_{x,n})^\T \in \mathbb{R}^{n \times p}$.

The target quantity can be decomposed as follows:
\begin{equation}
\label{eqn:guX decompose}
    \left|\left|\frac{g(\bm{U})^\T \bm{X}}{n} - \gamma^\T \Lambda^\T\right|\right|_\infty \leq \left|\left|\left(\frac{g(\bm{U})^\T \bm{U}}{n} - \gamma^\T\right) \Lambda^\T\right|\right|_\infty + \left|\left|\frac{g(\bm{U})^\T E_x}{n}\right|\right|_\infty.
\end{equation}

We will control the first and second terms on the RHS of Equation \eqref{eqn:guX decompose} separately.

For the first term on RHS of equation \eqref{eqn:guX decompose}, we have
\begin{equation*}
\begin{split}
    &\left|\left|\left(\frac{g(\bm U)^\T \bm U - \gamma^\T}{n}\right)\Lambda^\T\right|\right|_\infty\\ =& \max_{1 \leq j\leq p} \left|\left(\frac{g(\bm U)^\T \bm U }{n}- \gamma^\T \right)\Lambda_{j,\cdot}\right|\\
    \leq& \left|\left|\frac{g(\bm U)^\T \bm U }{n}- \gamma^\T\right|\right|_2 \max_{1\leq j\leq p}||\Lambda_{j,\cdot}||_2,
\end{split}
\end{equation*}
where $\Lambda_{j,\cdot}\in \mathbb{R}^q$ is the jth row of $\Lambda$. The inequality follows the Cauchy-Schwarz inequality. We observed that $\frac{g(\bm U)^\T \bm U }{n}- \gamma^\T$ is the empirical average minus expectation for a $q\times 1$ vector. Define vector $a\in \mathbb{R}^{q\times 1}$, whose jth element is given by $a_j = g(\bm U)^\T \bm U_{\cdot,j}/n - \gamma_j$. Note that $\mathbb{E}(a_j) = 0$ and $\text{Var}(a_j) = \text{Var}(U_jg(U))/n=\Gamma_{j,j}/n$, where $\Gamma$ is defined as condition \ref{C3}. Let $t_n = \sqrt{\log(p)/n}$ for some positive constant $A$,
we have the following result:
\begin{equation*}
\begin{split}
    &P(\left|\left|\frac{g(\bm U)^\T \bm U }{n}- \gamma^\T\right|\right|_2 \geq t_n)\\
    =&P(\left|\left|\frac{g(\bm U)^\T \bm U }{n}- \gamma^\T\right|\right|_2^2 \geq t_n^2)\\
    =&P(\left|\left|a^\T\right|\right|_2^2 \geq t_n^2)\\
    \leq& \frac{\sum^q_{j=1}\mathbb{E}(a_j^2)}{t_n^2} \;\;(Markov's\;\; inequality)\\
    \leq &\frac{\sum_{j=1}^q\Gamma_{j,j}}{nt_n^2}\\
    \leq & \frac{C_3}{\log p}.\;\;(Condition\;\; \ref{C2})
\end{split}
\end{equation*}
The quantity ${C_3}/{\log p}$ goes to $0$ as $p$ goes to infinity. So far, we have shown that 
$$
\left|\left|\frac{g(\bm U)^\T \bm U }{n}- \gamma^\T\right|\right|_2 = O_p(\sqrt{\log(p)/{n}}).
$$
We observe that $\max_{1\leq j\leq p}||\Lambda_{j,\cdot}||_2^2 \leq \max_{1\leq j\leq p}\text{Var}(X_j)  \leq C_6$ by Condition \ref{C3}. Combining the above equations, we have
\begin{equation}
\label{eqn:guX decompose:part1}
    \left|\left|\left(\frac{g(\bm U)^\T \bm U }{n}- \gamma^\T\right)\Lambda^\T\right|\right|_\infty = O_p(\sqrt{\frac{\log(p)}{n}})
\end{equation}
We now control the second term on the RHS of Equation \eqref{eqn:guX decompose}. The proof approach is very similar to that used for proving Lemma \ref{lem:tailboundforepsilon}. Let $\zeta = g(\bm{U})^\T E_x/n \in \mathbb{R}^p$. We aim to show that:
$$
||\zeta ||_\infty = O_p\left(\sqrt{\frac{\log(p)}{n}}\right).
$$

First, we demonstrate that $||g(\bm{U})||_2/\sqrt{n}$ is bounded with high probability. For a constant $A_1$, consider the event $\Omega_g: \left\{\frac{||g(\bm{U})||_2^2}{n} < A_1\sigma_g^2\right\}$. We apply Markov's inequality to obtain the following result:
\begin{equation}
\label{eqn:macovomegag}
    \mathbb{P}(\Omega_g^c) \leq \frac{1}{A_1}.
\end{equation}

Note that $\epsilon_{x,1}, \ldots, \epsilon_{x,n}$ are i.i.d. sub-Gaussian random vectors, and $g(\bm{U})$ is independent of $\epsilon_{x,i}$. Consider the \(j\)th element of the vector $\zeta$: 
$$
\zeta_j = \sum_{i=1}^n \frac{g(U_i)\epsilon_{x,ij}}{n}.
$$

Conditional on $\bm{U}$, $\zeta_j$ is a linear combination of $n$ independent sub-Gaussian random variables $\epsilon_{x,1j}, \epsilon_{x,2j}, \ldots, \epsilon_{x,nj}$, which suggests that $\zeta_j$ is sub-Gaussian with mean 0 and parameter $\theta_j = \widetilde{\sigma}_j ||g(\bm{U})/n||_2$.

Recall that $t_n = \sqrt{\log(p)/n}$. For a positive constant $A_2$, from the tail bound of sub-Gaussian random variables:
\begin{align*}
    P(||\zeta||_\infty \geq A_2 t_n \mid \bm{U}, \Omega_g)
    &\leq \sum_{j=1}^p P(|\zeta_j| \geq A_2 t_n \mid \bm{U}, \Omega_g)\\
    &\leq 2p \max_j \exp\left(-\frac{A_2^2 t_n^2}{\theta_j^2}\right)\\
    &\leq 2p \exp\left(-\frac{A_2^2 t_n^2}{\max_j \theta_j^2}\right)\\
    &\leq 2p \exp\left(-\frac{A_2^2 \log(p)/n }{C_3 A_1/n}\right)\\
    &\leq 2p^{1 - \frac{A_2^2}{C_3 A_1}}.
\end{align*}

The last inequality holds because under the event $\Omega_g$, we have 
$\theta_j^2 = \widetilde{\sigma}_g^2 \frac{||g(\bm{U})||_2^2}{n^2} \leq \frac{C_3 A_1}{n}.$ Integrating this quantity with respect to $\bm{U}$, conditional on $\Omega_g$, we derive
\begin{equation}
\label{eqn:zetainfity}
    P(||\zeta||_\infty \geq A_2 t_n \mid \Omega_g) \leq 2p^{1 - \frac{A_2^2}{C_3 A_1}}.
\end{equation}

Combining Equations \eqref{eqn:macovomegag} and \eqref{eqn:zetainfity}, we obtain
\begin{equation*}
    \begin{split}
        \mathbb{P}(||\zeta||_\infty \geq A_2 t_n)
        &= \mathbb{P}(||\zeta||_\infty \geq A_2 t_n \mid \Omega_g) \mathbb{P}(\Omega_g) + \mathbb{P}(||\zeta||_\infty \geq A_2 t_n \mid \Omega_g^c) \mathbb{P}(\Omega_g^c)\\
        &\leq \mathbb{P}(||\zeta||_\infty \geq A_2 t_n \mid \Omega_g) + \mathbb{P}(\Omega_g^c)\\
        &\leq 2p^{1-\frac{A_2^2}{C_3 A_1}} + \frac{1}{A_1}.
    \end{split}
\end{equation*}
We can select positive constants $A_1$ and $A_2$ to make the probability $\mathbb{P}(||\zeta||_\infty \geq A_2 t_n)$ arbitrarily small. Thus far, we have demonstrated that
\begin{equation}
\label{eqn:guX decompose:part2}
    \left|\left|\frac{g(\bm{U})^\T E_x}{n}\right|\right|_\infty = O_p\left(\sqrt{\frac{\log(p)}{n}}\right).
\end{equation} 

Combining Equations \eqref{eqn:guX decompose:part1} and \eqref{eqn:guX decompose:part2}, we conclude the proof of Lemma \ref{lem: guX}.

\paragraph{Proof of Lemma \ref{lem:O}}
Refer to Lemma C.10 in \cite{fan2013large_sup}. Our conditions \ref{C1}--\ref{C3} are sufficient to verify their assumptions, and $q$ is known in our setting.

\paragraph{Proof of Lemma \ref{lem:highdloadingestimation}}
Refer to Theorem 3.3 in \cite{fan2013large_sup}. Our conditions \ref{C1}--\ref{C3} are sufficient to verify their assumptions, and $q$ is known in our setting.

\paragraph{Proof of Lemma \ref{lem:rate guXFbeta}}

Recall that $\widehat{\Lambda} = (\sqrt{\widehat{\lambda}_1}\widehat{\xi}_1 \; \ldots \; \sqrt{\widehat{\lambda}_q}\widehat{\xi}_q)$ and $\widehat{\Lambda}^\T \widehat{F}^\T = 0$. Consider the following decomposition for the target quantity:
\begin{equation}
    \label{eqn:decomp_lem:rate guXFbeta}
    \begin{split}
    &\frac{g(\bm{U})^\T \bm{X}}{n} \widehat{F}^\T (\widehat{\beta} - \dot{\beta})\\
    &= \left[\left\{\frac{g(\bm{U})^\T \bm{X}}{n} - \gamma^\T \Lambda^\T\right\} + \gamma^\T \Lambda^\T \right]\widehat{F}^\T (\widehat{\beta} - \dot{\beta}) \\
    &= \left[\left\{\frac{g(\bm{U})^\T \bm{X}}{n} - \gamma^\T \Lambda^\T\right\} + \left\{\gamma^\T (I - O^\T O)\Lambda^\T\right\} + \gamma^\T O^\T (O \Lambda^\T  - \widehat{\Lambda}^\T) \right]\widehat{F}^\T (\widehat{\beta} - \dot{\beta}) \\
    &= (a+b+c)\widehat{F}^\T(\widehat{\beta} - \dot{\beta}),
    \end{split}
\end{equation}
where $a, b, c \in \mathbb{R}^{1 \times p}$, and $O \in \mathbb{R}^{p \times p}$ is defined in Lemma \ref{lem:O}. We now control the infinity norms of $a, b,$ and $c$ respectively. Note $||a||_\infty = O_p(\sqrt{\log(p)/n})$ given Lemma \ref{lem: guX}.

Let $\Lambda_{j,\cdot} \in \mathbb{R}^{q \times 1}$ be the $j$th row of $\Lambda \in \mathbb{R}^{p \times q}$. For $||b||_\infty$, we have the following result:
\begin{equation}
    \label{eqn:decomp_lem:rate guXFbeta,partb}
    \begin{split}
        ||b||_\infty &= \max_{1 \leq j \leq p} |b_j| \\
        &= \max_{1 \leq j \leq p} |\gamma^\T (I - O^\T O)\Lambda_{j,\cdot}| \\
        &\leq ||\gamma^\T (I - O^\T O)||_2  \max_{1 \leq j \leq p} ||\Lambda_{j,\cdot}||_2 \\
        &\leq ||\gamma^\T||_2 || (I - O^\T O)||_2  \max_{1 \leq j \leq p} ||\Lambda_{j,\cdot}||_2 \\
        &\leq C_3C_6 \; O_p\left(\frac{1}{\sqrt{n}}\right) \\
        &= O_p\left(\frac{1}{\sqrt{n}}\right),
    \end{split}
\end{equation}
where the first inequality follows from the Cauchy-Schwarz inequality. The last inequality is derived using conditions \ref{C2}, \ref{C3}, and Lemma \ref{lem:O}, while considering condition \ref{C1}: \( n = O(p) \).

For \( ||c||_\infty \), we first note that \( ||O^\T||_2 = \sqrt{||O O^\T||_2} = O_p(1) \). Consequently, we have the following bound for \( ||c||_\infty \):
\begin{equation}
\label{eqn:decomp_lem:rate guXFbeta,partc}
\begin{split}
        ||c||_\infty &= \max_{1\leq j\leq p}|c_j|\\
        &=  \max_{1\leq j\leq p} |\gamma^\T O^\T (O\Lambda_{j,\cdot} - \widehat\Lambda_{j,\cdot})|\\
        &\leq ||\gamma^\T O^\T ||_2\max_{1\leq j\leq p}||O\Lambda_{j,\cdot} - \widehat\Lambda_{j,\cdot}||_2\\
        &\leq O_p(  \frac{1}{\sqrt{p}}+ \sqrt{\frac{\log (p)}{n}}), 
 \end{split}
\end{equation}

The first inequality follows from the Cauchy-Schwarz inequality, and the last inequality follows from \( ||O^\T||_2 = O_p(1) \), Condition \ref{C2}, and Lemma \ref{lem:highdloadingestimation}. Note that Condition \ref{C1} gives \( n = O(p) \), which implies \( O_p\left( \frac{1}{\sqrt{p}} + \sqrt{\frac{\log p}{n}} \right) = O_p\left( \sqrt{\frac{\log p}{n}} \right) \). Combining Lemmas \ref{lem: guX} and equations \eqref{eqn:decomp_lem:rate guXFbeta,partb} and \eqref{eqn:decomp_lem:rate guXFbeta,partc}, we have:
$$
||a+b+c||_\infty = O_p(\sqrt{\frac{\log(p)}{n}})
$$
We now control $(a+b+c)F^\T(\widehat \beta - \dot\beta)$. Observe that $\widehat{F}^\T = {B}_{{\widehat\Lambda}^\perp}{B}_{{\widehat\Lambda}^\perp}^T = \sum^p_{i=q+1}\widehat{\xi}_i\widehat{\xi}_i^\T$ and $I_p -\widehat{F}^\T =  \sum^q_{i=1}\widehat{\xi}_i\widehat{\xi}_i^\T$, we have 
\begin{equation}
    \label{eqn:finalresultguxfbeta}
    \begin{split}
            &(a+b+c)\widehat F^\T (\widehat{\beta}- \dot{\beta})\\
            =&(a+b+c)(\widehat{\beta}- \dot{\beta}) - (a+b+c)(\sum^q_{i=1}\widehat{\xi}_i\widehat{\xi}_i^\T)(\widehat{\beta}- \dot{\beta}) \\
            \leq& ||(a+b+c)||_\infty||(\widehat{\beta}- \dot{\beta})||_1 - \text{Trace}\{(a+b+c)(\sum^q_{i=1}\widehat{\xi}_i\widehat{\xi}_i^\T)(\widehat{\beta}- \dot{\beta})\}\\
            =& ||(a+b+c)||_\infty||(\widehat{\beta}- \dot{\beta})||_1 - \text{Trace}\{(\sum^q_{i=1}\widehat{\xi}_i\widehat{\xi}_i^\T)(\widehat{\beta}- \dot{\beta})(a+b+c)\}\\
            \leq& ||(a+b+c)||_\infty||(\widehat{\beta}- \dot{\beta})||_1 + |\text{Trace}(\sum^q_{i=1}\widehat{\xi}_i\widehat{\xi}_i^\T)|\;|\text{Trace}\{(\widehat{\beta}- \dot{\beta})(a+b+c)\}|\\
            =&||(a+b+c)||_\infty||(\widehat{\beta}- \dot{\beta})||_1 + |\text{Trace}(\sum^q_{i=1}\widehat{\xi}_i\widehat{\xi}_i^\T)|\;|\text{Trace}\{(a+b+c)(\widehat{\beta}- \dot{\beta})\}|\\
            =&||(a+b+c)||_\infty||(\widehat{\beta}- \dot{\beta})||_1 + |\text{Trace}(\sum^q_{i=1}\widehat{\xi}_i\widehat{\xi}_i^\T)|\;|\{(a+b+c)(\widehat{\beta}- \dot{\beta})\}|\\
            \leq & (1+q)||(a+b+c)||_\infty||(\widehat{\beta}- \dot{\beta})||_1.
    \end{split}
\end{equation}
Line 2 follows from \(\widehat{F} = I_p - \sum_{i=1}^q \widehat{\xi}_i \widehat{\xi}_i^\T\). Line 3 uses Hölder's inequality. Lines 4 and 6 employ the trace property: \(\operatorname{Trace}(AB) = \operatorname{Trace}(BA)\). Line 5 follows from the Cauchy–Schwarz inequality for the trace. The last line follows from the fact that the trace of a matrix equals the sum of its eigenvalues. Define \(d = a + b + c \in \mathbb{R}^{1 \times p}\). This concludes the proof.


\paragraph{Proof of Lemma \ref{lem:Ex}}
Since \(0 < C_1 \leq \lambda_{\min}(D) \leq \lambda_{\max}(D) \leq C_2 < \infty\), the matrix \(D\) satisfies Equations (1.12) and (1.16) with finite \(K(s, 1, D)\) and \(\rho_{\max}(s)\) of \cite{zhou2009restricted}.

Let \(\theta_{T_0}\) be the subvector of \(\theta\) confined to the locations of its \(s\) largest coefficients. Define \(E_s = \{\theta \in \mathbb{R}^p : \|D^{1/2}\theta\|_2 = 1, \|\theta_{T_0}\|_1 \leq \|\theta_{T_0^C}\|_1\}\). We observe that each column of \(E_x D^{-1/2}\) is an independent copy of an isotropic \(\phi_2\) random vector on \(\mathbb{R}^p\). It is evident that under our Condition \ref{C1}, \(n \gg s \log(p)\). From Theorem 1.6 of \cite{zhou2009restricted} (with \(k_0\) in this theorem taken as 1), with probability \(1 - \exp(-cn)\) for some \(c > 0\), we have for all \(\theta \in E_s\),
\begin{equation}
\label{eqn:Ex}
    0.9^{1/2} \leq \frac{\|(E_x D^{-1/2})D^{1/2}\theta \|_2}{\sqrt{n}} = \frac{\|E_x\theta \|_2}{\sqrt{n}} \leq 1.1^{1/2}.
\end{equation}

We notice that 
\begin{align*}
   &\max_{||D^{1/2}\theta||_2 = 1, ||\theta_{T_0}||_1 \leq ||\theta_{T_0^C}||_1} \frac{||E_x\theta||_2}{\sqrt{n}} =     \max_{||D^{1/2}\theta||_2 = 1, ||\theta_{T_0}||_1 \leq ||\theta_{T_0^C}||_1} \frac{||E_x\theta||_2}{\sqrt{n}||D^{1/2}\theta||_2} = \max_{ ||\theta_{T_0}||_1 \leq ||\theta_{T_0^C}||_1} \frac{||E_x\theta||_2}{\sqrt{n}||D^{1/2}\theta||_2}\\
      &\min_{||D^{1/2}\theta||_2 = 1, ||\theta_{T_0}||_1 \leq ||\theta_{T_0^C}||_1} \frac{||E_x\theta||_2}{\sqrt{n}} =     \min_{||D^{1/2}\theta||_2 = 1, ||\theta_{T_0}||_1 \leq ||\theta_{T_0^C}||_1} \frac{||E_x\theta||_2}{\sqrt{n}||D^{1/2}\theta||_2} = \min_{ ||\theta_{T_0}||_1 \leq ||\theta_{T_0^C}||_1} \frac{||E_x\theta||_2}{\sqrt{n}||D^{1/2}\theta||_2}.
\end{align*}

Equation \eqref{eqn:Ex} implies that 
\begin{align*}
   &\max_{\theta \in E_s^A} \frac{\theta^\T E_x^\T E_x\theta}{n} \leq 1.1\lambda_{\max}(D),\\
   &\min_{\theta \in E_s^A} \frac{\theta^\T E_x^\T E_x\theta}{n} \geq0.9\lambda_{\min}(D),
\end{align*}

where $E_s^A = \{\theta \in\mathbb{R}^p: ||\theta||_2 = 1, ||\theta_{T_0}||_1 \leq ||\theta_{T_0^C}||_1\}$. We let $E_s^B = \{\theta \in\mathbb{R}^p: ||\theta||_2 = 1, 0<||\theta||_0 \leq 2s\}$. Note that any $\theta \in E_s^A$, one must have $\theta \in E_s^B$ given the definition of $T_0$, which means $E_s^B\subset E_s^A $. Thus, with the same probability,
\begin{align*}
   &\max_{\theta \in E_s^B} \frac{\theta^\T E_x^\T E_x\theta}{n}\leq\max_{\theta \in E_s^A} \frac{\theta^\T E_x^\T E_x\theta}{n} \leq 1.1\lambda_{\max}(D),\\
   &\min_{\theta \in E_s^B}\frac{\theta^\T E_x^\T E_x\theta}{n}\geq\min_{\theta \in E_s^A} \frac{\theta^\T E_x^\T E_x\theta}{n} \geq 0.9\lambda_{\min}(D).
\end{align*}

\paragraph{Proof of Lemma \ref{lem:highdL_1error rate}}
Let $E =({\epsilon}_{y,i},\ldots,{\epsilon}_{y,n})^\T  \in \mathbb{R}^n$. We have 
$$
\bm Y = {\bf X}\dot{\beta} + g(\bm U) + E
$$

Due to the optimality of $\widehat{\beta}$, we have 
\begin{equation}\label{eqn:optimality beta hat}
\begin{split}
    &\qquad \frac{||\bm Y - \widehat {\bf{X}}\widehat{\beta}||^2_2}{2n} \leq
    \frac{||Y - \widehat {\bf{X}}\dot{\beta}||^2_2}{2n}\\
    \Longleftrightarrow&\qquad \frac{||\widehat {\bf{X}}(\widehat{\beta} - \dot{\beta})||^2_2}{2n}   \leq\frac{(Y-\widehat {\bf{X}}\dot{\beta})^
    \T\widehat {\bf{X}}(\widehat{\beta}-\dot{\beta})}{n} .
    \end{split}
\end{equation}
We can decompose the RSH of \eqref{eqn:optimality beta hat} as
\begin{equation}
    \begin{split}
    \label{eqn:RSH process final result}
        \frac{(Y-\widehat {\bf{X}}\dot{\beta})^
    \T\widehat {\bf{X}}(\widehat{\beta}-\dot{\beta})}{n} &= \frac{(\bm X\dot{\beta} + g(\bm U) +E - \bm X\widehat{F}^\T\dot{\beta})^\T \bm X\widehat{F}^\T(\widehat{\beta}-\dot{\beta})}{n}\\
    &=\frac{\{\bm X(I-\widehat{F}^\T)\dot{\beta} + g(\bm U) +E\}^\T \bm X\widehat{F}^\T(\widehat{\beta}-\dot{\beta})}{n}\\
    &=\frac{(g(\bm U) \bm X \widehat {F}^\T + E^\T \bm  X\widehat{F}^\T)(\widehat{\beta}-\dot{\beta})}{n}\\
    &= \frac{(g(\bm U)^\T\widehat {\bf{X}} + E^\T \widehat {\bf{X}})(\widehat{\beta}-\dot{\beta})}{n}\\
    &\leq (1+q)||d||_\infty||\widehat\beta - \dot\beta ||_1 + ||\frac{E^\T \widehat {\bm X}}{n}||_\infty||\widehat\beta - \dot\beta ||_1.
    \end{split}
\end{equation}
The third equality follows Lemma \ref{lem:property of hat F highd} and  $d\in\mathbb{R}^{1\times p} $ is defined at Lemma \ref{lem:rate guXFbeta}. 

Note that $||\widehat\beta -\dot\beta||_0\leq 2s$, we have $||\widehat\beta - \dot\beta ||_1\leq \sqrt{2s} ||\widehat\beta - \dot\beta||_2$. 

We now process the LHS of \eqref{eqn:optimality beta hat}. Given Lemma \ref{prop:REC_highd}, for any $\delta$, there exists an integer $n_1$ such that 
$\dot \Omega := \{||\widehat X(\widehat\beta - \dot{\beta})||^2_2/n \geq \pi_0^2||\widehat{\beta} - \dot{\beta}||_2^2\}$ with $\mathbb P(\dot \Omega) \geq 1-\delta$ for $n>n_1$.  

Conditional on event $\dot \Omega$, and combining with equation \eqref{eqn:RSH process final result}, we can rewrite \eqref{eqn:optimality beta hat} into
\begin{equation}
    \begin{split}
        \frac{\pi_0^2}{2}||\widehat{\beta} - \dot{\beta}||_2^2&\leq \frac{||\widehat X(\widehat\beta - \dot{\beta})||^2_2}{2n }\\
        &\leq  \left\{(1+q)||d||_\infty + ||\frac{E^\T \widehat {\bm X}}{n}||_\infty\right\}||\widehat\beta - \dot\beta ||_1\\
        &\leq \left\{(1+q)||d||_\infty + ||\frac{E^\T \widehat {\bm X}}{n}||_\infty\right\}\sqrt{2s}||\widehat\beta - \dot\beta ||_2.
    \end{split}
\end{equation}
Cancel a common factor $||\widehat\beta - \dot\beta ||_2$ and notice that $||\widehat\beta - \dot\beta ||_1\leq \sqrt{2s} ||\widehat\beta - \dot\beta||_2$, we have 
\begin{equation*}
\begin{split}
        ||\widehat\beta - \dot\beta ||_1 & \leq \sqrt{2s} ||\widehat\beta - \dot\beta||_2 \\
        &\leq \sqrt{2s}\; \frac{2}{\pi_0^2}\left\{(1+q)||d||_\infty + ||\frac{E^\T \widehat {\bm X}}{n}||_\infty\right\}\sqrt{2s}\\
        &= \frac{4s}{\pi_0^2}\left\{(1+q)||d||_\infty + ||\frac{E^\T \widehat {\bm X}}{n}||_\infty\right\}
\end{split}
\end{equation*}
with high probability, which is the desired result.

\paragraph{Proof of Lemma \ref{prop:REC_highd}}

For any $\delta>0$, we are going to show that there exist an integer $n_0$ such that 
$$
\inf_{n>n_0} \mathbb{P}\{ {||\widehat {\bf{X}}\theta||_2}\geq \pi_0\sqrt{n}{||\theta||_2}, \forall ||\theta||_0\leq 2s\} \geq 1 -\delta
$$

We have ${\bf X} = {\bf U} \Lambda + E_x $, where ${\bf U} = (U_1\;U_2\;\ldots\;U_n )^\T$, $E_x = (\epsilon_{x,1}\;\epsilon_{x,2}\;\ldots\;\epsilon_{x,n})^\T\in \mathbb{R}^{n\times p}$. We also notice that 
\begin{equation*}
    \begin{split}
                \widehat{\bf X} &= {\bf X}\widehat{F}^\T = {\bf X} - {\bf X}(I_p - \widehat{F}^\T)\\
                &= E_x + R,
    \end{split}
\end{equation*}

where $R = {\bf U} \Lambda - {\bf X}(I_p - \widehat{F}^\T)$. By \eqref{eqn:singular,pca} and Lemma \ref{lem:highdprojection}, we have   ${\bf X}(I_p - \widehat{F}^\T) =\sum^q_{i=1}\sqrt{\lambda_i(n-1)}\eta_i\xi_i^\T$. We further have the following decomposition:
$$
\frac{\bf \widehat X^\T  \widehat X}{n}  = \frac{E_x^\T E_x}{n}  + \frac{E_x^\T R}{n}+\frac{ R^\T E_x}{n} + \frac{R^\T R}{n},
$$
which implies
 \begin{equation}
\label{eqn:sec decomposition}
    \begin{split}
    &\min_{||\theta||_2=1,||\theta||_0\leq 2s} \theta^\T \frac{\bf \widehat X^\T \widehat X}{n}\theta \geq \min_{||\theta||_2=1,||\theta||_0\leq 2s} \theta^\T \frac{E_x^\T E_x}{n} \theta\\
        &- 2\max_{||\theta||_2=1,||\theta||_0\leq 2s} \theta^\T \frac{E_x^\T R}{n} \theta -  \max_{||\theta||_2=1,||\theta||_0\leq 2s} \theta^\T \frac{R^\T R}{n} \theta.\\
    \end{split}
\end{equation}
We now control each term one by one: 

For $\theta^\T {R^\T R} \theta/n$, we have 
\begin{equation}
\label{eqn:SEC,2}
    \begin{split}
        \max_{||\theta||_2=1,||\theta||_0\leq 2s}\theta^\T \frac{R^\T R}{n} \theta &=         \max_{||\theta||_2=1,||\theta||_0\leq 2s}\frac{1}{n} \sum^n_{i = 1}(\sum^{p}_{j=1}R_{i,j} \theta_j)^2\\
        &\leq \max_{||\theta||_2=1,||\theta||_0\leq 2s}\max_{i} (\sum^{p}_{j=1}R_{i,j} \theta_j)^2;\\
        &\leq \max_{||\theta||_2=1,||\theta||_0\leq 2s} (||R||_\infty ||\theta||_1)^2\\
        &\leq 2s||R||_\infty^2.
    \end{split}
\end{equation}

Given equation \eqref{eqn:SEC,1.2}, we have 
\begin{equation}
\label{eqn:SEC,3}
    \begin{split}
        \max_{||\theta||_2=1,||\theta||_0\leq 2s}\theta^\T \frac{R^\T E_x}{n} \theta &\leq          \max_{||\theta||_2=1,||\theta||_0\leq 2s}\sqrt{\theta^\T \frac{R^\T R}{n} \theta}\sqrt{\theta^\T \frac{E_x^\T E_x}{n} \theta}\\
        &\leq \sqrt{1.1s\lambda_{\max}(D)}||R||_\infty.
    \end{split}
\end{equation}

Combining equations \eqref{eqn:SEC,1.1}, \eqref{eqn:SEC,1.2}, \eqref{eqn:SEC,2}, and \eqref{eqn:SEC,3}, we can rewrite \eqref{eqn:sec decomposition} as 

$$
\min_{||\theta||_2=1,||\theta||_0\leq 2s} \theta^\T \frac{\bf \widehat X^\T \widehat X}{n}\theta \geq 0.9\lambda_{\min} (D) - 2s||R||^2_\infty - \sqrt{2.2s\lambda_{\max}(D)}||R||_\infty,
$$
with probability $1-\exp(-cn)$.

Hence, we only need to show that $s||R||^2_\infty \overset{p}{\to} 0$, which has been guaranteed by Lemma 6 of \cite{guo2022doubly_sup}.

\section{Proof of Theorems}
\label{sec:thmproof}




\subsection{Proof of Theorem \ref{thm:l0 plurality}}
We aim to demonstrate that if there exists a vector $${\beta}^* \in \underset{{\widetilde\beta}\in \mathbb{R}^{p}}{\argmin} \;\mathbb{E}\{Y - \widetilde X^\T\widetilde\beta\}^2 \text{, such that }||{\widetilde\beta}||_0\leq p-q-1,$$ and ${\beta}^*\not= \dot{\beta}$, then a contradiction arises.


Given Lemma \ref{lem:l0 structure}, there exists an $\alpha \in \mathbb{R}^q$ such that ${\beta}^* = \dot{\beta} + \Sigma_X^{-1}\Lambda\alpha$.
Define $C := \{ j \mid {\beta}_j^* = 0 \}$ and $M := \{ j \mid {\beta}_j^* \not= 0 \}$. We have $|M| \leq p-q-1$ and $|C| \geq q+1$ because $||{\beta}^*||_0 \leq p-q-1$. We first establish the following claim:
\begin{enumerate}
    \item [{\bf Claim:}] $|C \cap \mathcal{A}| \geq 2$
\end{enumerate}

If not, we would have $|C \cap \mathcal{A}| \leq 1$. Given that $|C| \geq q+1$, it must follow that $|C \cap \mathcal{A}^c| \geq q$. Consider $\widetilde{C} \subset C \cap \mathcal{A}^c$ such that $|\widetilde{C}| = q$. Examining an element within $\widetilde{C}$, we find:
\begin{equation*}
    \begin{split}
        0 = {\beta}^*_{\widetilde{C}} & = [\dot{\beta} + \Sigma_X^{-1}\Lambda\alpha]_{\widetilde{C}}\\
        & = \dot{\beta}_{\widetilde{C}} + [\Sigma_X^{-1}\Lambda\alpha]_{\widetilde{C}}\\
        & = [\Sigma_X^{-1}\Lambda\alpha]_{\widetilde{C}}.
    \end{split}
\end{equation*}

By the invertibility Assumption \ref{A1}, we consequently have $\alpha = 0$, which implies ${\beta}^* = \dot{\beta} + \Sigma_X^{-1}\Lambda\alpha = \dot{\beta}$.

We have $|C| \geq q+1$, and $|C \cap A| \geq 2$. Let $\{C^{(i)}\}_{i=1}^{q+1}$ be such that each $C^{(i)}$ is a proper subset of $C$, $|C^{(i)}| = q$, and $C^{(i)} \cap A \neq \emptyset$. Define ${\beta}^{(i)}$ as follows: 
$$
    \beta^{(i)} = \underset{\widetilde\beta_{C^{(i)}}=0, \widetilde\beta \in \mathbb{R}^p}{\argmin}\mathbb{E}\{Y - \widetilde X^\T\widetilde\beta\}^2, \quad i=1,2,\ldots,q+1.
$$

From Lemma \ref{lem:Expert has unique sol}, the set $\{\beta^{(i)}\}_{i=1}^{q+1}$ is uniquely defined. Given that $\emptyset \subsetneqq C^{(i)} \subsetneqq C$, we have:
\begin{equation}
\label{eqn:betaitilebeta}
        \underset{\widetilde\beta \in \mathbb{R}^p}{\min}\mathbb{E}\{Y - \widetilde X^\T\widetilde\beta\}^2 \leq \underset{\widetilde\beta_{C^{(i)}}=0, \widetilde\beta \in \mathbb{R}^p}{\min}\mathbb{E}\{Y - \widetilde X^\T\widetilde\beta\}^2 \leq \underset{\widetilde\beta_{C}=0, \widetilde\beta \in \mathbb{R}^p}{\min}\mathbb{E}\{Y - \widetilde X^\T\widetilde\beta\}^2.
\end{equation}

Since ${\beta}^* \in \underset{\widetilde\beta \in \mathbb{R}^p}{\argmin}\;\mathbb{E}\{Y - \widetilde X^\T\widetilde\beta\}^2$ and ${\beta}^* \in \underset{\widetilde\beta_{C}=0, \widetilde\beta \in \mathbb{R}^p}{\argmin}\mathbb{E}\{Y - \widetilde X^\T\widetilde\beta\}^2$, 
the equation \eqref{eqn:betaitilebeta} can be rewritten as:
$$
        \underset{\widetilde\beta \in \mathbb{R}^p}{\min}\mathbb{E}\{Y - \widetilde X^\T\widetilde\beta\}^2 = \underset{\widetilde\beta_{C^{(i)}}=0, \widetilde\beta \in \mathbb{R}^p}{\min}\mathbb{E}\{Y - \widetilde X^\T\widetilde\beta\}^2 = \underset{\widetilde\beta_{C}=0, \widetilde\beta \in \mathbb{R}^p}{\min}\mathbb{E}\{Y - \widetilde X^\T\widetilde\beta\}^2,
$$

which implies that $\beta^* \in \underset{\widetilde\beta_{C^{(i)}}=0, \widetilde\beta \in \mathbb{R}^p}{\argmin}\mathbb{E}\{Y - \widetilde X^\T\widetilde\beta\}^2 = \{\beta^{(i)}\}$, where the last equality is validated by Lemma \ref{lem:Expert has unique sol}. Thus, we have $\beta^* = \beta^{(i)}$ for $i \in \{1, 2, \ldots, q+1\}$, which violates Condition \ref{A4}.

This contradiction indicates that the set $\{\underset{{\widetilde\beta} \in \mathbb{R}^{p}}{\argmin} \;\mathbb{E}\{Y - \widetilde X^\T\widetilde\beta\}^2 \text{ s.t. } ||{\widetilde\beta}||_0 \leq p-q-1\} = \{\dot{\beta}\}$ if $s \leq p-q-1$, while $\{\underset{{\widetilde\beta} \in \mathbb{R}^{p}}{\argmin} \;\mathbb{E}\{Y - \widetilde X^\T\widetilde\beta\}^2 \text{ s.t. } ||{\widetilde\beta}||_0 \leq p-q-1\} = \emptyset$ if $s \geq p-q$, corresponding to the two cases in Theorem \ref{thm:l0 plurality}.

\subsection{Proofs of Theorem \ref{thm:lowd thm} }

\paragraph{Proof of the first part of Theorem \ref{thm:lowd thm}}

We are going to show that for any $\delta >0 $, there exist $A_\delta$ and $n_0$ such that $$
||\widehat \beta - \dot{\beta}||_1 \leq A_\delta/\sqrt{n}
$$
with probability at least $1-\delta$ for all $n>n_0$.


Due to the optimality of $\widehat{\beta}$, we have 
\begin{equation}\label{eqn:optimality beta hat lowd}
\begin{split}
    &\qquad \frac{||Y - \widehat {\bf{X}}\widehat{\beta}||^2_2}{2n} \leq
    \frac{||Y - \widehat {\bf{X}}\dot{\beta}||^2_2}{2n}\\
    \Longleftrightarrow&\qquad \frac{||\widehat {\bf{X}}(\widehat{\beta} - \dot{\beta})||^2_2}{2n}   \leq\frac{(Y-\widehat {\bf{X}}\dot{\beta})^
    \T\widehat {\bf{X}}(\widehat{\beta}-\dot{\beta})}{n} .
    \end{split}
\end{equation}
By the models \eqref{eqn:model1} and \eqref{eqn:model2}, we can decompose the first term of RSH of \eqref{eqn:optimality beta hat lowd} as
\begin{equation*}
    \begin{split}
        \frac{(Y-\widehat {\bf{X}}\dot{\beta})^
    \T\widehat {\bf{X}}(\widehat{\beta}-\dot{\beta})}{n} &= \frac{(\bm X\dot{\beta} +g(\bm U)+ E  - \bm X\widehat{F}^\T\dot{\beta})^\T\bm  X\widehat{F}^\T(\widehat{\beta}-\dot{\beta})}{n}\\
    &=\frac{\{\bm X(I-\widehat{F}^\T)\dot{\beta} + g(\bm U) + E \}^\T \bm X\widehat{F}^\T(\widehat{\beta}-\dot{\beta})}{n}\\
    &=\frac{(g(\bm U)^\T { \bf{X}}\widehat{F}^\T +  E ^\T{ \bf{X}}\widehat{F}^\T )(\widehat{\beta}-\dot{\beta})}{n}\\
    &= \frac{(g(\bm U)^\T \widehat {\bf{X}} +  E ^\T \widehat {\bf{X}})(\widehat{\beta}-\dot{\beta})}{n}. 
    \end{split}
\end{equation*}
The third equality follows by Lemma \ref{lem:property of hat F}.

Note that $||\widehat \beta - \dot{\beta}||_0\leq 2s$, we have 
$||\widehat \beta - \dot{\beta}||_1 \leq \sqrt{2s}||\widehat \beta - \dot{\beta}||_2$. Given Lemma \ref{prop:REC_lowd}, there exists an integer $n_1$ such that 
$||\widehat {\bm X}(\widehat\beta - \dot{\beta})||^2_2/n \geq \pi_0^2||\widehat{\beta} - \dot{\beta}||_2^2$ with probability at least $1-\delta/2$ for $n>n_1$. Equation \eqref{eqn:optimality beta hat lowd} yields.

\begin{equation}
\label{eqn:eqn:optimality beta hat lowd,2}
\begin{split}
 \pi_0^2||\widehat{\beta} - \dot{\beta}||_2^2 &\leq ||{\bf \widehat X}(\widehat\beta - \dot{\beta})||^2_2/n\\ 
 &\leq \frac{(g(\bm U)^\T \widehat {\bf{X}} +  E ^\T \widehat {\bf{X}})(\widehat{\beta}-\dot{\beta})}{n} \\
&\leq ||\widehat \beta - \beta||_1\left( \left|\left|\frac{{g(\bm U)^\T \widehat {\bf{X}}} }{n}\right|\right|_\infty + \left|\left|\frac{{\bf \widehat X}^\T E}{n}\right|\right|_\infty\right)\\
&\leq \sqrt{2s}||\widehat \beta -\dot{\beta}||_2\left( \left|\left|\frac{{g(\bm U)^\T \widehat {\bf{X}}} }{n}\right|\right|_\infty + \left|\left|\frac{{\bf \widehat X}^\T E}{n}\right|\right|_\infty\right),
\end{split}
\end{equation}

with probability at least $1-\delta/2$, where $ \left|\left|{{\bf \widehat X}^\T E}/{n}\right|\right|_\infty = O_p(1/\sqrt{n})$ by Lemma \ref{lem:tailboundforepsilon,lowd}. Besides, given Lemma \ref{lem: Order of ||FB|| lowd}, we have
$$
 \left|\left|\frac{{g(\bm U)^\T \widehat {\bf{X}}} }{n}\right|\right|_\infty \leq   \left|\left|\frac{{ g(\bm U)^\T \widehat {\bf{X}}} }{n}\right|\right|_2 = O_p(\frac{1}{\sqrt{n}}).
$$

Cancelling a factor $||\widehat \beta - \dot{\beta} ||_2$ of \eqref{eqn:eqn:optimality beta hat lowd,2} and noting that $||\widehat \beta - \dot{\beta} ||_1 \leq ||\widehat \beta - \dot{\beta} ||_2\sqrt{2s}$ yield the claim of first part of Theorem \ref{thm:lowd thm}.

\paragraph{Proof of the second part of Theorem \ref{thm:lowd thm}}
Let $C_{16} = \min_{j \in \mathcal{A}} |\beta_j| \geq 0$. Considering $k = s$, the event $\{\widehat{\mathcal{A}} \neq \mathcal{A}\} \subset \{||\widehat{\beta} - \dot{\beta}||_1 \geq C_{16}\}$. From the first part of Theorem \ref{thm:lowd thm}, we know that $||\widehat{\beta} - \dot{\beta}||_1 = O_p\left(\frac{1}{\sqrt{n}}\right)$, which means that for any $\delta > 0$, there exists a constant $M_\delta$ and an integer $n_1 > 0$ such that
$$
\sup_{n > n_1} \mathbb{P}\left(||\widehat{\beta} - \dot{\beta}||_1 \geq \frac{M_\delta}{\sqrt{n}}\right) \leq \delta.
$$ 

For $n \geq n_2 := \max(n_1, \frac{M_\delta^2}{C_{16}^2})$, $C_{16} \geq \frac{M_\delta}{\sqrt{n}}$, so $\{\widehat{\mathcal{A}} \neq \mathcal{A}\} \subset \{||\widehat{\beta} - \dot{\beta}||_1 \geq C_{16}\} \subset \{||\widehat{\beta} - \dot{\beta}||_1 \geq \frac{M_\delta}{\sqrt{n}}\}$.

Thus, we have 
\begin{equation*}
    \sup_{n \geq n_2} \mathbb{P}(\widehat{\mathcal{A}} \neq \mathcal{A}) \leq \sup_{n \geq n_2} \mathbb{P}\left(||\widehat{\beta} - \dot{\beta}||_1 \geq \frac{M_\delta}{\sqrt{n}}\right) \leq \sup_{n \geq n_1} \mathbb{P}\left(||\widehat{\beta} - \dot{\beta}||_1 \geq \frac{M_\delta}{\sqrt{n}}\right) \leq \delta.
\end{equation*}
Given the arbitrariness of $\delta$, we conclude that $\mathbb{P}(\widehat{\mathcal{A}} \neq \mathcal{A}) \rightarrow 0$ as $n \rightarrow \infty$.

\subsection{Proofs of Theorem \ref{thm:highd error rate}}

\paragraph{Proof of the first part of Theorem \ref{thm:highd error rate}}

By Lemma \ref{lem:highdL_1error rate}, we have 
$$
||\widehat{\beta}-\dot{\beta}||_1=O_p\left(s\left\{(1+q)||d||_\infty + ||\frac{E^\T \widehat {\bm X}}{n}||_\infty\right\}\right). 
$$

Since $||d||_\infty = O_p(\sqrt{{\log(p)}/{n}})$ by Lemma \ref{lem:rate guXFbeta}, $||E^\T \widehat{\bm X}/n||_\infty = O_p(\sqrt{{\log(p)}/{n}}) $ by Lemma \ref{lem:tailboundforepsilon}, we have shown the desired result:
$$
||\widehat{\beta}-\dot{\beta}||_1=O_p\left(s(1+q)\sqrt{\frac{\log(p)}{n}}\right). \qquad  \square 
$$

\paragraph{Proof of the second part of Theorem \ref{thm:highd error rate}}
Recall from Assumption \ref{C4}, there exist constants $C_7, C_8 > 0$ such that $\underset{i \in \mathcal{A}}{\min} |\dot{\beta}_i| \geq n^{C_7 - 1/2}$ and $s^2(1 + q)^2\log{p} \leq n^{2C_7 - C_8}$. Considering $k = s$ and the event $\{\widehat{\mathcal{A}} \neq \mathcal{A}\} \subset \{||\widehat{\beta} - \dot{\beta}||_1 \geq n^{C_7 - 1/2}\}$. From Theorem \ref{thm:highd error rate} (a), we know that $||\widehat{\beta} - \dot{\beta}||_1 = O_p(s \sqrt{\log(p)/n})$, which means $\forall \delta > 0$, there exists a constant $M_\delta$ and an integer $n_1 > 0$ such that
$$
\sup_{n > n_1} \mathbb{P}\left(||\widehat{\beta} - \dot{\beta}||_1 \geq M_\delta s(q+1) \sqrt{\frac{\log(p)}{n}}\right) \leq \delta.
$$ 

For $n \geq n_2 := \max(n_1, M_\delta^{2/C_8})$, we have
\begin{align*}
n^{C_7 - 1/2} &= n^{C_7 - C_8/2} n^{-1/2} n^{C_8/2} \\
&\geq \{s^2(q+1)^2\log(p)\}^{1/2} n^{-1/2} M_\delta \\
&\geq M_\delta s(q+1) \sqrt{\log(p)/n},    
\end{align*}
which implies $\{\widehat{\mathcal{A}} \neq \mathcal{A}\} \subset \{||\widehat{\beta} - \dot{\beta}||_1 \geq n^{C_7 - 1/2}\} \subset \{||\widehat{\beta} - \dot{\beta}||_1 \geq M_\delta s(q+1) \sqrt{\log(p)/n}\}$.

We thus have 
\begin{equation*}
    \sup_{n \geq n_2} \mathbb{P}(\widehat{\mathcal{A}} \neq \mathcal{A}) \leq \sup_{n \geq n_2} \mathbb{P}\left(||\widehat{\beta} - \dot{\beta}||_1 \geq M_\delta s \sqrt{\frac{\log(p)}{n}}\right) \leq \sup_{n \geq n_1} \mathbb{P}\left(||\widehat{\beta} - \dot{\beta}||_1 \geq M_\delta s \sqrt{\frac{\log(p)}{n}}\right) \leq \delta.
\end{equation*}
Let $\delta \rightarrow 0$, we conclude $\mathbb{P}(\widehat{\mathcal{A}} \neq \mathcal{A}) \rightarrow 0$ as $n \rightarrow \infty$.

\subsection{Proof of Theorem \ref{thm:l0 plurality gmm}}
The proof is very similar to the one we provided for Theorem \ref{thm:l0 plurality}.

We are going to show that if there exists a vector $${\beta}^* \in \underset{{\widetilde\beta}\in \mathbb{R}^{p}}{\argmin} \;G(\widetilde\beta)\text{, such that }||{\widetilde\beta}||_0\leq p-q-1,$$ and ${\beta}^*\not= \dot{\beta}$, we will have a contradiction.

(i.) We first show that $\beta^* = \dot\beta +  \mathbb{E}^{-1}[ X \frac{\partial f}{\partial \beta^\T}|_{\beta = \widetilde \beta^*}]\Lambda \alpha^*$ for some $\alpha^* \in \mathbb{R}^q$ and $\widetilde \beta^*$ be a vector between $\beta^*$ and $\dot\beta$.

Since $G(\dot\beta) = 0$, we must have $G(\beta^*) = 0$. We have the following equation:
\begin{align*}
    G(\beta^*) &= ||\mathbb{E}[SIV\{Y-f(X;\beta^*)\}]||_2^2 \\&= ||\mathbb{E}[SIV\{f(X;\dot\beta)+g(U)+\epsilon_x-f(X;\beta^*)\}]||_2^2 \\
    &=||\mathbb{E}[SIV\{f(X;\dot\beta)-f(X;\beta^*)\}]||_2^2\\
    &=||\mathbb{E}[B_{\Lambda^\perp}^\T X \frac{\partial f}{\partial \beta^\T}(\dot\beta - \beta^*)\}]||_2^2\\
    &=||B_{\Lambda^\perp}^\T\mathbb{E}( X \frac{\partial f}{\partial \beta^\T}|_{\beta = \widetilde \beta^*})(\dot\beta - \beta^*)\}||_2^2,
\end{align*}
where $\widetilde \beta^*$ is a vector between $\beta^*$ and $\dot\beta$. The third equality holds because $SIV \ind g(U) + \epsilon_y$, and the last equality follows the mean value theorem. Since $0=G(\beta^*) = ||B_{\Lambda^\perp}^\T\mathbb{E}( X \frac{\partial f}{\partial \beta^\T}|_{\beta = \widetilde \beta^*})(\dot\beta - \beta^*)\}||_2^2$ and $\mathbb{E}(X\frac{\partial f}{\partial \beta^\T}|_{\beta = \widetilde \beta^*})$ is an invertible matrix by assumption \ref{D1}, we must have (i.)  $$ (i.) \;\;\beta^* = \dot\beta +  \mathbb{E}^{-1}( X \frac{\partial f}{\partial \beta^\T}|_{\beta = \widetilde \beta^*})\Lambda \alpha^*$$ for some $\alpha^* \in \mathbb{R}^q$.

Let $C :=\{ j\mid{\beta}_j^* = 0\}$, $M :=\{j\mid {\beta}_j^* \not= 0\}$. We have $|M|\leq p-q-1$ and $|C| \geq q+1$ by $||{\beta}^*||_0\leq p-q-1$.

(ii.) We next show the following inequality:
$|C\cap \mathcal{A}|\geq 2$.

Otherwise, we have $|C\cap \mathcal{A}| \leq 1$. Since $|C| \geq q+1$, we must have $|C\cap \mathcal{A}^c| \geq q$. Let $\widetilde{C}\subset C\cap \mathcal{A}^c$ such that $|\widetilde{C}|=q$. We consider an element inside $\widetilde{C}$:
\begin{equation*}
    \begin{split}
        0 = {\beta}^*_{\widetilde{C}} & = [\dot\beta +  \mathbb{E}^{-1}( X \frac{\partial f}{\partial \beta^\T}|_{\beta = \widetilde \beta^*})\Lambda \alpha^*]_{\widetilde{C}}\\
        & =\dot{\beta}_{\widetilde{C}}+[\mathbb{E}^{-1}( X \frac{\partial f}{\partial \beta^\T}|_{\beta = \widetilde \beta^*})\Lambda \alpha^*]_{\widetilde{C}}\\
        & = [\mathbb{E}^{-1}( X \frac{\partial f}{\partial \beta^\T}|_{\beta = \widetilde \beta^*})\Lambda \alpha^*]_{\widetilde{C}}.
    \end{split}
\end{equation*}
By the invertibility condition \ref{D1}, we hence have $\alpha^* = 0$, which implies ${\beta}^* = \dot{\beta} + \Sigma_X^{-1}\Lambda\alpha = \dot{\beta}$ and yields a contradition.

(iii.) Finally, we construct a contradiction.

We have $|C| \geq q+1$, and $|C\cap A|\geq 2$. Let $\{C^{(i)}\}^{q+1}_{i=1}$ such that $C^{(i)}\subsetneqq C$, $|C^{(i)}| = q$, $C^{(i)}\cap A \not= \emptyset$. Define ${\beta}^{(i)}$ as follow: 
 $$
    \beta^{(i)} = \underset{\widetilde\beta_{C^{(i)}}=0,\widetilde\beta \in \mathbb{R}^p }{\argmin}G(\widetilde\beta), i=1,2,\ldots,q+1.
$$

From Condition \ref{D2}, $\{\beta^{(i)}\}^{q+1}_{i=1}$ are uniquely defined. 
We observe that $\emptyset \subsetneqq C^{(i)}\subsetneqq C$, we have 
\begin{equation}
\label{eqn:betaitilebeta_thm4}
        \underset{\widetilde\beta \in \mathbb{R}^p }{\min}G(\widetilde \beta)\leq\underset{\widetilde\beta_{C^{(i)}}=0,\widetilde\beta \in \mathbb{R}^p }{\min}G(\widetilde \beta)\leq \underset{\widetilde\beta_{C}=0,\widetilde\beta \in \mathbb{R}^p }{\min}G(\widetilde \beta).
\end{equation}

Since ${\beta}^* \in \underset{\widetilde\beta \in \mathbb{R}^p }{\argmin}\;G(\widetilde\beta)$ and ${\beta}^* \in \underset{\widetilde\beta_{C}=0,\widetilde\beta \in \mathbb{R}^p }{\argmin}G(\widetilde\beta)$, 
 the equation \eqref{eqn:betaitilebeta_thm4} can be rewritten as 
$$
        \underset{\widetilde\beta \in \mathbb{R}^p }{\min}\;G(\widetilde \beta)=\underset{\widetilde\beta_{C^{(i)}}=0,\widetilde\beta \in \mathbb{R}^p }{\min}G(\widetilde \beta)= \underset{\widetilde\beta_{C}=0,\widetilde\beta \in \mathbb{R}^p }{\min}G(\widetilde \beta),
$$

which means $\beta^* \in \underset{\widetilde\beta_{C^{(i)}}=0,\widetilde\beta \in \mathbb{R}^p }{\argmin}G(\widetilde \beta)  = \{\beta^{(i)}\}$, where the last equality holds given Lemma \ref{lem:Expert has unique sol}. Now we get $\beta^* = \beta^{(i)} $ for $i \in \{ 1,2,\ldots,q+1\}$, which violates the Condition \ref{A4}.

This contradiction implies the set $\{\underset{{\widetilde\beta}\in \mathbb{R}^{p}}{\argmin} \;G(\widetilde \beta) s.t. ||{\widetilde\beta}||_0\leq p-q-1\} = \{\dot{\beta}\}$ if $p\geq q+s+1$, while  $ \{\underset{{\widetilde\beta}\in \mathbb{R}^{p}}{\argmin} \;G(\widetilde \beta)\; s.t.\; ||{\widetilde\beta}||_0\leq p-q-1\} = \emptyset$ if $p\leq q+s$, corresponding to the two cases in the Theorem \ref{thm:l0 plurality gmm}.

\section{Theoretical result for nonlinear outcome model}
\label{sec:appthmnonlinear}
We now provide a theoretical result for our estimator \eqref{eqn:nonlinear2SLS}. We focus on the low-dimensional setting where \( p \) is fixed, and leave the high-dimensional case for future investigation. 

We first clarify the notation and setting. Suppose we have \( p \) treatments and \( q \) latent confounders, and we observe \( n \) i.i.d. samples generated from models \eqref{eqn:model1} and \eqref{eqn:nonlinearmodel}, where the function \( f \) is unknown and the parameter \( \beta \) is to be estimated. Our goal is to investigate the properties of the estimator \( \widehat{\beta} \) obtained from \eqref{eqn:nonlinear2SLS}.

Let $\Sigma_X = \mathrm{Cov}(X)$ and $D = \mathrm{Cov}(\epsilon_x)$.  
Let \( \bm{X} \in \mathbb{R}^{n \times p} \) denote the design matrix, and let \( \bm{Y} = (Y_1, \ldots, Y_n)^\top \in \mathbb{R}^{n \times 1} \) be the response vector. Define
\[
f(\bm{X}; \beta) = \left(f(X_1; \beta), \ldots, f(X_n; \beta)\right)^\top \in \mathbb{R}^{n \times 1}
\]
as the vector of nonlinear responses, and let \( \bm{SIV} \in \mathbb{R}^{n \times (p-q)} \) be the matrix of synthetic instrumental variables. The projection matrix defined by \( \bm{SIV} \) is given by
\[
P_{\bm{SIV}} = \bm{SIV} \left(\bm{SIV}^\top \bm{SIV}\right)^{-1} \bm{SIV}^\top \in \mathbb{R}^{n \times n}.
\]

\subsection{Assumptions and discussion}
We make the following assumptions:

\begin{assumption}(Assumptions for nonlinear outcome models)
\label{ass:Assumptions for nonlinear }
\begin{enumerate}[label={E\arabic*}]

    \item \label{E1} The coefficients ${\Lambda}$ and the measurable functions $g(\cdot)$ and $f(\cdot;\beta)$ in models \eqref{eqn:model1} and \eqref{eqn:nonlinearmodel} are fixed and do not change as $n \to \infty$.
    
    \item \label{E2} $U_i$, $\epsilon_{x,i}$, and $\epsilon_{y,i}$ are independent random draws from the joint distribution of $(U,\epsilon_x,\epsilon_y)$ such that $E(\epsilon_x) = \bm{0}$, $E(U) = \bm{0}$, $\mathrm{Cov}(\epsilon_x) = D$, $\mathrm{Cov}(U) = I_q$, and $(U,\epsilon_x,\epsilon_y)$ are mutually independent. Furthermore, assume that $\mathrm{Var}({\epsilon}_y) = {\sigma}^2$ and $\max_{1 \leq j \leq p}\mathrm{Var}(X_j) = \sigma_x^2$; these parameters are fixed and do not change as $n \to \infty$.
    
    \item \label{E3} For the maximum likelihood estimator $\widehat{\Lambda}$, there exists an orthogonal matrix ${O} \in \mathbb{R}^{q \times q}$ such that $\| \widehat{\Lambda} - \Lambda O \|_2 = O_p(1/\sqrt{n})$.
    
    \item \label{E4} $\frac{\bm{X}^\T}{n}\frac{\partial f}{\partial \beta}\big|_{\beta =\bar\beta} \in \mathbb{R}^{p\times p}$ converges in probability to a matrix $M_{\bar\beta}$ uniformly in $\bar\beta$, and $\|M_{\bar\beta}\|_2$ is bounded from above for all $\bar\beta$.  
    
    \item \label{E5} Let $\bar{\Sigma} = M_{\bar{\beta}}^\T B_{\Lambda^{\perp}}(B_{\Lambda^{\perp}}^\T \Sigma_X B_{\Lambda^{\perp}} )^{-1}B_{\Lambda^{\perp}}^{\T}M_{\bar{\beta}}$.
    We assume
    \[
    \min_{\theta \in \mathbb{R}^p,\ 0 < \|\theta\|_0 \leq 2s}\frac{\theta^\T \bar\Sigma \theta}{\|\theta\|_2^2} > c
    \]
    for some positive constant $c$.
\end{enumerate}
\end{assumption}

Assumptions \ref{E1}--\ref{E3} and \ref{E5} are standard in low-dimensional settings and are similar to those made in Assumptions \ref{B1}--\ref{B4}. Assumption \ref{E4} is specifically required for nonlinear IV models \citep{amemiya1974nonlinear}. When \( f \) is linear, we have  
\[
\frac{\bm X^\T}{n} \left. \frac{\partial f}{\partial \beta} \right|_{\beta = \bar \beta} = \widehat{\Sigma}_X,
\]  
where \( \widehat{\Sigma}_X \) is the sample covariance matrix of \( \bm{X} \). In this case, it converges to the population covariance matrix as \( n \to \infty \).

\subsection{Theoretical results}
\begin{theorem}\label{thm:lowd thm nonlinear}
Under the conditions of Theorem \ref{thm:l0 plurality gmm} and Assumptions \ref{E1}--\ref{E5}, if the tuning parameter satisfies $\widehat{k} = s$, then the estimator $\widehat{\beta}$ obtained from \eqref{eqn:nonlinear2SLS} satisfies:

\begin{enumerate}
    \item [1.] ($\ell_1$-error rate) $\|\widehat{\beta} - \dot{\beta}\|_1 = O_p(n^{-1/2})$.
    \item [2.] (Variable selection consistency) Let $\mathcal{A} = \{j: \dot{\beta}_j \neq 0\}$ and $\widehat{\mathcal{A}} = \{j: \widehat{\beta}_j \neq 0\}$. Then $\mathbb{P}(\widehat{\mathcal{A}} = \mathcal{A}) \rightarrow 1$ as $n \rightarrow \infty$.
\end{enumerate}
\end{theorem}

\subsection{Lemmas and their proof}
\begin{lemma}[Convergence of a Key Matrix]\label{lem:nonlinear matrix convergency}
    Under Assumptions E1--E5, we have
    $$
    \left\|B_{\widehat{\Lambda}^{\perp}}\left(\frac{B_{\widehat{\Lambda}^{\perp}}^\T \bm X^\T \bm X B_{\widehat{\Lambda}^{\perp}}}{n}\right)^{-1}B_{\widehat{\Lambda}^{\perp}}^\T
    - B_{\Lambda^{\perp}}\left(B_{\Lambda^{\perp}}^\T D B_{\Lambda^\perp}\right)^{-1}B_{\Lambda^{\perp}}^\T \right\|_2 = O_p\left(\frac{1}{\sqrt{n}}\right).
    $$
\end{lemma}

\paragraph{Proof of Lemma \ref{lem:nonlinear matrix convergency}.}
To simplify notation, define
\[
M_1 = B_{\widehat{\Lambda}^{\perp}}\left(\frac{B_{\widehat{\Lambda}^{\perp}}^\T \bm X^\T \bm X B_{\widehat{\Lambda}^{\perp}}}{n}\right)^{-1}B_{\widehat{\Lambda}^{\perp}}^\T, \quad
M_2 = B_{\Lambda^{\perp}}\left(B_{\Lambda^{\perp}}^\T D B_{\Lambda^\perp}\right)^{-1}B_{\Lambda^{\perp}}^\T.
\]
Then,
\begin{equation}
    \begin{split}
        \|M_1 - M_2\|_2 
        &= \|\widehat{F} \widehat{D}^{-1} - F D^{-1} \|_2 \\
        &\leq \|(\widehat{F} - F)\widehat{D}^{-1}\|_2 + \|F(\widehat{D}^{-1} - D^{-1})\|_2 \\
        &\leq \|\widehat{F} - F\|_2 \cdot \|\widehat{D}^{-1}\|_2 + \|F\|_2 \cdot \|\widehat{D}^{-1}\|_2 \cdot \|\widehat{D} - D\|_2 \cdot \|D^{-1}\|_2 \\
        &= O_p\left(\frac{1}{\sqrt{n}}\right),
    \end{split}
\end{equation}
where \( F \) and \( \widehat{F} \) are defined in Lemma \ref{lem:order of hatF}, and their convergence is established therein. The final equality follows from the rate in Equation \ref{eqn:rateofD} together with Lemma \ref{lem:order of hatF}.

\begin{lemma}[Sparse Eigenvalue Condition, Nonlinear Setting]
\label{prop:REC_nonlinear}
Under Conditions \ref{E1}--\ref{E5}, there exists a constant \(\pi_0 > 0\) such that
\[
\liminf_n \mathbb{P}\left\{ \left\| P_{\mathrm{SIV}}\frac{\partial f}{\partial \beta}\big|_{\beta = \bar\beta}\theta \right\|_2 \geq \pi_0 \sqrt{n} \|\theta\|_2, \; \forall \theta \in \mathbb{R}^p, \|\theta\|_0 \leq 2s \right\} = 1.
\]
\end{lemma}

\paragraph{Proof of Lemma \ref{prop:REC_nonlinear}.}
The proof closely follows the argument used in Lemma \ref{prop:REC_lowd}.

Note that
\[
\frac{1}{n} \left( \frac{\partial f}{\partial \beta}\bigg|_{\beta = \bar\beta} \right)^\T P_{\mathrm{SIV}} \frac{\partial f}{\partial \beta}\bigg|_{\beta = \bar\beta}
= \left( \frac{1}{n} \left( \frac{\partial f}{\partial \beta} \bigg|_{\beta = \bar\beta} \right)^\T \bm X^\T \right) 
B_{\widehat{\Lambda}^{\perp}} \left( \frac{B_{\widehat{\Lambda}^{\perp}}^\T \bm X^\T \bm X B_{\widehat{\Lambda}^{\perp}}}{n} \right)^{-1} B_{\widehat{\Lambda}^{\perp}}^\T 
\left( \frac{1}{n} \bm X \frac{\partial f}{\partial \beta} \bigg|_{\beta = \bar\beta} \right).
\]

By Condition \ref{E4} and Lemma \ref{lem:nonlinear matrix convergency}, we have
\begin{equation}
\label{eqn:nonlinear,errorbound}
\left\| \frac{1}{n} \left( \frac{\partial f}{\partial \beta}\bigg|_{\beta = \bar\beta} \right)^\T P_{\mathrm{SIV}} \frac{\partial f}{\partial \beta}\bigg|_{\beta = \bar\beta} - \bar{\Sigma} \right\|_2 = O_p\left( \frac{1}{\sqrt{n}} \right).
\end{equation}

This implies that there exists a constant \( A_\delta > 0 \) and an integer \( n_1 \) such that
\[
\inf_{n > n_1} \mathbb{P}\left( \left\| \frac{1}{n} \left( \frac{\partial f}{\partial \beta}\bigg|_{\beta = \bar\beta} \right)^\T P_{\mathrm{SIV}} \frac{\partial f}{\partial \beta}\bigg|_{\beta = \bar\beta} - \bar{\Sigma} \right\|_2 \leq \frac{A_\delta}{\sqrt{n}} \right) \geq 1 - \delta.
\]

Define
\[
\pi_1 = \inf\left\{ \frac{\theta^\T \bar{\Sigma} \theta}{\|\theta\|_2^2} : \theta \in \mathbb{R}^p, \|\theta\|_0 \leq 2s \right\},
\]
which is strictly positive by Condition \ref{E5}. Set \(\pi_0 = \sqrt{\pi_1/2}\), \(n_2 = 4A_\delta^2 / \pi_1^2\), and let \(n_0 = \max(n_1, n_2)\). Then, with probability at least \(1 - \delta\), for all \(n > n_0\) and all \(\theta \in \mathbb{R}^p\) with \(\|\theta\|_0 \leq 2s\), we have
\[
\begin{split}
    \theta^\T \left( \frac{1}{n} \left( \frac{\partial f}{\partial \beta}\bigg|_{\beta = \bar\beta} \right)^\T P_{\mathrm{SIV}} \frac{\partial f}{\partial \beta}\bigg|_{\beta = \bar\beta} \right) \theta 
    &= \theta^\T \bar{\Sigma} \theta 
    + \theta^\T \left( \frac{1}{n} \left( \frac{\partial f}{\partial \beta}\bigg|_{\beta = \bar\beta} \right)^\T P_{\mathrm{SIV}} \frac{\partial f}{\partial \beta}\bigg|_{\beta = \bar\beta} - \bar{\Sigma} \right) \theta \\
    &\geq \|\theta\|_2^2 \left( \pi_1 - \left\| \frac{1}{n} \left( \frac{\partial f}{\partial \beta}\bigg|_{\beta = \bar\beta} \right)^\T P_{\mathrm{SIV}} \frac{\partial f}{\partial \beta}\bigg|_{\beta = \bar\beta} - \bar{\Sigma} \right\|_2 \right) \\
    &\geq \|\theta\|_2^2 \pi_0^2.
\end{split}
\]

This completes the proof.

\begin{lemma}
\label{lem:rightsidenonlinear}
    Under Conditions \ref{E1}--\ref{E4}, we have 
    \[
    \left\| \frac{1}{n}(\bm{Y} - f(\bm{X}; \dot\beta))^\T 
    P_{\bf SIV} \left. \frac{\partial f}{\partial \beta} \right|_{\beta = \bar\beta} \right\|_2 = O_p\left(\frac{1}{\sqrt{n}}\right),
    \]
    where \( \bar{\beta} \in \mathbb{R}^p \) is an element of the parameter space.
\end{lemma}

\paragraph{Proof of Lemma \ref{lem:rightsidenonlinear}.}
Consider the data-generating mechanism \( \bm{Y} = f(\bm{X}; \dot{\beta}) + g(\bm{U}) + E_y \). We decompose the target quantity as follows:
\begin{equation*}
\begin{split}
    \frac{1}{n}(\bm{Y} - f(\bm{X}; \dot{\beta}))^\T P_{\bf SIV} \left. \frac{\partial f}{\partial \beta} \right|_{\beta= \bar{\beta}} 
    &= \frac{(g(\bm{U}) + E_y)^\T \text{SIV}}{n} \left( \frac{\text{SIV}^\T \text{SIV}}{n} \right)^{-1} \frac{\text{SIV}^\T \left. \frac{\partial f}{\partial \beta} \right|_{\beta = \bar{\beta}}}{n} \\
    &= A B C,
\end{split}
\end{equation*}
where 
\[
A = \frac{(g(\bm{U}) + E_y)^\T \bm X}{n}, \quad 
B = B_{\widehat{\Lambda}^\perp}\left( \frac{\text{SIV}^\T \text{SIV}}{n} \right)^{-1} B_{\widehat{\Lambda}^\perp}^\T, \quad 
C = \frac{\bm{X}^\T \left. \frac{\partial f}{\partial \beta} \right|_{\beta = \bar{\beta}}}{n}.
\]
We now bound \( \|AB\|_2 \) and \( \|C\|_2 \) separately.

For the first term, using the definition \( \text{SIV} = \bm{X} B_{\widehat{\Lambda}^\perp} \), we have
\begin{equation}
\label{eqn:control_of_A}
\begin{split}
    \|AB\|_2 &= \left\| \frac{(g(\bm{U}) + E_y)^\T \bm{X} }{n} B \right\|_2 \\
    &= \left\| \left( \frac{(g(\bm{U}) + E_y)^\T \bm{X}}{n} - \text{Cov}(g(U), X) \right)B \right\|_2 
    + \left\| \text{Cov}(g(U), X)B \right\|_2 \\
    &= O_p\left(\frac{1}{\sqrt{n}}\right) \| B\|_2
    + \left\| \text{Cov}(g(U), U) \left( \Lambda^\T - O^\T \widehat{\Lambda}^\T \right)  B \right\|_2 \\
    &= O_p\left(\frac{1}{\sqrt{n}}\right) \| B\|_2 \\
    &= O_p\left(\frac{1}{\sqrt{n}}\right).
\end{split}
\end{equation}
The second line follows from decomposing the empirical covariance. The third line uses the identity 
\(
\text{Cov}(g(U), X) = \text{Cov}(g(U), U) \Lambda^\T
\)
and the orthogonality condition 
\(
\widehat{\Lambda}^\T B_{\widehat{\Lambda}^\perp} = 0.
\)
The fourth line follows from Condition~\ref{E3}, which ensures that \( \text{Cov}(g(U), U) \) is bounded and that \( B_{\widehat{\Lambda}^\perp} \) is an orthogonal matrix. The final line uses Lemma~\ref{lem:nonlinear matrix convergency}.

For the second term, we have
\begin{equation}
\label{eqn:control_of_C}
\begin{split}
    \|C\|_2 &= \left\| \frac{ \bm X^\T}{n} \left. \frac{\partial f}{\partial \beta} \right|_{\beta = \bar\beta} \right\|_2 \\
    &\leq \left\| \left( \frac{\bm X^\T}{n} \left. \frac{\partial f}{\partial \beta} \right|_{\beta = \bar\beta} - M_{\bar\beta} \right) \right\|_2 + \left\| M_{\bar\beta} \right\|_2 \\
    &= O_p(1),
\end{split}
\end{equation}
where the decomposition and bound follow directly from Condition~\ref{E4}.

Combining equations \eqref{eqn:control_of_A} and \eqref{eqn:control_of_C}, we conclude the proof of Lemma~\ref{lem:rightsidenonlinear}.

\subsection{Proof of Theorem}

\paragraph{Proof of the first part of Theorem \ref{thm:lowd thm nonlinear}.}

We focus on equation \eqref{eqn:nonlinear2SLS}. Due to the optimality of $\widehat{\beta}$ compared to $\dot{\beta}$, we have 
\begin{equation}\label{eqn:optimalitynonlinear}
\begin{split}
    &\quad \frac{\|P_{\bf SIV}(\bm Y -  f(\bm X;\widehat{\beta}))\|^2_2}{2n} \leq
    \frac{\|P_{\bf SIV}(\bm Y -  f(\bm X;\dot{\beta}))\|^2_2}{2n}\\
    \Longleftrightarrow&\quad \frac{\|P_{\bf SIV}(f(\bm X;\widehat{\beta}) - f(\bm X;\dot{\beta}))\|^2_2}{2n}   \leq\frac{(\bm Y-f(\bm X;\dot\beta))^
    \T P_{\bf SIV}(f(\bm X;\widehat{\beta}) - f(\bm X;\dot{\beta}))}{n}\\
    \Longleftrightarrow &\quad \frac{1}{2n}\big\|P_{\bf SIV} \left.\frac{\partial f}{\partial \beta}\right|_{\beta = \bar\beta}(\widehat\beta-\dot\beta) \big\|^2_2   \leq\frac{1}{n}(\bm Y-f(\bm X;\dot\beta))^
    \T P_{\bf SIV}\left.\frac{\partial f}{\partial \beta}\right|_{\beta = \bar\beta}(\widehat\beta-\dot\beta).
    \end{split}
\end{equation}
The last transformation follows from a Taylor expansion: there exists some \( \bar{\beta} \) between \( \widehat{\beta} \) and \( \dot{\beta} \) such that
\[
f(\bm{X}; \widehat\beta) - f(\bm{X}; \dot{\beta}) = \left.\frac{\partial f}{\partial \beta}\right|_{\beta = \bar{\beta}} (\widehat{\beta} - \dot{\beta}).
\]

Since $\|\widehat \beta - \dot{\beta}\|_0\leq 2s$, we have 
\[
\|\widehat \beta - \dot{\beta}\|_1 \leq \sqrt{2s}\,\|\widehat \beta - \dot{\beta}\|_2.
\]
By Lemma \ref{prop:REC_nonlinear}, there exists an integer $n_1$ such that 
\[
\frac{\|P_{\bf SIV}\left.\frac{\partial f}{\partial \beta}\right|_{\beta = \bar\beta}(\widehat\beta - \dot{\beta})\|^2_2}{2n} \geq \pi_0^2\|\widehat{\beta} - \dot{\beta}\|_2^2
\]
with probability at least $1-\delta/2$ for $n>n_1$. Substituting into \eqref{eqn:optimalitynonlinear} gives
\begin{equation}
\label{eqn:eqn:optimality beta hat lowd,2nonlinear}
\begin{split}
 \pi_0^2\|\widehat{\beta} - \dot{\beta}\|_2^2 &\leq \frac{\|P_{\bf SIV}\left.\frac{\partial f}{\partial \beta}\right|_{\beta = \bar\beta}(\widehat\beta - \dot{\beta})\|^2_2}{2n} \\ 
 &\leq \frac{1}{n}(\bm Y-f(\bm X;\dot\beta))^
    \T P_{\bf SIV}\left.\frac{\partial f}{\partial \beta}\right|_{\beta = \bar\beta}(\widehat\beta-\dot\beta) \\
&\leq \|\widehat \beta - \dot\beta\|_1 \left\|\frac{1}{n}(\bm Y-f(\bm X;\dot\beta))^
    \T P_{\bf SIV}\left.\frac{\partial f}{\partial \beta}\right|_{\beta = \bar\beta} \right\|_\infty\\
&\leq \sqrt{2s}\,\|\widehat \beta -\dot{\beta}\|_2 \left\|\frac{1}{n}(\bm Y-f(\bm X;\dot\beta))^
    \T P_{\bf SIV}\left.\frac{\partial f}{\partial \beta}\right|_{\beta = \bar\beta} \right\|_\infty.
\end{split}
\end{equation}

Canceling one factor of $\|\widehat \beta - \dot{\beta} \|_2$ in \eqref{eqn:eqn:optimality beta hat lowd,2nonlinear} and using Lemma \ref{lem:rightsidenonlinear}, we obtain
\[
  \left\|\frac{1}{n}(\bm Y-f(\bm X;\dot\beta))^     
    \T P_{\bf SIV}\left.\frac{\partial f}{\partial \beta}\right|_{\beta = \bar\beta} \right\|_\infty 
    = O_p\!\left(\frac{1}{\sqrt{n}}\right),
\]
which proves the first part of Theorem \ref{thm:lowd thm nonlinear}.

\paragraph{Proof of the second part of Theorem \ref{thm:lowd thm nonlinear}.}

Let $C_{16} = \min_{j \in \mathcal{A}} |\beta_j| \geq 0$. Considering $k = s$, the event $\{\widehat{\mathcal{A}} \neq \mathcal{A}\}$ is contained in $\{\|\widehat{\beta} - \dot{\beta}\|_1 \geq C_{16}\}$. From the first part of Theorem \ref{thm:lowd thm nonlinear}, we know that 
\[
\|\widehat{\beta} - \dot{\beta}\|_1 = O_p\!\left(\frac{1}{\sqrt{n}}\right),
\]
which means that for any $\delta > 0$, there exists a constant $M_\delta$ and an integer $n_1 > 0$ such that
\[
\sup_{n > n_1} \mathbb{P}\left(\|\widehat{\beta} - \dot{\beta}\|_1 \geq \frac{M_\delta}{\sqrt{n}}\right) \leq \delta.
\] 

For $n \geq n_2 := \max(n_1, M_\delta^2 / C_{16}^2)$, we have $C_{16} \geq M_\delta/\sqrt{n}$, so
\[
\{\widehat{\mathcal{A}} \neq \mathcal{A}\} \subset \{\|\widehat{\beta} - \dot{\beta}\|_1 \geq C_{16}\} \subset \{\|\widehat{\beta} - \dot{\beta}\|_1 \geq M_\delta/\sqrt{n}\}.
\]

Thus,
\[
    \sup_{n \geq n_2} \mathbb{P}(\widehat{\mathcal{A}} \neq \mathcal{A}) 
    \leq \sup_{n \geq n_2} \mathbb{P}\!\left(\|\widehat{\beta} - \dot{\beta}\|_1 \geq \frac{M_\delta}{\sqrt{n}}\right) 
    \leq \sup_{n \geq n_1} \mathbb{P}\!\left(\|\widehat{\beta} - \dot{\beta}\|_1 \geq \frac{M_\delta}{\sqrt{n}}\right) 
    \leq \delta.
\]
Since $\delta$ is arbitrary, we conclude that $\mathbb{P}(\widehat{\mathcal{A}} \neq \mathcal{A}) \to 0$ as $n \to \infty$.

\section{Results of comparison methods in the real data example}
\label{sec:comparison}

We include the genes identified by various comparison methods in the mouse obesity dataset described in Section \ref{sec:data}.

The Lasso method identifies the following genes: Igfbp2, Ankhd1, Rab27a, Dct, Gck, Tex15, Wfdc15b, Rab6b, Avpr1a, Abca8a, F12, Arx, Gna14, Vwf, C4b, Zar1, Taf7, B4galnt4, Upk3a, Tiam2, Pex11a, Mmp1b, Cd36, Bglap-rs1, Prdm16, Olfr378, G6pc, Ccnl2, Ccnb1, Clstn3, Smok3a, Meox1, Fras1, Gstm2, Cfd, Gpx6, Efemp1, Osbpl6, Dok2, Plcl2, Cebpe, Plxnb1, Myl10, Tmem174, Insl6, Ifitm7, PqlC2, Oas1e, Itgad, Gldc, Rxfp1, Pgf, Adh7, Msr1, Vil1, Cyp26a1, Zfp30, Ggta1, Fanca, Xpo4, Doxl2, Sall2, Gprc6a, Pet2, Otop2, Epb4.2, BC029214, Frem1, Dcx, Xcl1, Olfr1033, Sntg2, Copz2, Angpt2, Il13, Dnase1l3, Olfr1501, Xdh, Rbm3, Il5ra, Galns, Nme2, Fbxo16, Egr2, Dhrs7b, Lpar2, and Npm3. 

The 2SR method \citep{lin2015regularization_sup} identifies the following genes: Igfbp2, Lamc1, Sirpa, Gstm2, Ccnl2, Glcci1, Vwf, Irx, Apoa4, Socs2, Avpr1a, Abca8a, Gpld1, Fam105a, Dscam, Slc22a3, and 2010002N04Rik.

The Auxiliary Variable method \citep{miao2023identifying_sup} identifies the following genes: Gstm2, 2010002N04Rik, Igfbp2, and Avpr1a.

The Null Variable method \citep{miao2023identifying_sup} identifies the following genes: Gstm2 and Dscam.

The Trim method \citep{cevid2020spectral_supp} identifies the following genes: Igfbp2, Ankhd1, Rab27a, and Dct. 

The IV-Lasso method identifies the following genes: Igfbp2, Rab27a, Ankhd1, Hao2, Dct, Fras1, Gck, Tex15, Nox4, Insl6, Vwf, Txk, Padi2, and Gstm2. 

\section{Additional simulation results}
\label{sec:additional_simulation}

\subsection{Comparison between simulation and real data features}
\label{sec:comparison_sim&data}

We define the signal-to-noise ratio (SNR) as
\[
\text{SNR} = \frac{\text{Var}(X \beta)}{\text{Var}(Y - X\beta)}.
\]

If $\beta$ is unknown, we estimate the signal-to-noise ratio from finite samples as
\[
\widehat{\text{SNR}} = \frac{\widehat{\text{Var}}_n (X \widehat{\beta})}{\widehat{\text{Var}}_n(Y - X \widehat{\beta})}.
\]

Table \ref{tab:appandsim} presents a comparison between the application data and the simulated data, where $\sigma_y = 5$ and $\text{Var}(\epsilon_y) = \sigma_y^2$. Sparsity is represented by the norms $||\beta||_0$ for the simulation and $||\widehat{\beta}||_0$ for the data. The ``Number of Confounders'' refers to $q$ in the simulation and $\widehat{q}$ in the data. The definition of the signal-to-noise ratio (SNR) is provided above. In the simulation, the reported SNR is the average over 1,000 Monte Carlo replications with different seeds to account for the randomness of $\gamma$ and $\Lambda$. As shown in the table, sparsity, the number of confounders, and SNR exhibit strong similarities between the two settings.

\begin{table}[!htbp]
\begin{center}
\caption{Comparison between application and simulated data. Sparsity is indicated by the norms \( ||\dot{\beta}||_0 \) for the simulation and \( ||\widehat{\beta}||_0 \) for the data. ``Number of Confounders'' refers to \( q \) in the simulation and \( \widehat{q} \) in the data. The SNR denotes the signal-to-noise ratio.}
\label{tab:appandsim}
\begin{tabular}{lll}
\toprule
                          & Simulation & Application \\ \hline
Sparsity                  &  5          &    5         \\ \hline
Number of confounders     &  3          & 3            \\ \hline
SNR                       &  $0.965$    & $0.969$      \\ \hline
\end{tabular}
\end{center}
\end{table}

\subsection{Simulation for weak effects}
\label{sec:simuweakeffect}
\begin{figure}[!htbp]
    \centering{
    \begin{subfigure}{0.49\textwidth}
    \includegraphics[scale = 0.15]{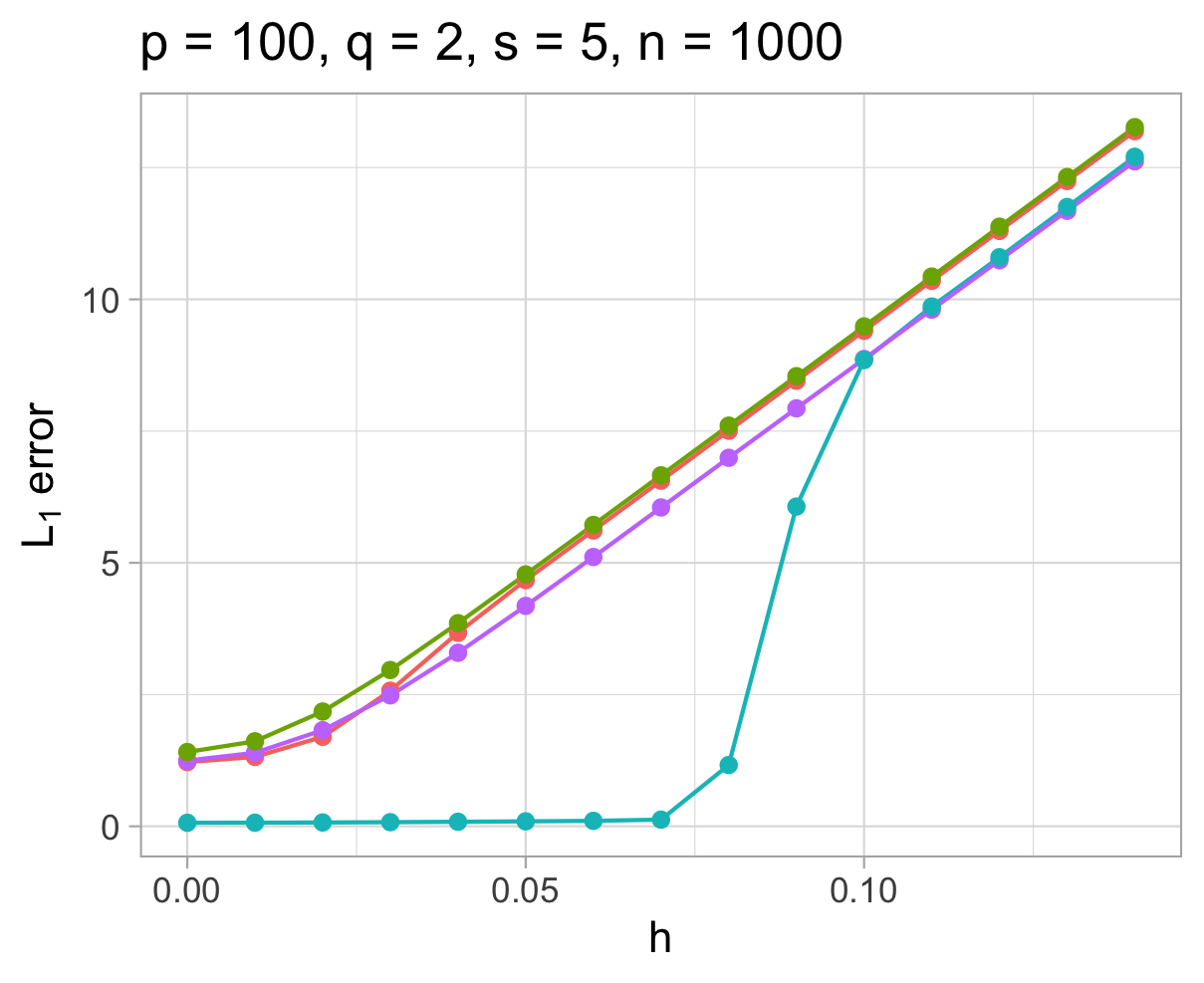}
    \centering
    \caption{Estimation errors $||\widehat{\beta} - \beta^\#||$. }
  \end{subfigure}}
  \hfill
  \centering{
  \begin{subfigure}{0.49\textwidth}
  \centering
    \includegraphics[scale = 0.15]{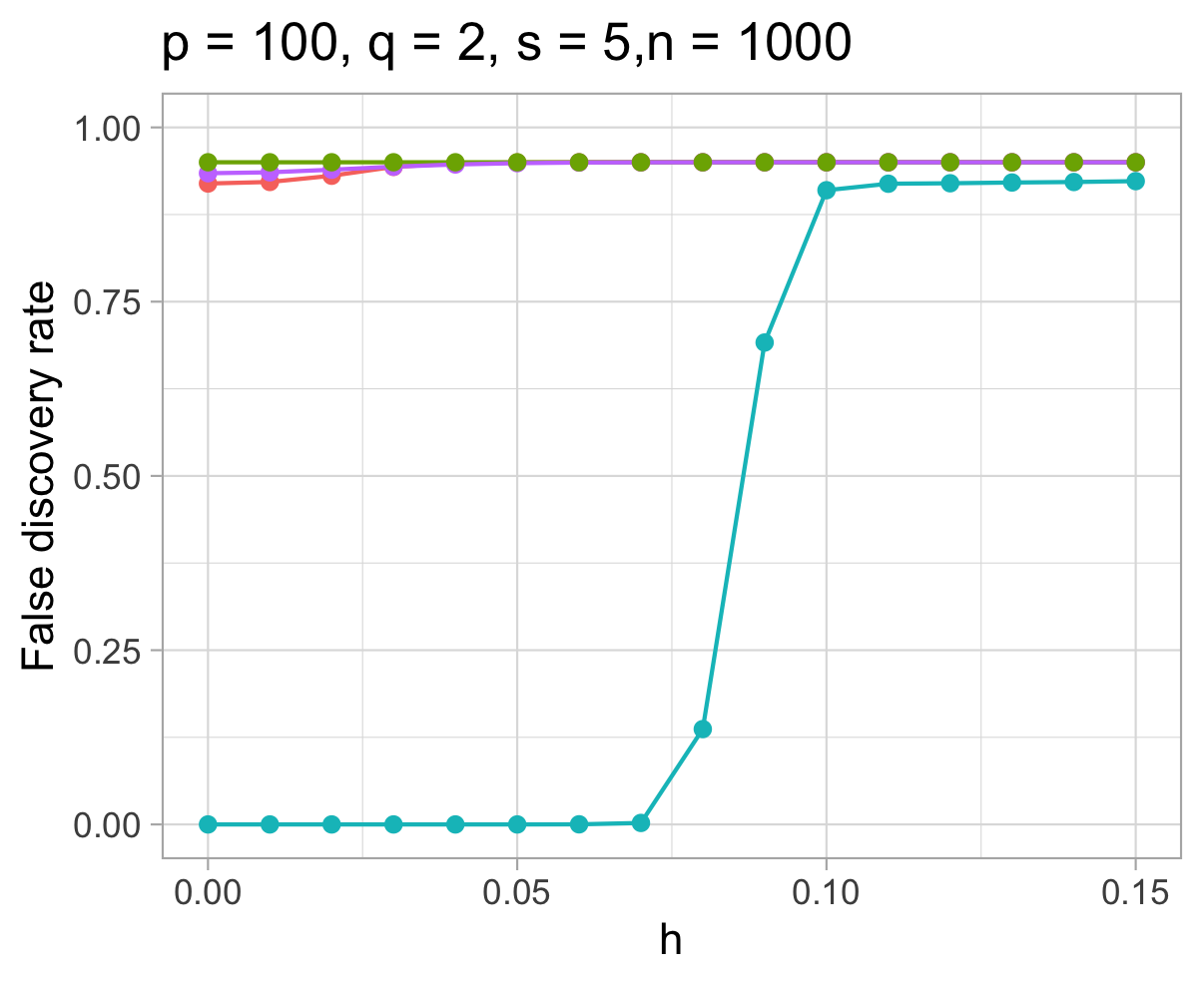}
    \caption{False discovery rate. }
  \end{subfigure}
  \hspace*{\fill}
  }
    \caption{Simulation results for SIV (blue), Lasso (red), Trim (purple), and Null (green), based on 1,000 Monte Carlo runs.}
    \label{fig:weakeffects}
\end{figure}

We present simulation results for dense confounding with many weak effects. In our simulations, we set \( n = 1000 \), \( p = 100 \), \( q = 2 \), \( \beta_1 = \beta_2 = \ldots = \beta_5 = 1 \), and \( \beta_6 = \beta_7 = \ldots = \beta_p = h \), where \( h \) varies from 0 to 0.15. The other parameters and variables are generated as in the simulation setting described in Section \ref{sec:simulation} of the main paper. Since the true causal parameter \( \beta \) is not identifiable in this setting, we report the \( \ell_1 \)-difference between \( \widehat{\beta} \) and \( \beta^\# \) for each method, where \( \beta^\#_1 = \beta^\#_2 = \ldots = \beta^\#_5 = 1 \) and \( \beta^\#_6 = \beta^\#_7 = \ldots = \beta^\#_p = 0 \). The simulation results are presented in Figure \ref{fig:weakeffects}.

The findings in Figure \ref{fig:weakeffects} highlight the impact of weak effects. Specifically, when weak effects are present but their magnitude is very small, our method performs comparably to its performance in sparse settings, showing superior accuracy relative to alternatives. However, as the magnitude of the weak effects increases and $h$ exceeds a certain threshold, their influence becomes dominant, and all methods converge to similar performance.

\subsection{Comparison with the moment selection estimator}
\label{sec:MS}

One reviewer suggested that we could apply the algorithm in \citet{andrews1999consistent_sup} to estimate $\beta$. We discuss this method in this section.

\subsection*{Comparison between our procedure and the procedure in \cite{andrews1999consistent_sup}}

\citet{andrews1999consistent_sup} focus on the selection of true moment conditions. Suppose there are \(r\) moment conditions, of which \(r_0\) are correct, and assume that the number of parameters, denoted by \(p\), is less than \(r_0\). They introduce a method that identifies the \(r_0\) correct moment conditions, based on which the true parameters can be consistently estimated. Their results are applied to the selection of invalid instrumental variables and to addressing the over-identification problem.

In contrast, our method focuses on a different task: selecting true causal variables. In our proposal, there are \(p-q\) instrumental variables, which provide \(p-q\) moment conditions, and importantly, \textbf{all of these are valid}. Thus, we work with exactly \(p-q\) correct moment conditions. Since there are \(p\) parameters to identify, our situation corresponds to ``under-identification.'' However, identification and estimation become feasible once sparsity constraints are imposed on the treatment effects. To guarantee unique identification, the number of nonzero parameters must be fewer than the number of instrumental variables, that is, \(s := \|\beta\|_0 < p-q\).

In the following, we first review the method of \citet{andrews1999consistent_sup} in the classical IV setting and discuss the pitfalls of extending their approach to our context. We then present simulations comparing their method against ours. The results show that our proposed estimator outperforms theirs in the scenario we evaluated.

\subsubsection*{Review of \cite{andrews1999consistent_sup}'s method in the classical IV setting}

Consider the classical IV setting with one treatment \(X\) and three instruments $Z := (Z^{(1)}, Z^{(2)}, Z^{(3)})$. These instruments yield three moment conditions:
\begin{equation*}
    \begin{split}
        g_1(\beta) &= \mathbb{E}\{(Y-X\beta)Z^{(1)}\},\\
        g_2(\beta) &= \mathbb{E}\{(Y-X\beta)Z^{(2)}\},\\
        g_3(\beta) &= \mathbb{E}\{(Y-X\beta)Z^{(3)}\}.
    \end{split}
\end{equation*}

If \(Z^{(i)}\) is a valid IV, then \(g_i(\dot{\beta}) = 0\) at the true value \(\dot{\beta}\). In finite samples, empirical averages replace expectations, and the generalized method of moments (GMM) is used:
\begin{equation*}
   \widehat{\beta} = \argmin_{\beta \in \mathbb{R}} (g_1, g_2, g_3)^\T W (g_1, g_2, g_3),
\end{equation*}
where \(W \in \mathbb{R}^{3 \times 3}\) is a weight matrix (assume \(W = I_3\) for simplicity). Let \(g(\beta) = (g_1(\beta), g_2(\beta), g_3(\beta))^\T\). For a subset \(A \subset \{1,2,3\}\), denote by \(g^A(\beta)\) the moment conditions using only indices in \(A\), and define \(\widehat{\beta}^A = \argmin_{\beta \in \mathbb{R}} (g^A(\beta))^\T g^A(\beta)\).

The moment selection estimator selects the ``correct" moment constraints \(\widehat{A}\) by solving
$$
\widehat{A} = \argmin_{A \subset \{1,2,3\}} n(g^A(\widehat{\beta}^A))^\T g^A(\widehat{\beta}^A) - h(|A|)k_n,
$$
where \(|A|\) is the cardinality of \(A\), \(h(\cdot)\) is a strictly increasing function, and \(k_n \to \infty\) with \(k_n = o(n)\). The penalty term \(h(|A|)k_n\) rewards the use of more moment conditions.

\subsubsection*{Potential pitfalls of the extension}

\begin{enumerate}
    \item \textbf{Coherence issue:} The moment selection estimator becomes substantially more complex when \(q \geq 2\), as it requires accounting for logical dependencies among moment constraints. For example, with two latent confounders and index sets \(A=\{1,2\}\), \(B=\{2,3\}\), and \(C=\{1,3\}\), we can construct moment constraints \(g_A\), \(g_B\), and \(g_C\). If both \(g_A\) and \(g_B\) are accepted, then \(g_C\) must also be valid since \(C \subset A \cup B\). Such coherence relationships complicate the selection process dramatically as \(q\) grows.

    \item \textbf{Computational issue:} In high-dimensional settings (\(p > n\)), directly solving the moment selection estimator is infeasible, as it requires computing GMM estimators for all possible subsets of moment conditions. To our knowledge, no efficient algorithm addresses this in high dimensions. In contrast, our estimator can be implemented efficiently using the ``abess'' package.
\end{enumerate}

\subsubsection*{Simulation results}

We compared our estimator with the moment selection estimator in a simple setting with three treatments and one unmeasured confounder (\(p=3, q=1\)). In this case, coherence issues do not arise, and computation is feasible. Even here, our estimator outperforms the moment selection estimator.

Specifically, consider the structural model:
\begin{equation*}
    \begin{split}
        X &= \Lambda U + \epsilon_x,\\
        Y &= X^\T \beta + U^\T \gamma + \epsilon_y,
    \end{split}
\end{equation*}
with \(X = (X_1, X_2, X_3)\), one confounder \(U\), and parameters \(\Lambda = (1, -1, 2)^\T\), \(\gamma = 1\), and \(\beta = (1,0,0)^\T\). The errors satisfy \(U_i \sim \mathbb{N}(0,1)\), \(\epsilon_{y,i} \sim \mathbb{N}(0,1)\), and \(\epsilon_{x,i} \sim \mathbb{N}(0, I_3)\). We ran simulations with \(n \in \{500, 1000, 1500, \ldots, 5000\}\). 

\begin{figure}[!htbp]
    \centering
    \includegraphics[width = 7cm]{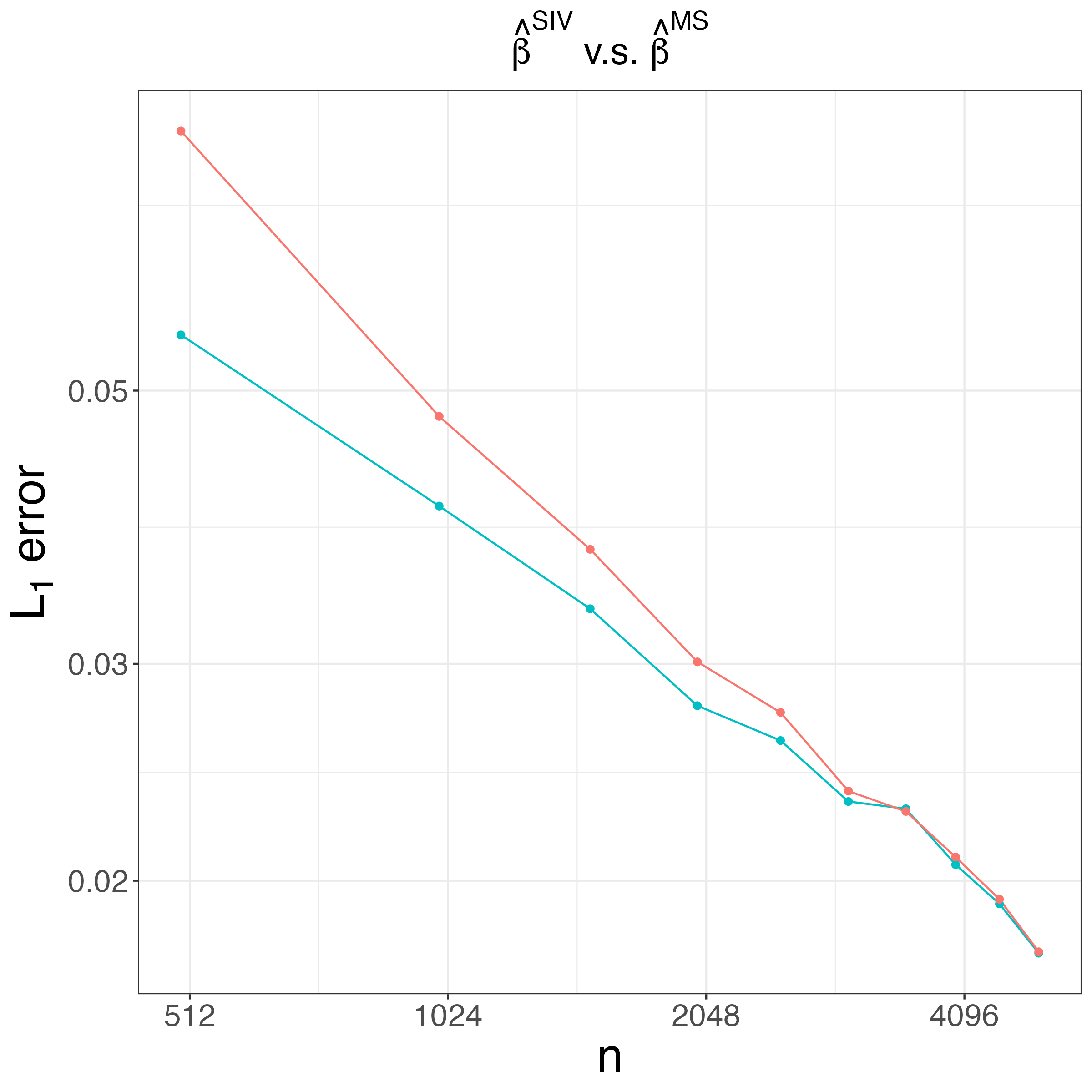}
    \caption{Comparison between the SIV estimator (blue line) and the moment selection estimator (red line) \citep{andrews1999consistent_sup}, based on 1000 Monte Carlo simulations.}
    \label{fig:MS estimator_sup}
\end{figure}

The results, shown in Figure \ref{fig:MS estimator_sup}, report the $\ell_1$ error \(||\widehat{\beta} - \dot{\beta}||_1\). Our estimator consistently outperforms the moment selection estimator of \citet{andrews1999consistent_sup}.

\subsection{Details of Simulation Settings for Nondiagonal $\text{Cov}(\epsilon_x)$}
\label{sec:detail:nondiagonal}
In the simulation setup for nondiagonal $\text{Cov}(\epsilon_x)$, we randomly selected 20 pairs from $i, j \in \{1, 2, \ldots, p\}$ and assigned $D_{i,j} = D_{j,i} = 1$. The list of these pairs is provided below.

 $( 5, 87)$, 
 $(14,38 )$,
 $(15,85)$,
 $(25,   50)$,
$(32,   46)$,
$(37,   75)$,
$(44,   37)$,
$(45,   10)$,
$(52,   33)$,
$(52,   37)$,
$(60,   92)$,
$(66,   88)$,
$(66,  100)$,
$(73,   55)$,
$(74,   34)$,
$(86,   77)$,
$(87,   31)$,
$(89,   53)$,
$(91,   82)$, and 
$( 97,   96)$.

\subsection{An Alternative Cross-Validation Strategy for the IV-Lasso Estimator}
\label{sec:ivlasso1se}

As discussed in Section~\ref{sec:simu_linear}, the IV-Lasso estimator performs suboptimally in our simulation setting. This is primarily because standard cross-validation tends to select overly complex models, with the Lasso estimator often including more variables than necessary. To address this issue, we consider the one-standard-error (1-se) rule \citep{hastie2009elements,kang2016instrumental}, which selects the most regularized model whose cross-validation error lies within one standard error of the minimum. This approach favours simpler models that perform comparably to the best model identified by standard cross-validation.

To evaluate the potential benefit of the 1-se rule, we conduct a series of simulation studies. The outcome model is $f(X;\beta) = X^\T\beta$, and the hidden variable model is $g(U) = U^\T\gamma$. We set $q = 3$, $s = 5$, and define the true coefficient vector as $\beta = (1,1,1,1,1,0,\ldots,0)^\T \in \mathbb{R}^p$. The elements of both $\Lambda_{j,k}$ and $\gamma_k$ are independently drawn from the uniform distribution on $[-1,1]$ for $j = 1,\ldots,p$ and $k = 1,\ldots,q$. The latent variables $U_{i,k}$ are generated independently from the standard normal distribution for $i = 1,\ldots,n$ and $k = 1,\ldots,q$. The noise terms are generated as $\epsilon_x \sim \mathcal{N}(0, \sigma^2_x I_p)$ and $\epsilon_y \sim \mathcal{N}(0, \sigma^2)$, with $\sigma_x = 2$ and $\sigma = 1$.

We assess estimator performance under two regimes: (i) low-dimensional, with $p = 100$ and $n \in \{200,600,1000,\ldots,5000\}$; and (ii) high-dimensional, with $n = 500$ and $p \in \{500,750,1000,\ldots,3000\}$. All results are averaged over 1000 Monte Carlo replications.

We compare the following estimators:
\begin{itemize}
    \item[\textbf{SIV:}] The original sparse IV estimator defined in~\eqref{L0 optimization}.
    \item[\textbf{IV-Lasso:}] The IV-Lasso estimator (Section~\ref{sec:simu_linear}), with tuning selected by standard cross-validation.
    \item[\textbf{IV-Lasso-1SE:}] The IV-Lasso estimator, with tuning selected by the 1-se rule.
\end{itemize}

\begin{figure}[ht]
 \centering
 \begin{subfigure}{0.45\textwidth}
     \includegraphics[width=\textwidth]{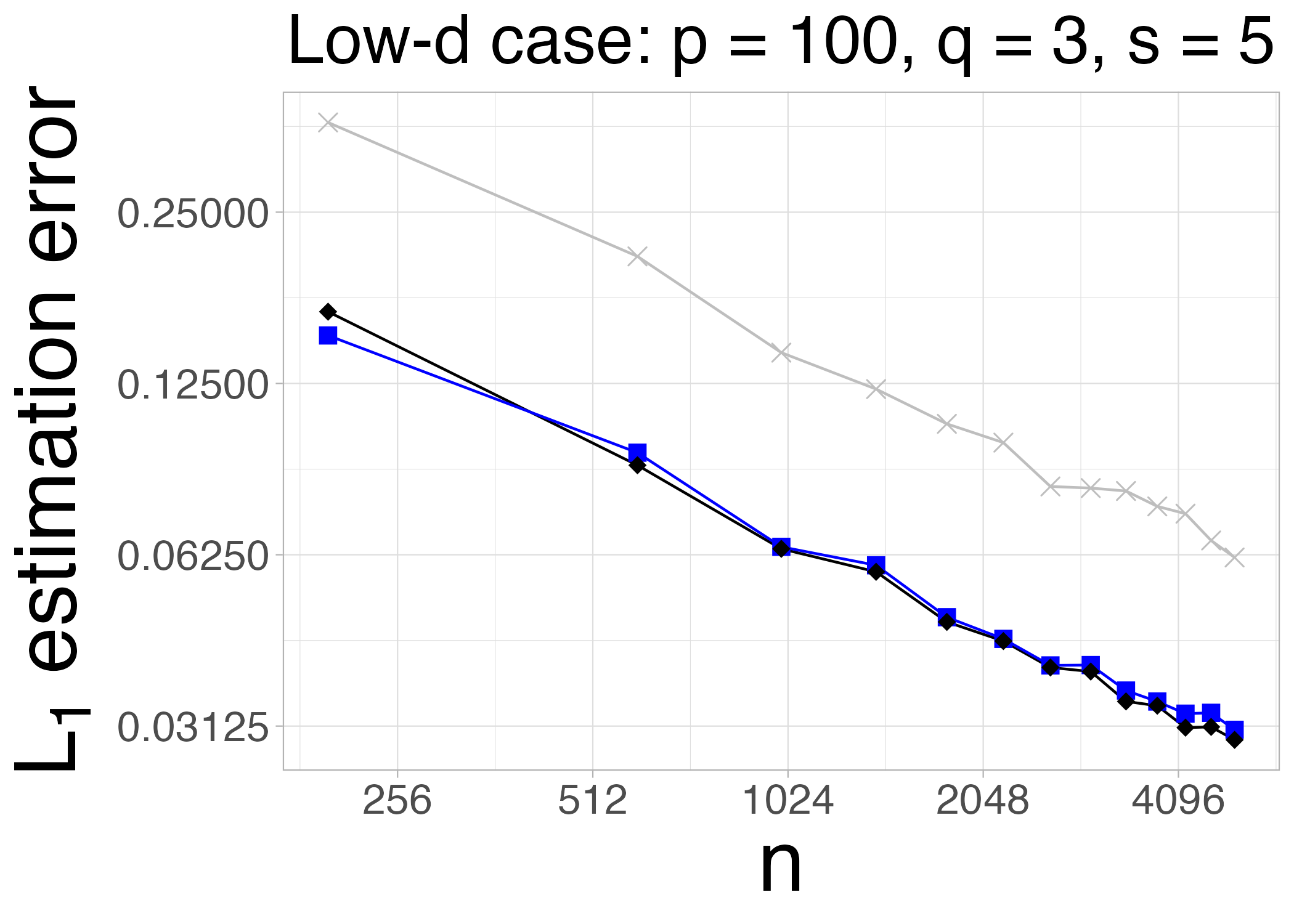}
     \caption{Low-dimensional case: $p = 100$, $n$ varies from $200$ to $5000$.}
     \label{fig:1sea_sup}
 \end{subfigure}
 \hfill
 \begin{subfigure}{0.45\textwidth}
     \includegraphics[width=\textwidth]{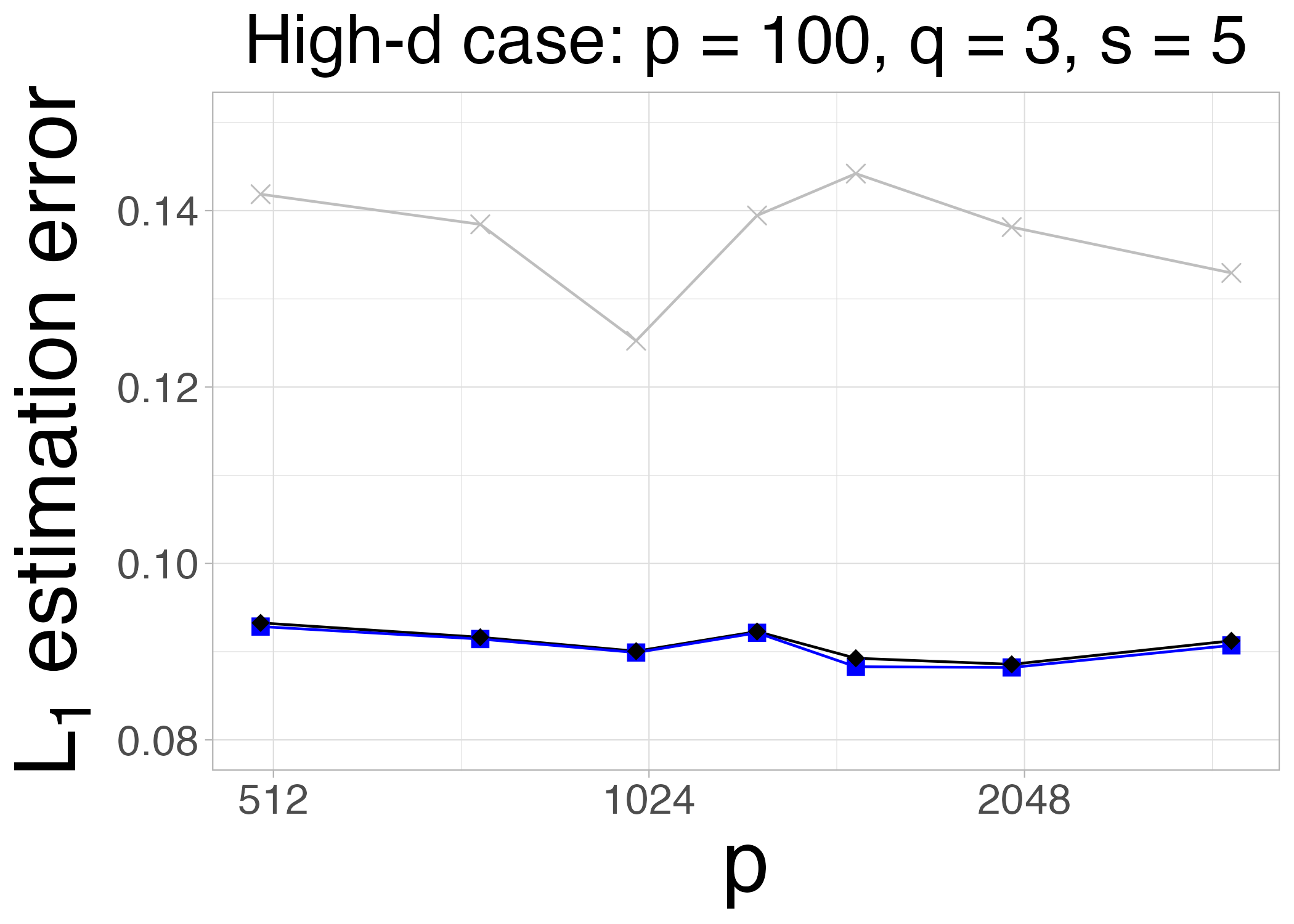}
     \caption{High-dimensional case: $n = 500$, $p$ varies from $500$ to $3000$.}
     \label{fig:1sec_sup}
 \end{subfigure}
 \caption{$L_1$ estimation errors of SIV (\textcolor{blue}{$\blacksquare$}), IV-Lasso (\textcolor{gray}{$\times$}), and IV-Lasso-1SE (\textcolor{black}{$\blacklozenge$}), based on 1000 Monte Carlo replications.}
 \label{fig:1seestimation_sup}
\end{figure}

Figure~\ref{fig:1seestimation_sup} reports the $L_1$ estimation errors across both regimes. The original IV-Lasso method performs worse than the SIV estimator. In contrast, IV-Lasso-1SE, by applying the 1-se rule, achieves estimation accuracy comparable to that of SIV in both low- and high-dimensional settings.

\subsection{Simulation Results for Statistical Inference}
\label{sec:CI_sup}

\begin{figure}[ht]
 \centering
 \begin{subfigure}{0.45\textwidth}
     \includegraphics[width=\textwidth]{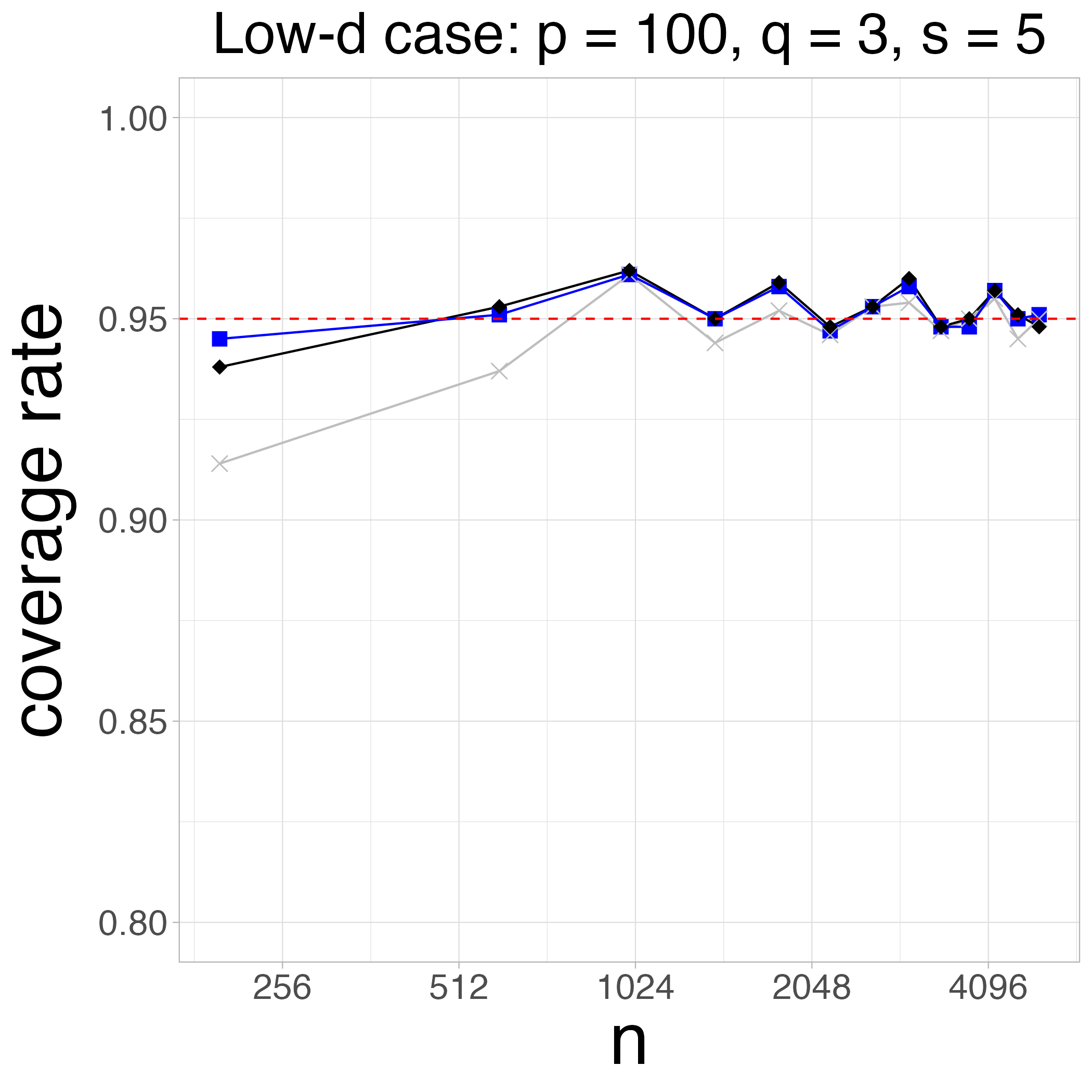}
     \caption{Low-dimensional case: $p = 100$, $n$ varies from $200$ to $5000$.}
     \label{fig:1sea_inference_sup}
 \end{subfigure}
 \hfill
 \begin{subfigure}{0.45\textwidth}
     \includegraphics[width=\textwidth]{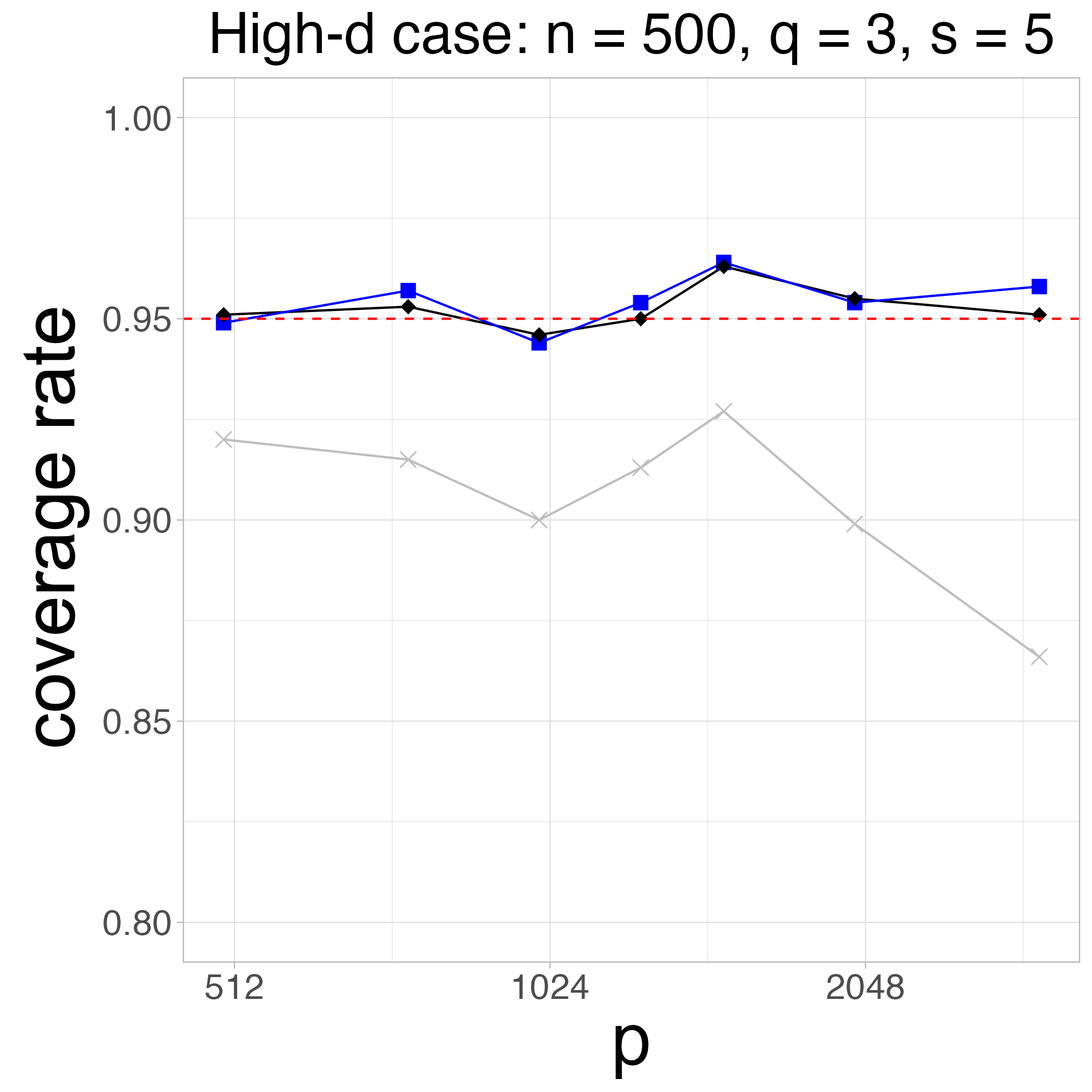}
     \caption{High-dimensional case: $n = 500$, $p$ varies from $500$ to $3000$.}
     \label{fig:1seb_inference_sup}
 \end{subfigure}
 
 \caption{Inference results for SIV ($\blacksquare$, blue), IV-Lasso ($\times$, grey), and IV-Lasso-1SE ($\blacklozenge$, black), based on 1000 Monte Carlo runs.}
 \label{fig:1se_inference_sup}
\end{figure}

We include additional simulation results to evaluate the performance of statistical inference procedures. Specifically, under the original setting described in Section~\ref{sec:simu_linear}, we assess the empirical coverage of confidence intervals for $\beta_1$. To this end, we apply various methods to select the set of causal variables, denoted by $\widehat{A}$, and construct 95\% confidence intervals using the \texttt{ivreg} function.  Specifically, to construct a confidence interval for $\beta_1$, we use \texttt{ivreg} with the outcome specified as $Y$, the treatment as $X_{\{1\} \cup \widehat{A}}$ (where $\widehat{A}$ denotes the set of causal variables selected by a given algorithm, such as SIV, IV-Lasso with cross-validation, or IV-Lasso-1se), and the instrument as the constructed synthetic instrument $SIV$.  and obtain $95\%$ confidence interval for $\beta_1$ The simulation results are summarised in Figure~\ref{fig:1se_inference_sup}.

Our findings indicate that both the IV-Lasso-1SE and SIV methods yield reasonably accurate inference results. In contrast, the original IV-Lasso method performs poorly in the high-dimensional setting, primarily due to inconsistent variable selection when the tuning parameter is chosen via cross-validation.

\subsection{SIV Method for Count Data}
\label{sec:countdata}

We extend the SIV method to accommodate count data models \citep{mullahy1997instrumental}. The cited work considers a Poisson regression model with unmeasured confounders, where \( Y \in \{0,1,2,\ldots\} \) and
\[
\mathbb{E}(Y \mid X, U) = \exp(X^\top\beta + g(U)).
\]
Because the response is count-valued, a direct logarithmic transformation is not applicable. If the confounder \( U \) were observed and \( g(U) \) were linear, then \( \beta \) could be consistently estimated via standard Poisson regression using the \texttt{glm} function with a log link. However, if \( U \) is unobserved, \citet{mullahy1997instrumental} propose using an instrumental variable \( Z \), in which case the following moment condition holds:
\[
\mathbb{E}\left\{ \frac{Y}{\exp(X^\top \beta)} - 1 \mid Z \right\} = 0,
\]
provided that \( \beta \) equals the true parameter value.

In the absence of observed instruments, a synthetic instrument can be constructed, allowing us to proceed analogously to Equation~\eqref{eqn:nonlinear2SLS} in the main manuscript. Specifically, we consider the optimization problem:
\begin{equation}
\label{eqn:nonlinear2SLS_possion_sup}
    \argmin_{\beta \in \mathbb{R}^p} \left\| \bm{SIV}({\bm{SIV}^\top \bm{SIV}})^{-1} \bm{SIV}^\top 
 \left\{\frac{\bm Y}{\exp (\bm X \beta)} - 1\right\} \right\|_2^2 \quad \text{subject to } \|\beta\|_0 \leq k,
\end{equation}
where \( \frac{\bm Y}{\exp(\bm X \beta)} - 1 \) is an \( n \times 1 \) vector with the \( i \)th element given by \( Y_i / \exp(X_i^\top \beta) - 1 \). Note that Equation~\eqref{eqn:nonlinear2SLS} is not directly applicable in the Poisson setting, since the residual \( Y - \exp(X^\top \beta) \) remains dependent on the instrument under the Poisson data-generating process.

To illustrate the effectiveness of Equation~\eqref{eqn:nonlinear2SLS_possion_sup}, we conduct a simulation study under a confounded Poisson regression framework. We set \( q = 2 \), \( s = 2 \), and \( p = 10 \). The treatment model is given by \( X = \Lambda U + \epsilon_x \), and the outcome \( Y \) is generated from a Poisson distribution:
\[
Y_i \sim \text{Poisson}(\lambda_i), \quad \lambda_i = \exp(X_i^\top \beta + U_i^\top \gamma),
\]
where \( \beta = (0.3, 0.3, 0, 0, \ldots, 0)^\top \in \mathbb{R}^{10} \). Each element of \( \Lambda_{j,k} \) and \( \gamma_k \) is independently drawn from \( \mathcal{N}(0,1) \) for \( j = 1, \ldots, p \) and \( k = 1, \ldots, q \). The latent variables \( U_{i,k} \) are i.i.d.\ standard normal, and the noise terms are generated as \( \epsilon_x \sim \mathcal{N}(0, \sigma_x^2 I_p) \) and \( \epsilon_y \sim \mathcal{N}(0, \sigma^2) \), with \( \sigma_x = 2 \) and \( \sigma = 1 \).

\begin{figure}[ht]
    \centering
    \includegraphics[width=0.5\linewidth]{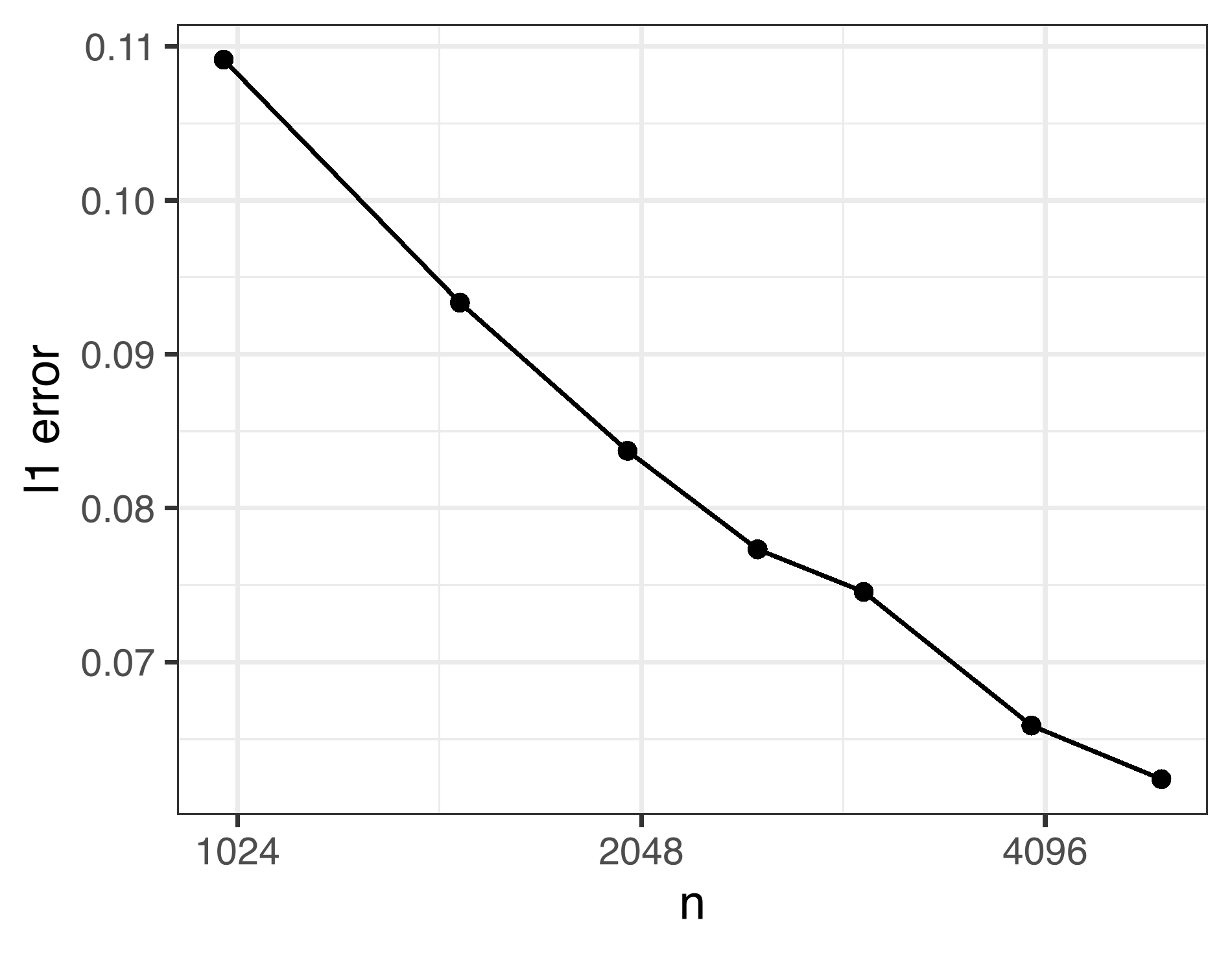}
    \caption{SIV method for confounded Poisson regression.}
    \label{fig:poisson_reg_sup}
\end{figure}

We assess estimation performance for \( n \in \{1000, 2000, \ldots, 5000\} \). All results are based on 1,000 Monte Carlo replications. The \( \ell_1 \) estimation errors are reported in Figure~\ref{fig:poisson_reg_sup}. The results indicate that the proposed algorithm performs well in the confounded Poisson regression setting.

\subsection{Additional Discussion of the Trim Method}
\label{sec:trim_sup}

\subsubsection{An update on the implementation of the Trim method}
\label{sec:update_supp}
Before discussing the performance of the Trim method, we note an update in the implementation code, which was adapted from \url{https://github.com/zijguo/Doubly-Debiased-Lasso/blob/main/R/utils.R}. On line 33 of their code, the coefficient is extracted as  
\[
\texttt{{betahat} = \text{as.matrix}\big(\text{coef}(fit, S = fit\$lambda.min)[-1]\big),}
\]  
where \texttt{fit} is the \texttt{cv.glmnet} object. However, the argument should be written with a lowercase \texttt{s} rather than an uppercase \texttt{S}. When specified as \texttt{S}, {\tt R} treats the argument \texttt{s} as missing and defaults to \texttt{s = "lambda.1se"} within the \texttt{cv.glmnet} object.  

In our revised implementation, we corrected this line to  
\[
\texttt{betahat = as.matrix(coef(fit, s = fit\$lambda.min)[-1])},
\]  
consistent with the description in \cite{cevid2020spectral}: ``In all simulations, unless stated otherwise, the penalty level is chosen by cross-validation.'' Accordingly, we updated the simulation results in Section~\ref{sec:simu_linear} of the manuscript. This correction leads to a higher observed false discovery rate for the Trim method.

\subsubsection*{The Trim method}

For the \emph{Trim method}, the reasons for its poor performance differ between low- and high-dimensional settings:  
\begin{itemize}
    \item \textbf{Low-dimensional setting:} The Trim method is inconsistent because its consistency requires a stringent assumption, namely $\|b\|_2 = O(1/\sqrt{n})$ \citep[Remark 5]{cevid2020spectral}, where $b$ denotes the bias from unmeasured confounding variables. This condition typically holds in high-dimensional settings but not in fixed-dimensional (low-dimensional) cases, where $\|b\|_2$ remains constant.

  \item \textbf{High-dimensional setting:} The SIV method consistently outperforms the Trim method, mainly due to two factors:
\begin{itemize}
    \item \textbf{Improved variable selection} – As shown in Figure~\ref{fig:fixnsparsity} of the manuscript, the SIV estimator identifies causal variables more accurately.  
    \item \textbf{Reduced shrinkage bias} – The Trim method employs $\ell_1$-penalization, which induces shrinkage and biases estimates toward zero. In contrast, the SIV method uses $\ell_0$-optimization, which mitigates shrinkage and better preserves signal strength.
\end{itemize}
\end{itemize}

Motivated by your Comment 2, we note that the variable selection with the \emph{Trim} estimator may also be improved by replacing cross-validation with the one-standard-error (1se) rule. In addition, the penalization effect can be mitigated by refitting. To illustrate these improvements, we consider three variants of the Trim estimator:  

\begin{itemize}
    \item [Trim:] The original Trim transformation with the tuning parameter selected by cross-validation.
    \item [Trim-1se:] The Trim transformation with the tuning parameter $\lambda$ selected using the 1se rule.
    \item [Trim-1se-refit:] The Trim-1se estimator with an additional refitting step on the selected variables: 
    \begin{equation}
    \begin{split}
        &\widehat{\mathcal{A}} = \{j : \widehat{\beta}_j^{\;\text{Trim-1se}} \neq 0,\;\; \widehat{\beta}^{\;\text{Trim-1se}} \text{ is the Trim estimator with the 1se rule}\}, \\
        &\widehat{\beta}^{\;\text{Trim-1se-refit}} := \argmin_{\beta \in \mathbb{R}^p,\; \beta_{\widehat{\mathcal{A}}^c}=0} \;\; \|F_{\text{Trim}}(Y - X\beta)\|_2^2,
    \end{split}
    \end{equation}
    where $F_{\text{Trim}} \in \mathbb{R}^{n \times n}$ is the trimming transformation defined in \citet{cevid2020spectral}.
\end{itemize}

\begin{figure}[!ht]
    \centering
    \begin{subfigure}[b]{0.45\textwidth}
        \centering
        \includegraphics[width=\textwidth]{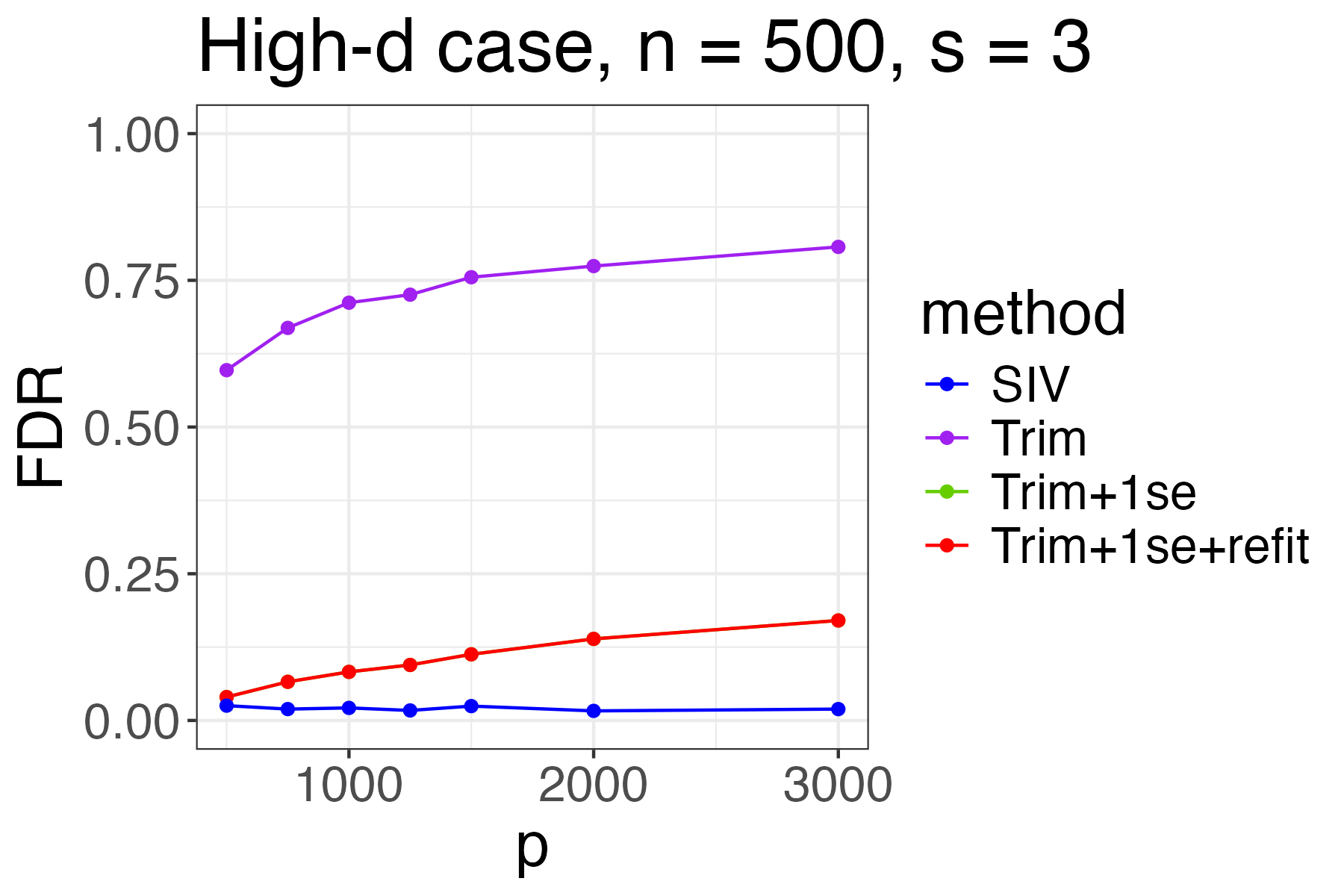} 
        \caption{$n=500$, $p$ varies from 500 to 3000.}
        \label{fig:sub1_supp}
    \end{subfigure}
    \hfill
    \begin{subfigure}[b]{0.45\textwidth}
        \centering
        \includegraphics[width=\textwidth]{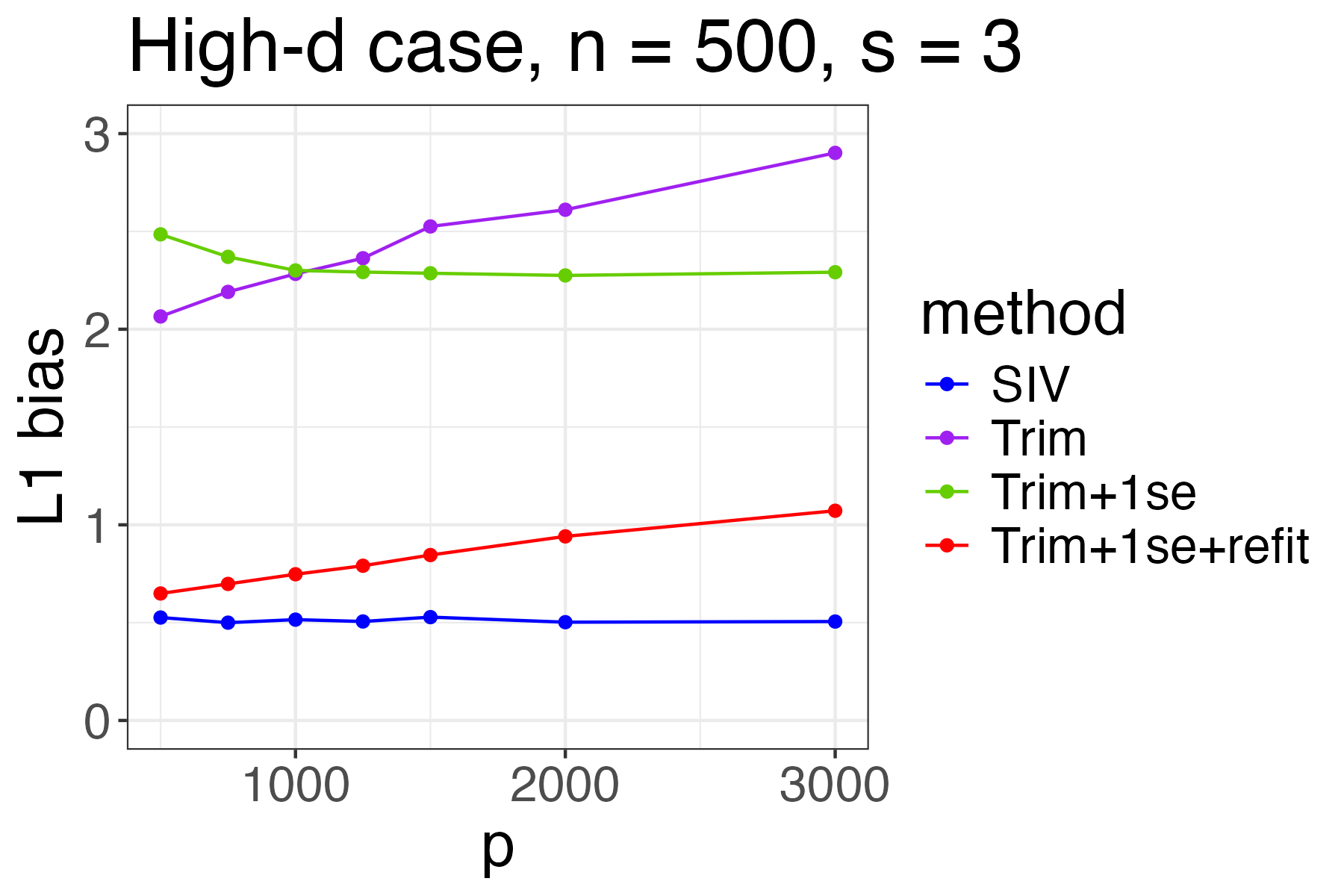} 
        \caption{$n=500$, $p$ varies from 500 to 3000.}
        \label{fig:sub2_supp}
    \end{subfigure}
    \caption{FDR and estimation results comparing the SIV method with various Trim methods. The false discovery rates of Trim-1se and Trim-1se-refit coincide because Trim-1se-refit uses the same subset of variables for refitting.}
    \label{fig:sivandtrim_supp}
\end{figure}

We evaluated the performance of the SIV, Trim, Trim-1se, and Trim-1se-refit estimators in the high-dimensional setting described in Section~\ref{sec:simu_linear} of the manuscript. Figure~\ref{fig:sivandtrim_supp} summarizes the simulation results. Figure~\ref{fig:sub1_supp} reports the false discovery rates (FDRs). The original Trim estimator exhibits a high FDR, which is substantially reduced when combined with the 1se rule. Figure~\ref{fig:sub2_supp} presents the $\ell_1$-estimation errors. The Trim-1se-refit estimator performs significantly better than the original version, demonstrating that its performance can be empirically improved through the 1se rule and refitting. Nevertheless, the proposed SIV method remains the most favorable among all estimators considered.



\subsubsection*{Why our method still performs better than Trim+1se+refit?}

 We believe that the superior performance of our estimator stems from the fact that the SIV method achieves identification of the causal parameter $\beta$ in the population sense, whereas the Trim transformation does not correspond to any identifiable population target. 

To illustrate this distinction, we examine several ``oracle'' variants of the SIV and Trim estimators, assuming that the active set $\mathcal{A} := \{j : \beta_j \neq 0\}$ is known.

\begin{itemize}
    \item[\textbf{SIV-oracle}] 
    Let $\widehat{X} = \widehat{\mathbb{E}}(X \mid SIV)$. 
    The SIV-oracle estimator is defined as
    \[
    \widehat{\beta} := \argmin_{\beta \in \mathbb{R}^p,\; \beta_{\mathcal{A}^c}=0} \|Y - \widehat{X}\beta\|_2^2.
    \]
    This estimator corresponds to the oracle version of the SIV method.

    \item[\textbf{Trim-oracle}]  
    Following \citet{cevid2020spectral}, let $\widetilde{X} := F_{\text{Trim}}X$ and $\widetilde{Y} := F_{\text{Trim}}Y$,  
    where $F_{\text{Trim}}X$ caps all singular values of $X$ that exceed the median singular value at that threshold, while leaving smaller singular values unchanged. 
    The Trim-oracle estimator is defined as
    \[
    \widehat{\beta} := \argmin_{\beta \in \mathbb{R}^p,\; \beta_{\mathcal{A}^c}=0} \|\widetilde{Y} - \widetilde{X}\beta\|_2^2.
    \]
    This estimator corresponds to the oracle version of the Trim method.

    \item[\textbf{Trim-oracle-top-$q$}]  
    We further introduce a new variant that directly targets the directions of unmeasured confounding. 
    Specifically, as discussed in \citet{cevid2020spectral}, the top $q$ singular values of $X$ correspond to the $q$ unmeasured confounders, while the remaining singular values capture the signal of the causal variables. 
    Intuitively, setting the top $q$ singular values to zero removes the influence of unmeasured confounders while retaining the signal from the causal variables. 
    Formally, let $\widetilde{X} := F_{\text{Trim},q}X$ and $\widetilde{Y} := F_{\text{Trim},q}Y$,  
    where $F_{\text{Trim},q}X$ sets the top $q$ singular values of $X$ to zero and leaves the remaining singular values unchanged. 
    The Trim-oracle-top-$q$ estimator is defined as
    \[
    \widehat{\beta} := \argmin_{\beta \in \mathbb{R}^p,\; \beta_{\mathcal{A}^c}=0} \|\widetilde{Y} - \widetilde{X}\beta\|_2^2.
    \]
\end{itemize}

We focus on the high-dimensional setting described in Section~\ref{sec:simu_linear}, with a minor adjustment to the outcome model:
\[
Y_i = X_i \beta + U_i \gamma + \epsilon_{y,i}.
\]
Here, we set $\epsilon_{y,i} = 0$ to isolate the impact of unmeasured confounders, excluding the influence of independent noise. 
The confounding parameters are specified as $\gamma_1 = \cdots = \gamma_q = \sqrt{p/500}$, so that the strength of confounding grows slightly with the number of treatments $X$. 
Let $b$ denote the bias for $\beta$ introduced by the unmeasured confounder $U$. 
We can show that the bias for the oracle variable satisfies 
\(
\|b_\mathcal{A}\|_2^2 = o\!\left(\tfrac{1}{p}\right)
\)
under the scenario considered here. 
All other aspects of the data-generating mechanism remain the same as in Section~\ref{sec:simu_linear}.

\begin{figure}[!ht]
    \centering
    \includegraphics[width=0.5\linewidth]{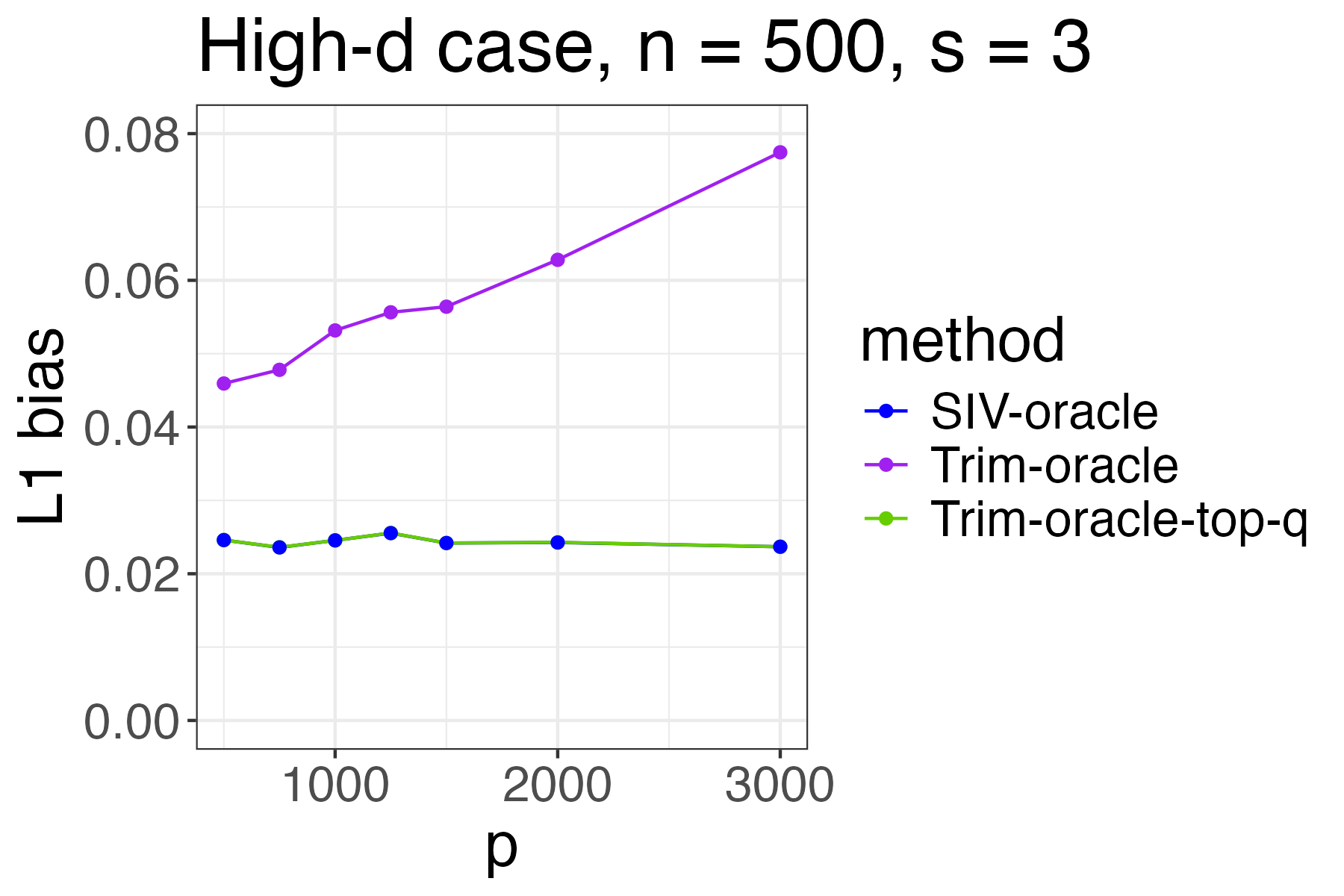}
    \caption{Estimation results comparing various oracle estimators. 
    The SIV-oracle and Trim-oracle-top-$q$ estimators perform nearly identically, so their lines overlap.}
    \label{fig:oracle_supp}
\end{figure}

Figure~\ref{fig:oracle_supp} summarizes the simulation results for the oracle estimators. 
As shown, when $\mathcal{A}$, the set of causal variables, is known a priori, the SIV estimator better recovers the true causal relationship \(Y \sim X_{\mathcal{A}}\) compared to the Trim transformation. 
The comparison between Trim-oracle-top-$q$ and SIV-oracle demonstrates that our estimator is equivalent to removing the top $q$ singular values of $X$, corresponding to the unmeasured confounders. 
This simulation also suggests that the Trim estimator’s performance can be improved by modifying its singular value adjustment strategy, as implemented in Trim-oracle-top-$q$. These results reinforce our interpretation that the SIV method's superior performance arises from its population-level identification of the causal parameter, rather than from arbitrary spectral regularization.

\subsection{Simulation results for nonlinear outcome models with nondiagonal $\text{Cov}(\epsilon_x)$}
\label{sec:nonlinear,nondigonal}

We provide additional simulation results to evaluate the performance of the proposed estimator in \eqref{eqn:nonlinear2SLS} under nonlinear outcome models with nondiagonal covariance structures for $\text{Cov}(\epsilon_x)$. Notably, the GMM procedure does not require $\text{Cov}(\epsilon_x)$ to be diagonal. Even when $\text{Cov}(\epsilon_x)$ is nondiagonal, the moment condition $SIV \ind Y - f(X; \beta)$ continues to hold, allowing valid application of the GMM framework for estimating the nonlinear causal function.

When $\text{Cov}(\epsilon_x)$ is nondiagonal, it is necessary to estimate the latent factor loading matrix $\Lambda$ using alternative methods. In low-dimensional settings where $\text{Cov}(\epsilon_x)$ is assumed sparse, we apply the stable principal component pursuit approach \citep{zhou2010stable}. For high-dimensional scenarios, the POET estimator \citep{fan2013large} provides a viable alternative.

In this simulation, we induce a nondiagonal covariance structure by setting $D_{i,j} = D_{j,i} = 1$ for four selected pairs $(i, j) \in \{(2,4),(5,6),(5,9),(6,10)\}$, and $D_{i,i} = 4$ for $i = 1, \dots, 10$. All other aspects of the data-generating mechanism remain unchanged. The low-rank structure $\widehat{\Lambda} \widehat{\Lambda}^\top$ is estimated via stable principal component pursuit, from which we recover $\widehat{\Lambda}$. We then implement the SIV method from \eqref{eqn:nonlinear2SLS}, along with the U-hat1 and U-hat2 methods described in Section~\ref{sec:sim_nonlinear}.

\begin{figure}[ht]
    \centering
    \begin{subfigure}{0.49\textwidth}
        \includegraphics[scale=0.4]{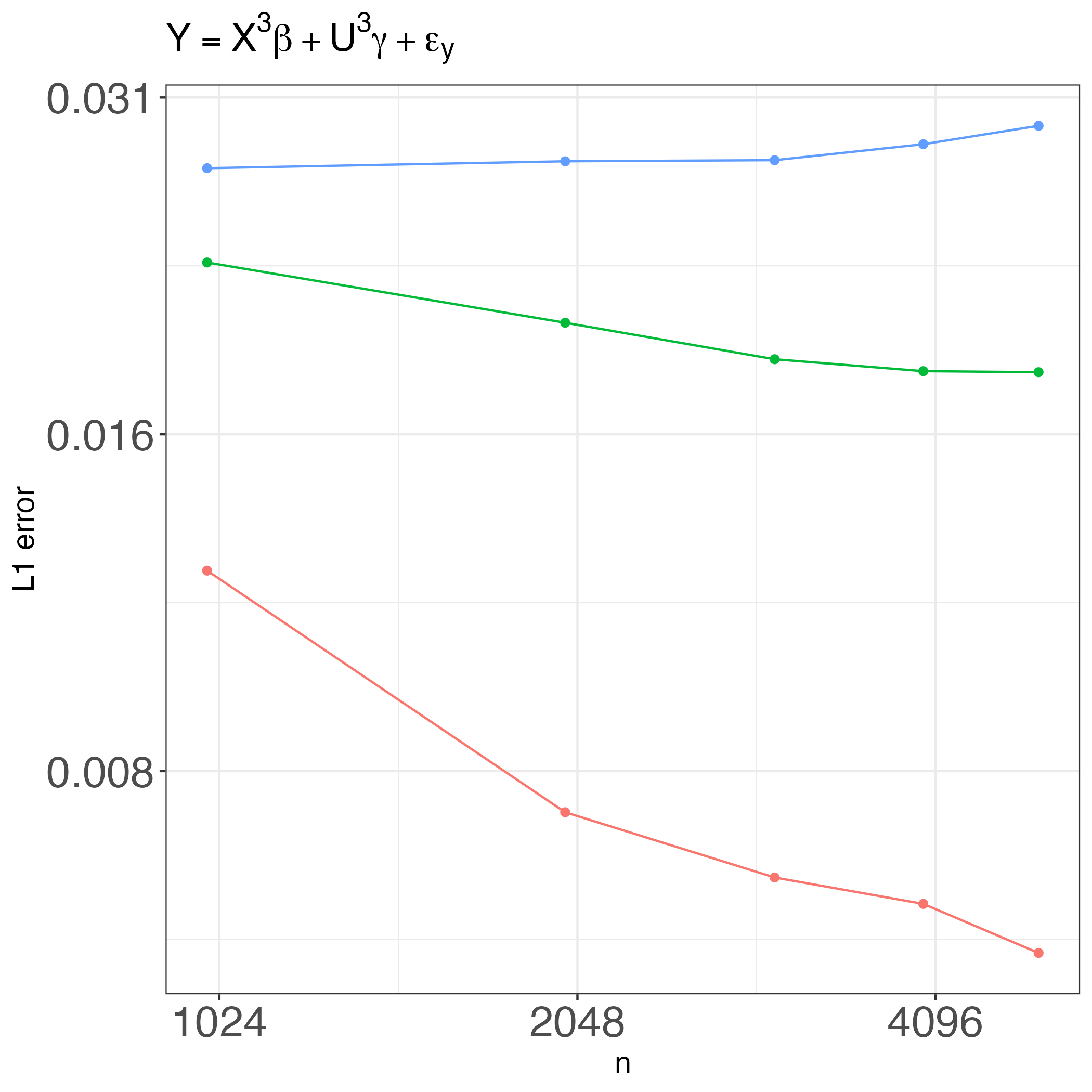}
        \caption{Nonlinear setting 1.}
    \end{subfigure}
    \hfill
    \begin{subfigure}{0.49\textwidth}
        \includegraphics[scale=0.4]{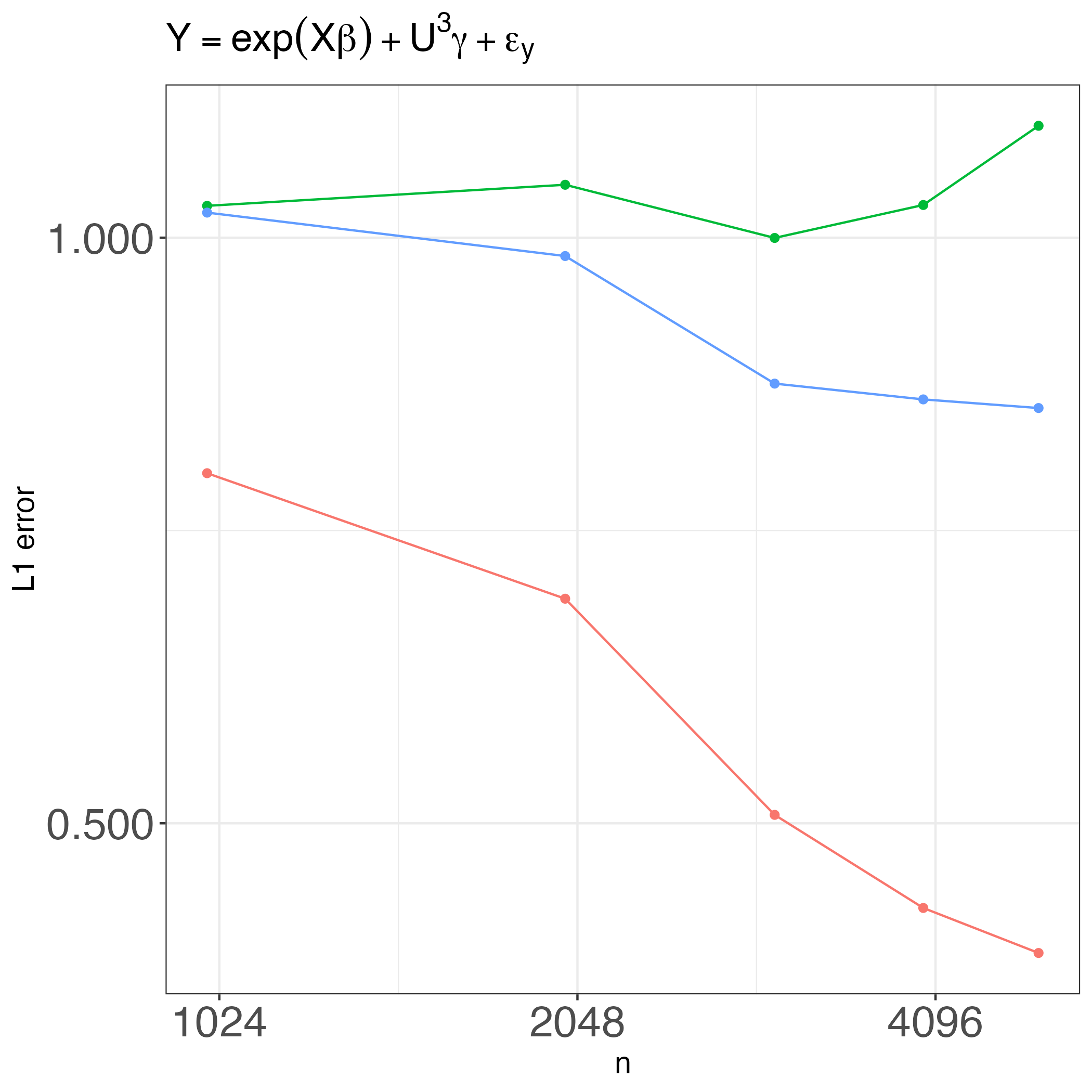}
        \caption{Nonlinear setting 2.}
    \end{subfigure}
    \caption{Simulation results for nonlinear models with $p = 10$ and $n = 1000, 2000, \dots, 5000$. Methods shown: SIV (red), U-hat1 (green), U-hat2 (blue).}
    \label{fig:nonlinear_response}
\end{figure}

Figure~\ref{fig:nonlinear_response} presents the results. Across both nonlinear settings, only the SIV method yields consistent estimates of $\beta$. In contrast, U-hat1 and U-hat2 exhibit substantial bias, particularly under the exponential outcome model, where their $\ell_1$ errors remain large even as the sample size increases.

\section{Further discussions on the U-hat1 Method}
\label{sec:Uhat}
In Section \ref{sec:sim_nonlinear} of our manuscript, we considered the so-called U-hat1 method as comparison procedures in our simulation study. Recall that the U-hat1 method for the linear outcome model ($Y = X^\T \beta + U^\T\gamma + \epsilon_y$) proceeds as follows:
\begin{itemize}
\item[(1)] Estimate \( U \) by \(\widehat{U} = X\widehat{\gamma}\), where \(\widehat{\gamma} = \widehat{\Sigma}_X^{-1}\widehat{\Lambda}\).
\item[(2)] Run the regression \( Y \sim X + \widehat{U} \) subject to the constraint \( ||\beta||_0 \leq k \), where $k$ is a tuning parameter.
\end{itemize}

In the linear outcome model, the U-hat1 method coincides with our proposed SIV method. However, as shown in Section \ref{sec:sim_nonlinear}, under more general and realistic nonlinear models, our approach enables both the identification and estimation of \( f \), whereas the U-hat1 method does not. In what follows, we explain why the U-hat1 method aligns with our proposed approach in the linear setting and why it fails in the nonlinear setting.




\subsection*{Comparison of U-hat1 and SIV Methods}

\subsubsection*{Equivalence under the Linear Outcome Model}

We first discuss the equivalence between the U-hat and SIV methods under the linear outcome model. Let \(\bm{Y} \in \mathbb{R}^n\) denote the vector of outcomes, \(\bm{X} \in \mathbb{R}^{n \times p}\) the matrix of treatments, and \(\widehat{\bm{U}} \in \mathbb{R}^{n \times q}\).  
The U-hat1 method regresses \(\bm{Y}\) on \(\bm{X}\) and \(\widehat{\bm{U}}\) simultaneously:
\begin{equation*}
\argmin_{\beta \in \mathbb{R}^p,\;\gamma \in \mathbb{R}^q} 
\|\bm{Y} - \bm{X}\beta - \widehat{\bm{U}}\gamma\|^2_2 
\quad \text{subject to } \|\beta\|_0 \leq k.
\end{equation*}
In contrast, the second-stage regression of the proposed method solves
\begin{equation*}
\argmin_{\beta \in \mathbb{R}^p} 
\|\bm{Y} - \widehat{\bm{X}}\beta\|^2_2 
\quad \text{subject to } \|\beta\|_0 \leq k.
\end{equation*}
It can be shown (see Section~\ref{sec:proof}) that
\begin{equation*}
\widehat{\bm{X}} 
= (I_n - \widehat{\bm{U}} (\widehat{\bm{U}}^\T \widehat{\bm{U}} )^{-1} \widehat{\bm{U}}^\T)\bm{X},
\end{equation*}
which coincides with the fitted residual of \(\bm{X}\) on \(\widehat{\bm{U}}\).  
Thus, the U-hat and SIV methods yield identical results under the linear outcome model.

\subsubsection*{Inequivalence under a Nonlinear Outcome Model}

We now explain why this equivalence does not extend to nonlinear outcome models. Consider the structural equation models:
\begin{flalign}
X &= \Lambda U + \epsilon_x, \label{eqn:model1_supp}\\
Y &= f(X;\beta) + g(U) + \epsilon_y. \label{eqn:model2_supp}
\end{flalign}

The U-hat1 method entails fitting the nonlinear regression
\[
\bm{Y} \sim f(\bm{X};\beta) + g(\widehat{\bm{U}}).
\]
Since \(U\) is unmeasured, the data contain no information about the function \(g\). One might impose a working model, such as \(g(U) = U\), leading to the regression  
\[
\bm{Y} \sim f(\bm{X};\beta) + \widehat{\bm{U}}, 
\]
or equivalently,  
\[
\bm{Y} \sim (I_n - \widehat{\bm{U}}(\widehat{\bm{U}}^\T \widehat{\bm{U}})^{-1} \widehat{\bm{U}}^\T) f(\bm{X};\beta) + \widehat{\bm{U}}.
\]

By contrast, the SIV method employs an estimating-equation approach. Since the synthetic instrument (SIV) is a linear combination of \(\epsilon_x\), it is independent of \(U\), and hence of any measurable function \(g(U)\):
\begin{equation}
\label{eqn:ind}
    SIV \ind U \;\;\Rightarrow\;\; SIV \ind g(U).
\end{equation}
Using this property, we construct the moment condition
\begin{equation}
\label{eqn:gmm_response}
\mathbb{E}[\,SIV \{Y - f(X;\beta)\}\,] = 0,
\end{equation}
which is equivalent to fitting the regression
\[
\bm{Y} \sim \bm{SIV}(\bm{SIV}^\T \bm{SIV})^{-1}\bm{SIV}^\T f(\bm{X};\beta).
\]

In general (see Section~\ref{sec:proof}),
\[
\bm{SIV}(\bm{SIV}^\T \bm{SIV})^{-1}\bm{SIV}^\T f(\bm{X};\beta) 
\;\neq\; (I_n - \widehat{\bm{U}}(\widehat{\bm{U}}^\T \widehat{\bm{U}})^{-1}\widehat{\bm{U}}^\T) f(\bm{X};\beta),
\]
whenever \(f\) is nonlinear. Hence, the equivalence established in the linear case does not hold in nonlinear models.

In our manuscript, we show numerically that the U-hat1 method is inconsistent under nonlinear outcome models, both when using a working specification \(g(U)=U\) and even in the unrealistic case where \(g(U)\) is correctly specified. In contrast, the proposed SIV method consistently estimates the treatment parameter \(\beta\).

\subsection{A Proposition and Its Proof}
\label{sec:proof}

\begin{proposition}
\label{prop:USIVXeq}
Consider the low-dimensional setting where $p < n$. Let $X \in \mathbb{R}^{n\times p}$ denote the design matrix of treatments, and let $f(X;\beta)\in \mathbb{R}^{n\times 1}$ be the vector of causal effects. Recall that
\begin{equation*}
    \begin{split}
        &SIV = X B_{\widehat \Lambda^\perp},\\
        &\widehat{U} = X\;\widehat{\text{Cov}}^{-1}(X)\widehat \Lambda.
    \end{split}
\end{equation*}
We have the following results:
\begin{align}
    &\{I_n - SIV(SIV^\T SIV)^{-1}SIV^\T - \widehat U(\widehat U^\T \widehat U)^{-1}\widehat U^\T \}X = 0, \label{eqn:decomp_residual}\\
    &\{I_n - SIV(SIV^\T SIV)^{-1}SIV^\T - \widehat U(\widehat U^\T \widehat U)^{-1}\widehat U^\T \}f(X;\beta) \neq 0 \quad \text{if $f(X;\beta)$ is nonlinear in $X$}. \label{eqn:noteq}
\end{align}
\end{proposition}

\paragraph{Proof of \eqref{eqn:decomp_residual}.}

Note that \((B_{\widehat{\Lambda}^\perp}, \widehat{\text{Cov}}^{-1}(X)\widehat{\Lambda}) \in \mathbb{R}^{p \times p}\) is invertible. The columns of \((B_{\widehat{\Lambda}^\perp}, \widehat{\text{Cov}}^{-1}(X)\widehat{\Lambda})\) therefore form a basis of \(\mathbb{R}^p\).  
Thus, any $\alpha \in \mathbb{R}^p$ can be written as
\[
\alpha = B_{\widehat \Lambda^\perp} \alpha_1 + \widehat{\text{Cov}}^{-1}(X)\widehat \Lambda \alpha_2,
\]
where $\alpha_1 \in \mathbb R^{p-q}$ and $\alpha_2 \in \mathbb R^q$. We now show that, for any \(\alpha \in \mathbb{R}^p\),
\[
\{I_n - SIV(SIV^\top SIV)^{-1}SIV^\top - \widehat{U}(\widehat{U}^\top \widehat{U})^{-1}\widehat{U}^\top\}X\alpha = 0,
\]
which establishes \eqref{eqn:decomp_residual}.

For the first term of \eqref{eqn:decomp_residual}, we have
\[
X\alpha = X\{B_{\widehat \Lambda^\perp} \alpha_1 + \widehat{\text{Cov}}^{-1}(X)\widehat \Lambda \alpha_2\} 
= SIV\alpha_1 + \widehat U \alpha_2.
\]

For the second term of \eqref{eqn:decomp_residual}, we compute
\begin{equation*}
\begin{split}
    &\{SIV(SIV^\T SIV)^{-1}SIV^\T  + \widehat U(\widehat U^\T \widehat U)^{-1}\widehat U^\T \}X \alpha \\
    &= \{SIV(SIV^\T SIV)^{-1}SIV^\T  + \widehat U(\widehat U^\T \widehat U)^{-1}\widehat U^\T \}\{SIV\alpha_1 + \widehat U \alpha_2\} \\
    &= SIV\alpha_1 + \widehat U \alpha_2,
\end{split}
\end{equation*}
where the last equality uses the orthogonality condition \(\widehat U^\T SIV = 0\).  
Thus, \eqref{eqn:decomp_residual} holds.

---

Before proving \eqref{eqn:noteq}, we establish the following claim:
\[
X(X^\T X)^{-1}X^\T 
= SIV(SIV^\T SIV)^{-1}SIV^\T + \widehat U(\widehat U^\T \widehat U)^{-1}\widehat U^\T.
\]

\paragraph{Proof of the claim.}  
Let \(A = X(X^\T X)^{-1}X^\T\) and  
\(B = SIV(SIV^\T SIV)^{-1}SIV^\T + \widehat U(\widehat U^\T \widehat U)^{-1}\widehat U^\T\).  
From \eqref{eqn:decomp_residual}, we have
\begin{equation}
\label{eqn:A=BA}
    (I_n-B)A = 0.
\end{equation}
Moreover,
\begin{align}
    &X(X^\T X)^{-1}X^\T SIV = SIV, \tag{a}\\
    &X(X^\T X)^{-1}X^\T \widehat{U} = \widehat{U}. \tag{b}
\end{align}
Combining (a) and (b), we obtain
\begin{equation}
\begin{split}
\label{eqn:resultofab}
    AB &= SIV(SIV^\T SIV)^{-1}SIV^\T + \widehat{U}(\widehat U^\T \widehat U)^{-1}\widehat U^\T \\
    &= B.
\end{split}
\end{equation}
Finally,
\begin{align*}
    (A-B)(A-B)^\T &= (A-B)(A-B) \\
    &= A + B - AB - BA \\
    &= A + B - A - B \\
    &= 0,
\end{align*}
where the first equality uses the symmetry of \(A\) and \(B\), the second follows from idempotence (\(A^2=A\), \(B^2=B\)), and the last follows from \eqref{eqn:A=BA} and \eqref{eqn:resultofab}.  
Thus, the claim is proved.

---

\paragraph{Proof of \eqref{eqn:noteq}.}  
Consider the decomposition of \(f(X;\beta) \in \mathbb{R}^{n\times 1}\):
\begin{equation}
\begin{split}
f(X;\beta) &= X(X^\T X)^{-1}X^\T f(X;\beta) + (I_n - X(X^\T X)^{-1}X^\T )f(X;\beta) \\
&= a + b,
\end{split}
\label{decom:f(xbeta)}
\end{equation}
where \(a\) and \(b\) denote the linear and nonlinear components of \(f(X;\beta)\), respectively.  

We first show that \(b \neq 0\). If \(b=0\), then
\[
f(X;\beta) = X(X^\T X)^{-1}X^\T f(X;\beta) = X\alpha,
\]
for some \(\alpha = (X^\T X)^{-1}X^\T f(X;\beta) \in \mathbb{R}^{p \times 1}\), implying that \(f(X;\beta)\) is linear in \(X\). This contradicts the assumption that \(f\) is nonlinear. Hence \(b \neq 0\).

Finally, using the claim above,
\[
(I_n - B)f(X;\beta) = (I_n - A)f(X;\beta) = b \neq 0,
\]
which proves \eqref{eqn:noteq}.

\subsection{A Necessary Condition}
\label{sesup:nec_condition}
To further explain why the U-hat1 method fail in the nonlinear setting, we have expanded Section~\ref{sesup:nec_condition} of the supplementary material to derive a necessary condition for identification and to provide additional analysis and simulation evidence. 
Specifically, we show that the U-hat1 method satisfy the identification condition when $f(X;\beta)$ is linear but may fail when $f(X;\beta)$ is nonlinear, even if the unmeasured confounder–outcome relationship $g(U)$ is linear. 
Below, we include the detailed derivation and results added to the supplementary material.

Specifically, we derive a necessary condition \eqref{eqn:identifiabilityequation_supp} that the U-hat1 method must satisfy to identify the causal parameter. We further demonstrate, through a counterexample, that this condition may fail when the treatment–outcome relationship is nonlinear, even if the unmeasured confounder–outcome relationship remains linear.

Suppose the data-generating mechanism is  
\(
Y = f(X;\beta^*) + g(U) + \epsilon_y,
\)  
where $\beta^*$ is the true parameter of interest with $\|\beta^*\|_0 = s$, and $f(X;\beta)$ denotes the causal function parameterized by $\beta$ (e.g., $\exp(X^\top \beta)$). We assume that the functional form of $f$ is known.  

We focus on the population version of the U-hat1 method, where $\widehat{\Lambda}$ and $\widehat{\text{Cov}}^{-1}(X)$ are replaced by their population counterparts, $\Lambda$ and $\text{Cov}^{-1}(X)$, respectively. We further assume that the sparsity level of $\beta^*$ is known to be $s$, and that the U-hat1 method is optimized under the constraint $\|\beta\|_0 = s$. The population version of the U-hat1 method is defined as  
\begin{equation}
\label{eqn:Uhat_popu_supp}
\widehat{U} := \Lambda^\top \text{Cov}^{-1}(X) X, 
\quad 
(\widehat{\beta}, \widehat{\gamma})
= \argmin_{\|\beta\|_0 = s,\;\gamma \in \mathbb{R}^p} \; 
\mathbb{E}\!\left[ \left( Y - f(X;\beta) - \widehat{U}^\top \gamma \right)^2 \right].
\end{equation}

To analyze the optimization problem of the U-hat1 method, we define the residuals $f_r$ and $Y_r$ as follows:  
\begin{equation}
\label{eqn:FrYr_a_supp}
f_r(X;\beta) = f(X;\beta) - \eta_f \widehat{U}, 
\quad 
Y_r = Y - \eta_Y \widehat{U},
\end{equation}
where  
\[
\eta_f = \text{Cov}\{f(X;\beta),\widehat{U}\}\,\text{Cov}^{-1}(\widehat{U})\in\mathbb{R}^{1\times q}, 
\quad 
\eta_Y = \text{Cov}(Y,\widehat{U})\,\text{Cov}^{-1}(\widehat{U})\in\mathbb{R}^{1\times q}.
\]  

Using \eqref{eqn:FrYr_a_supp}, we obtain  
\begin{equation}
\label{eqn:stepwise_regression_supp}
\begin{split}
\mathbb{E}\!\left[ \big( Y - f(X;\beta) - \widehat{U}^\top \gamma \big)^2 \right]
&= \mathbb{E}\!\left[ \big( Y_r - f_r(X;\beta) + \widehat{U}^\top (\eta_Y^\T - \eta_f^\T - \gamma) \big)^2 \right] \\
&= \mathbb{E}\!\left[ \big( Y_r - f_r(X;\beta) \big)^2 \right] 
+ \mathbb{E}\!\left[ \big( \widehat{U}^\top (\eta_Y - \eta_f - \gamma) \big)^2 \right],
\end{split}
\end{equation}
where the last equality follows from the orthogonality conditions  
$\text{Cov}(\widehat{U},Y_r) = 0$ and $\text{Cov}(\widehat{U},f_r(X;\beta))=0$.  

Recall that  
\begin{equation}
\label{eqn:Yr_supp}
\begin{split}
Y_r &= Y - \text{Cov}(Y,\widehat{U}) \, \text{Cov}^{-1}(\widehat{U}) \, \widehat{U} \\
&= \big(f(X;\beta^*) + g(U) + \epsilon_y\big) - \text{Cov}\!\big(f(X;\beta^*)+g(U),\widehat{U}\big)\,\text{Cov}^{-1}(\widehat{U}) \,\widehat{U} \\
&= f_r(X;\beta^*) + \widetilde{\epsilon}_y,
\end{split}
\end{equation}
where  
\[
\widetilde{\epsilon}_y := \epsilon_y + g(U) - \text{Cov}\{g(U),\widehat{U}\}\,\text{Cov}^{-1}(\widehat{U}) \,\widehat{U}.
\]  

We can further simplify $\widetilde{\epsilon}_y$:
\begin{equation}
\begin{split}
\label{eqn:simplifiedepsilon_y_supp}
\widetilde{\epsilon}_y &= \epsilon_y + g(U) - \text{Cov}(g(U), \Lambda^\top \text{Cov}^{-1}(X) X) \, \text{Cov}^{-1}(\widehat{U}) \, \Lambda^\top \text{Cov}^{-1}(X) X \\
&= \epsilon_y + g(U) - \text{Cov}(g(U), U) \, \Lambda^\top \text{Cov}^{-1}(X) X,
\end{split}
\end{equation}
where the simplification uses $X = \Lambda U+\epsilon_x$, $U \perp \epsilon_x$, and the linearity of covariance.

Equations \eqref{eqn:stepwise_regression_supp} and \eqref{eqn:Yr_supp} suggest that the optimization problem \eqref{eqn:Uhat_popu_supp} can be rewritten as
\begin{equation}
\label{eqn:optimization_reply_to_reviewer_supp}
    \widehat{\beta} = \argmin_{\|\beta\|_0 = s} \; 
    \mathbb{E}\!\left[ \left( Y_r - f_r(X;\beta) \right)^2 \right] 
    = \argmin_{\|\beta\|_0 = s} \; 
    \mathbb{E}\!\left[ \left( f_r(X;\beta^*) + \widetilde{\epsilon}_y - f_r(X;\beta) \right)^2 \right].
\end{equation}

A necessary condition for $\widehat{\beta} = \beta^*$ is  
\begin{equation}
\label{eqn:identifiabilityequation_supp}
\mathbb{E} \left[ \widetilde{\epsilon}_y \, 
\frac{\partial f_r(X;\beta)}{\partial \beta} \bigg|_{\beta = \beta^*} \right] = 0.
\end{equation}

If \eqref{eqn:identifiabilityequation_supp} is violated, then there exists a $\widetilde{\beta}$ in the neighborhood of $\beta^*$ such that $\widetilde{\beta}$ achieves a smaller loss than $\beta^*$, in the sense that
\[
\mathbb{E}\!\left[ \left( Y_r - f_r(X;\widetilde\beta) \right)^2 \right] 
< \mathbb{E}\!\left[ \left( Y_r - f_r(X;\beta^*) \right)^2 \right].
\]

We analyze whether \eqref{eqn:identifiabilityequation_supp} holds in two scenarios:  
\begin{enumerate}
    \item[Case 1:] $f(X;\beta)$ is linear;  
    \item[Case 2:] $f(X;\beta)$ is nonlinear.
\end{enumerate}

\textbf{Case 1: $f(X;\beta) = X^\top \beta$ (linear).}  
When $f(X;\beta) = X^\top \beta$, condition \eqref{eqn:identifiabilityequation_supp} always holds, providing additional justification for why the U-hat1 method is valid in the linear setting, as discussed in Section~\ref{sec:Uhat} of the supplementary material. We now verify this condition explicitly.

Given $f(X;\beta) = X^\top \beta$ and $\widehat{U} = \Lambda^\top \text{Cov}^{-1}(X)X$, we obtain the following expression after a straightforward (though tedious) calculation:
\begin{equation}
\begin{split}
f_r(X;\beta) 
&= X^\top \beta - \beta^\top \Lambda (\Lambda^\top \text{Cov}^{-1}(X) \Lambda)^{-1} \Lambda^\top \text{Cov}^{-1}(X) X \\
&= X_r^\top \beta,
\end{split}
\end{equation}
where 
\[
X_r = X - \Lambda (\Lambda^\top \text{Cov}^{-1}(X) \Lambda)^{-1} \Lambda^\top \text{Cov}^{-1}(X) X.
\]

The optimization problem \eqref{eqn:optimization_reply_to_reviewer_supp} then becomes
\[
\argmin_{\beta \in \mathbb{R}^p,\, \|\beta\|_0 = s} \; \mathbb{E}\!\left\{ X_r^\top \beta^* + \widetilde{\epsilon}_y - X_r^\top \beta \right\}^2,
\]
which is a least squares problem. In this setting, $\mathbb{E}[\widetilde{\epsilon}_y X_r] = 0$ is both \textbf{necessary and sufficient} for $\widehat{\beta} = \beta^*$. Moreover, since 
\(
\partial f_r / \partial \beta = X_r,
\)
condition \eqref{eqn:identifiabilityequation_supp} reduces to
\(
\mathbb{E} \big[ \widetilde{\epsilon}_y X_r \big] = 0.
\)

We now verify that \eqref{eqn:identifiabilityequation_supp} holds in this case through direct calculation:
\begin{equation} 
\label{eqn:suplinear_supp} 
\begin{split} 
\mathbb{E}(\widetilde{\epsilon}_yX_r)&=\text{Cov}\!\left(\widetilde{\epsilon}_y, X - \Lambda( \Lambda^\top \text{Cov}^{-1}(X)\Lambda)^{-1}\Lambda^\top \text{Cov}^{-1}(X)X\right) \\
&= \text{Cov}(g(U),X) - \text{Cov}(g(U),U)\Lambda^\top \text{Cov}^{-1}(X)\text{Cov}(X) \\
& \quad - \text{Cov}(g(U),X)\text{Cov}^{-1}(X)\Lambda( \Lambda^\top \text{Cov}^{-1}(X)\Lambda)^{-1}\Lambda^\top \\
& \quad + \text{Cov}(g(U), U )\Lambda^\top \text{Cov}^{-1}(X)\text{Cov}(X)\text{Cov}^{-1}(X)\Lambda( \Lambda^\top \text{Cov}^{-1}(X)\Lambda)^{-1}\Lambda^\top \\
&= 0.
\end{split} 
\end{equation} 
The first line follows from \eqref{eqn:simplifiedepsilon_y_supp}. In the second line, the first and second terms cancel, and the third and fourth terms also cancel. Thus, in the linear setting, condition \eqref{eqn:identifiabilityequation_supp} is satisfied. This necessity and sufficiency explain why the U-hat1 method estimates $\beta^*$ consistently when \(f\) is linear.

\textbf{Case 2: $f(X;\beta)$ nonlinear.}  
In this case, the necessary condition \eqref{eqn:identifiabilityequation_supp} may fail, even if $g(U)$ is linear in $U$. We illustrate this with a counterexample. We set $q = 1$, $s = 2$, and $p = 10$. The functions are defined as
\[
f(X;\beta) = \sum_{j=1}^{10} \cos^3(X_j)\beta_j, 
\quad g(U) = U\gamma.
\]  
We set $\Lambda_1 = \cdots = \Lambda_{10} = 1$ and $\gamma = 1$. The latent variables $U_{i,1}$ are i.i.d.\ from the uniform distribution $U(0,3)$ for $i = 1, \ldots, n$. The random errors $\epsilon_{x,i,j}$ are i.i.d.\ from the uniform distribution $U(0,5)$, and $\epsilon_{y,i}$ are i.i.d.\ from the standard uniform distribution. In this example, after a lengthy calculation, we obtain 
\[
\mathbb{E} \left[ \widetilde{\epsilon}_y \, 
\frac{\partial f_r(X;\beta)}{\partial \beta} \bigg|_{\beta = \beta^*} \right] 
= (-0.013, -0.013, \ldots, -0.013)^\top,
\]
indicating that the necessary condition \eqref{eqn:identifiabilityequation_supp} fails under a nonlinear $f$.

We further evaluate the finite-sample performance of our estimator and the U-hat1 method for sample sizes $n \in \{1000, \ldots, 5000\}$. All simulation results are based on 1000 Monte Carlo replications. The estimators we considered are
\begin{enumerate}
   \item[(1) \textbf{SIV:}] We obtain $\widehat{\beta}$ by solving 
   \[
    \widehat{\beta} =\argmin_{\beta \in \mathbb{R}^{10} } \left\| \bm{SIV}({{\bm{SIV}^\top \bm{SIV}}})^{-1}{{\bm{SIV}^\top (\bm{Y} - \cos^3(\bm{X})\beta)}} \right\|^2_2 
    \quad \text{subject to } \|\beta\|_0 \leq k.
   \]
   
   \item[(2) \textbf{U-hat:}] First, we obtain \(\widehat{\bm{U}} \in \mathbb{R}^{n \times q}\) using 
   \(\widehat{\bm{U}} = \bm{X} \widehat{\text{Cov}}(X)^{-1} \widehat{\Lambda}\). 
   Next, we obtain \(\widehat{\beta}\) by solving
   \[
   \widehat{\beta} = \underset{\beta \in \mathbb{R}^{10}, \, \gamma \in \mathbb{R}}{\argmin} 
   \|\bm{Y} - \cos^3(\bm X)\beta - \widehat{\bm{U}} \gamma\|_2^2 
   \quad \text{subject to} \quad \|\beta\|_0 \leq k.
   \]
\end{enumerate}

\begin{figure}[!ht]
    \centering
    \includegraphics[width=0.5\linewidth]{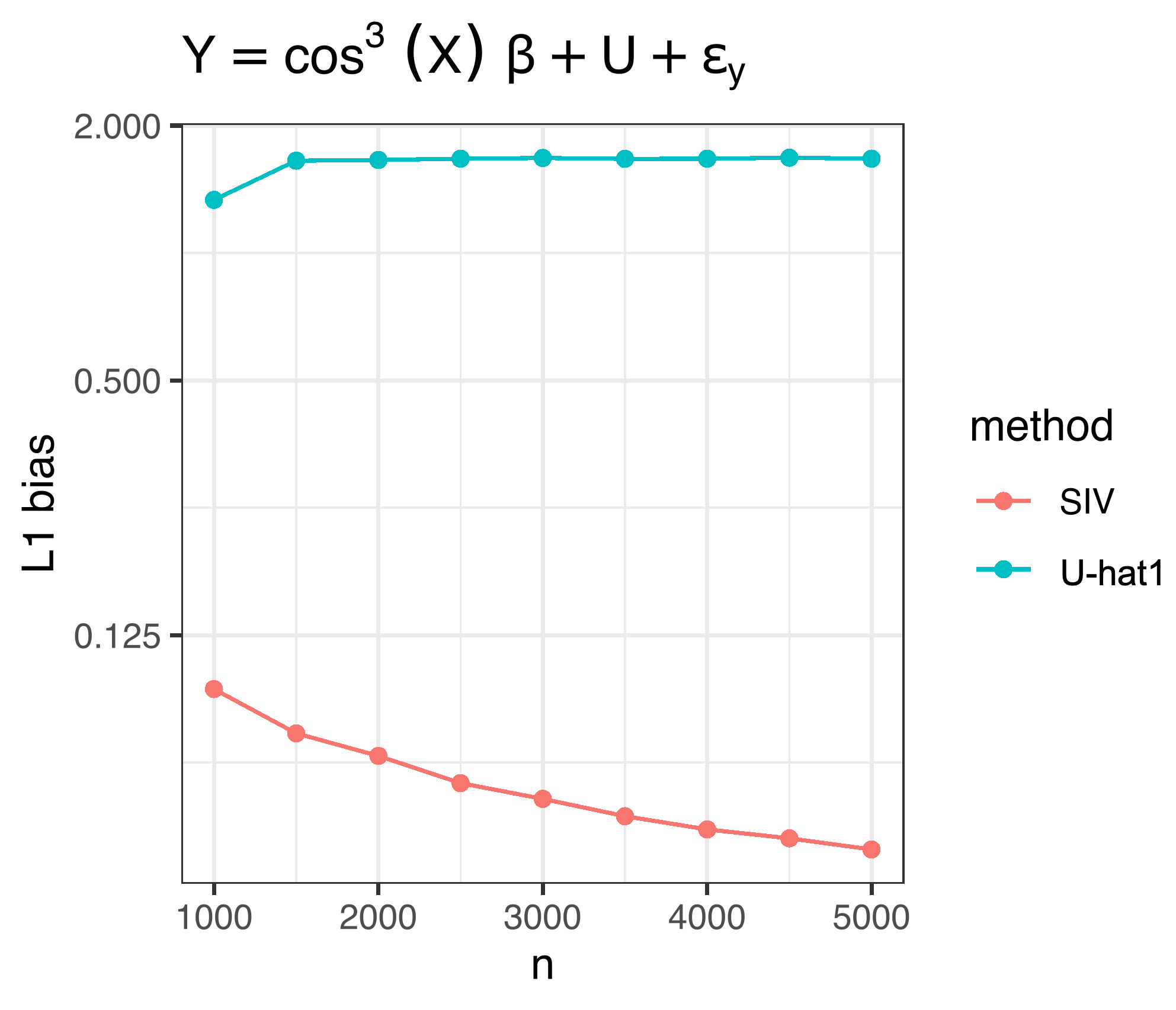}
    \caption{Performance of the U-hat1 and SIV methods under nonlinear $f(X; \beta)$.}
    \label{fig:uhatnew_supp}
\end{figure}

The simulation results in Figure~\ref{fig:uhatnew_supp} show that the U-hat1 method can fail even if $g(U)$ is linear, whereas our proposed method achieves consistent estimation. Moreover, the estimation error decreases with larger sample sizes.

\bibliographystyle{./asa}
\clearpage
\end{document}